\documentclass[a4paper, 12pt]{article}

\usepackage{a4wide}
\usepackage{graphicx}
\usepackage{amsmath}
\usepackage{amssymb}

\usepackage{subfigure}
\usepackage{cite}
\usepackage{xcolor}
\usepackage{soul}
\usepackage{cancel}
\usepackage[toc,page]{appendix}
\usepackage{mathtools}
\usepackage{hyperref}
\usepackage{enumitem}
\newcommand{\HB}{{\text{HB}}}
\newcommand{\half}{{\textstyle\frac{1}{2}}}
\newcommand{\threehalf}{{\textstyle\frac{3}{2}}}

\newcommand\mycom[2]{\genfrac{}{}{0pt}{}{#1}{#2}}

\renewcommand{\Im}{\mbox{Im\thinspace}}
\def\lsim{\mathrel{\rlap{\raise 2.5pt \hbox{$<$}}\lower 2.5pt\hbox{$\sim$}}}

\newcommand{\hc}{\text{h.c.}}

\numberwithin{equation}{section}

\newcommand\ztwo{\mathbb{Z}_2}

\allowdisplaybreaks

\begin{document}
\begin{titlepage}
\begin{center}

{\large \bf {CP-violation in the Weinberg 3HDM potential}}

\vskip 1cm

O. M. Ogreid,$^{a,}$\footnote{E-mail: omo@hvl.no}
P. Osland,$^{b,}$\footnote{E-mail: Per.Osland@uib.no} and
M. N. Rebelo$^{c,}$\footnote{E-mail: rebelo@tecnico.ulisboa.pt} 

\vspace{1.0cm}

$^{a}$Western Norway University of Applied Sciences,\\ Postboks 7030, N-5020 Bergen, 
Norway, \\
$^{b}$Department of Physics and Technology, University of Bergen, \\
Postboks 7803, N-5020  Bergen, Norway,\\
$^{c}$Centro de F\'isica Te\'orica de Part\'iculas -- CFTP and Dept de F\' \i sica\\
Instituto Superior T\'ecnico -- IST, Universidade de Lisboa, Av. Rovisco Pais, \\
P-1049-001 Lisboa, Portugal \\
\end{center}

\begin{abstract}

We explore the phenomenology of Weinberg's $\ztwo\times\ztwo$ symmetric three-Higgs-doublet potential, allowing for spontaneous violation of CP due to complex vacuum expectation values. An overview of all possible ways of satisfying the stationary-point conditions is given, with one, two or three non-vanishing vacuum expectation values, together with conditions for CP conservation in terms of basis invariants. All possible ways of satisfying the conditions for CP conservation are given. Scans of allowed parameter regions are given, together with measures of CP violation, in terms of the invariants. The light states identified in an earlier paper are further explored in terms of their CP-violating couplings. Loop-induced CP violation in $WWZ$ couplings, as well as charge-asymmetric scattering are also commented on.

\end{abstract}

\end{titlepage}

\setcounter{footnote}{0}

\section{Introduction}

Like any three-Higgs-doublet potential, the Weinberg 3HDM potential \cite{Weinberg:1976hu} has five neutral scalars and two pairs of charged ones. In a recent paper \cite{Plantey:2022jdg} we explored this potential, by scanning over its parameters, showing that if we require the existence of an SM-like state, $h_\text{SM}$ (based on the couplings to $WW$ and $ZZ$), then the potential tends to yield one or two neutral states below $m_\text{SM}=125~\text{GeV}$. This is caused by the fact that by breaking the $\ztwo\times\ztwo$ symmetry, one also breaks a U(1)$\times$U(1) symmetry, which is more restrictive than that of the Weinberg potential, and in parameter space is located close to the $\ztwo\times\ztwo$ symmetry.

Since the breaking of the continuous U(1)$\times$U(1) symmetry would lead to one or two massless Goldstone bosons, and since the $\ztwo\times\ztwo$-symmetric potential in some parameter domain is ``close'' to the U(1)$\times$U(1)-symmetric potential, some features will be similar. It was found \cite{Plantey:2022jdg} that the Weinberg potential, when we impose a near-SM coupling for one of the scalars to the electroweak gauge bosons, frequently yields one or two  scalars with masses below that of the SM Higgs.

While complex vacuum expectation values (vevs) of fields in the potential generally lead to CP violation, there are exceptions. We are identifying these by first establishing a set of basis-invariant quantities, having the property that if any one of them is non-zero, then CP is violated by the vacuum. Next, we identify every solution, for which all these invariants simultaneously vanish, and CP is conserved. Some of these cases allow for complex vevs.

Here, we  also further explore the CP-violating aspects of the potential, based on a parameter scan. 
After imposing the more important theoretical constraints, we also impose the more important experimental  constraints, namely those due to the electroweak precision observables ($S$, $T$, $U$),  $h_\text{SM}\to\gamma\gamma$, $\bar B\to X_s\gamma$ and the electron electric dipole moment (EDM).

Any 3HDM potential has two pairs of charged scalars. When the CP symmetry is broken in the scalar sector, as is the case here, there will be CP-violating processes in the scalar sector (not involving couplings to fermions), most easily illustrated in the charged-scalar sector.

With the neutral states labelled $h_1$, $h_2$, $h_3$, $h_4$ and $h_5$ (with increasing masses), we saw that the constraints imposed \cite{Plantey:2022jdg,Plantey:2022gwj} led to $h_2$ or $h_3$ predominantly being identified as the SM candidate. In the present paper we therefore focus on these two cases.

This paper is organized as follows.
In section~\ref{sect:notation} we define our notation, then, in section~\ref{sect:stationary-point} we present an overview of different ways of satisfying the stationary-point equations, identifying a minimum that may be new to the literature. This solution conserves a $\ztwo$ symmetry, while violating CP. In section~\ref{sect-CP-invariants} we present invariants that must vanish for CP to be conserved. Then, in section~\ref{sect:yukawa} we give an overview of possible Yukawa structures.  Section~\ref{sect:constraints} presents an overview of the most relevant experimental constraints, those due to the electroweak precision observables, the di-gamma signal strength, $\bar B\to X_s\gamma$ and the electron EDM. Other constraints, like the neutron EDM \cite{Chupp:2017rkp} were not taken into account, although we acknowledge the possibility that such constraints could severely limit the allowed parameter space of the model. This is beyond the scope of the present work. In section~\ref{sect:CP-violation-invariants} we review the impact the constraints have on the CP-violating invariants, and in section~\ref{sect:CP-processes} we discuss more exotic CP-violating processes. Results, in terms of allowed parameter space, are given in  section~\ref{sect:light-states}, and concluding remarks in section~\ref{sect:conclusions}.
Special cases of CP conservation are presented in appendix~\ref{app:CP-conserved}.

\section{Notation and definitions}
\label{sect:notation}

We parametrize the three Higgs doublets after electroweak symmetry breaking as
\begin{equation} \label{Eq:field-decomp}
\phi_i = e^{i\theta_i}\begin{pmatrix}
\varphi_i^+\\
\frac{1}{\sqrt{2}}(v_i+\eta_i+i\chi_i)
\end{pmatrix},\quad i=1,2,3.
\end{equation}
Here, the $v_i$ are real.
Imposing two $\ztwo$ symmetries, we automatically get a third one. The doublets may then be assigned the $\ztwo \times \ztwo \times \ztwo$ parities
\begin{equation} \label{Eq:Z2-charges}
\phi_1 : (+1,+1,-1) \qquad \phi_2 : (-1,+1,+1) \qquad \phi_3 : (+1,-1,+1).
\end{equation}
In this basis, which we will refer to as the symmetry basis, we write the most general $\ztwo\times\ztwo$-symmetric potential \cite{Weinberg:1976hu}, following the notation of Ivanov and Nishi \cite{Ivanov:2014doa}\footnote{Following tradition, we shall refer to the potential as being $\ztwo \times \ztwo$ symmetric.},
\begin{align} \label{Eq:potential}
 V=&-[m_{11}(\phi_1^\dagger \phi_1)+m_{22}(\phi_2^\dagger \phi_2)+m_{33}(\phi_3^\dagger \phi_3)] 
 +\lambda_{11}(\phi_1^\dagger\phi_1)^2 +\lambda_{22}(\phi_2^\dagger\phi_2)^2 +\lambda_{33}(\phi_3^\dagger\phi_3)^2  \nonumber \\
 &+\lambda_{12}(\phi_1^\dagger\phi_1)(\phi_2^\dagger\phi_2)
  +\lambda_{13}(\phi_1^\dagger\phi_1)(\phi_3^\dagger\phi_3)
  +\lambda_{23}(\phi_2^\dagger\phi_2)(\phi_3^\dagger\phi_3) \nonumber\\ 
  &+\lambda^\prime_{12}(\phi_1^\dagger\phi_2)(\phi_2^\dagger\phi_1)
  +\lambda^\prime_{13}(\phi_1^\dagger\phi_3)(\phi_3^\dagger\phi_1)
  +\lambda^\prime_{23}(\phi_2^\dagger\phi_3)(\phi_3^\dagger\phi_2) \nonumber\\
  &+\lambda_1\big[(\phi_2^\dagger\phi_3)^2+(\phi_3^\dagger\phi_2)^2\big]+\lambda_2\big[(\phi_3^\dagger\phi_1)^2+(\phi_1^\dagger\phi_3)^2\big]
+\lambda_3\big[(\phi_1^\dagger\phi_2)^2+(\phi_2^\dagger\phi_1)^2\big].
\end{align}
We shall assume real coefficients (i.e., we consider a CP-invariant potential), but allow for spontaneous CP violation via complex vevs \cite{Branco:1980sz} as discussed in Ref.~\cite{Plantey:2022jdg}.
Ivanov and Nishi  \cite{Ivanov:2014doa}  refer to a resulting symmetry $\ztwo\times\ztwo\times\ztwo^\ast$ due to the existence of explicit CP conservation, thus allowing for a transformation combining $\ztwo$ with the usual CP transformation.

The physical fields, $h_i$ and $h_k^+$, will be related to those of Eq.~(\ref{Eq:field-decomp}) via the rotation matrices $\mathcal O$ and $\mathcal U$ by
\begin{equation} \label{Eq:physicalfields}
h_i \equiv \sum_{m=1}^6 \mathcal O_{im}\, \varphi_m, \quad h_k^+ \equiv \sum_{n=1}^3 \mathcal U_{kn}\,\varphi_n^+,
\end{equation}
where $\varphi_m=(\eta_1,\eta_2,\eta_3,\chi_1,\chi_2,\chi_3)$. Here, $h_0 = G^0$ and $h_0^+=G^+$ are the Goldstone bosons, ${\cal O}$ is orthogonal and ${\cal U}$ is unitary. 

For some purposes the mixing matrices in the Higgs basis are very useful. We start by rewriting the neutral fields according to Eq.~(4.6) of Ref.~\cite{Plantey:2022jdg},
\begin{equation}
\begin{pmatrix}
\eta_1+i\chi_1\\
\eta_2+i\chi_2\\
\eta_3+i\chi_3
\end{pmatrix}
=\tilde R^\text{T}
\begin{pmatrix}
\eta_1^\HB+iG^0\\
\eta_2^\HB+i\chi_2^\HB\\
\eta_3^\HB+i\chi_3^\HB
\end{pmatrix},
\end{equation}
where the $\eta_i^\HB$ and $\chi_i^\HB$ refer to Higgs-basis fields, and (see Eq.~(2.15) of Ref.~\cite{Plantey:2022jdg})
\begin{equation}
\tilde R
=\frac{1}{vw}\begin{pmatrix}
|v_1|w & |v_2|w  & |v_3|w \\
-w^2 & |v_1||v_2| & |v_1||v_3|  \\
0 &-|v_3|v & |v_2|v
\end{pmatrix},
\end{equation}
with $v^2=v_1^2+v_2^2+v_3^2$ and $w^2=v_2^2+v_3^2$. Note that overall phases have been factored out in Eq.~(\ref{Eq:field-decomp}), where $\eta_i$ and $\chi_i$ are defined.
The Higgs-basis fields are related to the neutral mass eigenstates $h_j$ by Eqs.~(2.17) and (2.18) of Ref.~\cite{Plantey:2022jdg}, such that
\begin{subequations}
\begin{alignat}{3}
\eta_1&= \sum_{i=1}^3\tilde R_{i1}O_{ji} h_j, &\quad
\eta_2&= \sum_{i=1}^3\tilde R_{i2}O_{ji} h_j, &\quad
\eta_3&= \sum_{i=1}^3\tilde R_{i3}O_{ji} h_j,\\
\chi_1&= \sum_{i=2}^3\tilde R_{i1}O_{j\,2+i} h_j, &\quad
\chi_2&= \sum_{i=2}^3\tilde R_{i2}O_{j\,2+i} h_j, &\quad
\chi_3&= \sum_{i=2}^3\tilde R_{i3}O_{j\,2+i} h_j.
\end{alignat}
\end{subequations}

Likewise, for the charged fields, we have from Eq.~(A.2) of Ref.~\cite{Plantey:2022jdg},
\begin{equation}
\varphi_1^+=\sum_{i=2}^3\tilde R_{i1} (U^\dagger)_{i\text{-}1\,j}h_j^+, \quad
\varphi_2^+=\sum_{i=2}^3\tilde R_{i2} (U^\dagger)_{i\text{-}1\,j}h_j^+, \quad
\varphi_3^+=\sum_{i=2}^3\tilde R_{i3} (U^\dagger)_{i\text{-}1\,j}h_j^+.
\end{equation}

The couplings of physical neutral states to $ZZ$ or $WW$ are given by the first column of the Higgs-basis rotation matrix which can be expressed in terms of $\mathcal O$,
\begin{equation} \label{Eq:rot-matrx-55}
O_{i1}=\sum_{j=1}^3\tilde R_{1j}{\cal O}_{i+1\,j}=\sum_{j=1}^3\frac{v_j}{v}{\cal O}_{i+1\,j}.
\end{equation}

The $U(1)\times U(1)$ limit alluded to above, is obtained by setting $\lambda_1=\lambda_2=\lambda_3=0$. In that limit, with non-vanishing vevs breaking the symmetry, there are two massless states. For small, but nonzero values of $\lambda_1$, $\lambda_2$ and $\lambda_3$ the low-mass states which were the focus of our earlier work, may be considered analytic ``continuations'' of those massless ones.

\section{Solutions of stationary-point equations}
\label{sect:stationary-point}
Let $\{i,j,k\}$ be any permutation of $\{1,2,3\}$. We shall in the following interpret $\lambda_{ij}=\lambda_{ji}$  and $\lambda_{ij}'=\lambda_{ji}'$ whenever $i>j$. Here, we shall (for symmetry reasons) assume the most general form of the vevs, i.e.
\begin{equation} \label{Eq:vacuum-general}
	\phi_i = \frac{e^{i\theta_i}}{\sqrt{2}}\begin{pmatrix}
		0\\
		v_i
	\end{pmatrix},\quad i=1,2,3.
\end{equation}

Below, we list all solutions of the stationary-point equations for the real potential, indicating whether they conserve (CPC) or violate CP (CPV).\\

\noindent
{\it Solution 1 (CPC):}
\vspace*{-10pt}
\begin{eqnarray}
	v_i=0,\quad v_j=0,\quad m_{kk}=\lambda _{kk} v_k^2.
\end{eqnarray}

\noindent
{\it Solution 2 (CPV):}
\vspace*{-10pt}
\begin{eqnarray}
&v_i=0,\quad \lambda_i=0,\nonumber\\
&m_{jj}=\frac{1}{2} v_k^2 \left(\lambda _{jk}^{\prime}+\lambda _{jk}\right)+\lambda _{jj} v_j^2,\quad
m_{kk}=\frac{1}{2} v_j^2 \left(\lambda _{jk}^{\prime}+\lambda _{jk}\right)+\lambda _{kk} v_k^2.
\end{eqnarray}
This solution has a $\ztwo$ symmetry preserved by the vacuum. \\

\noindent
{\it Solution 3 (CPC):}
\vspace*{-10pt}
\begin{eqnarray}
&v_i=0,\quad \sin(\theta_k-\theta_j)=0,\nonumber\\
&m_{jj}=\frac{1}{2} v_k^2 \left(\lambda _{jk}^{\prime}+\lambda _{jk}+2 \lambda _i\right)+\lambda _{jj} v_j^2,\quad
m_{kk}=\frac{1}{2} v_j^2 \left(\lambda _{jk}^{\prime}+\lambda _{jk}+2\lambda _i\right)+\lambda _{kk} v_k^2.
\end{eqnarray}

\noindent
{\it Solution 4 (CPC):}
\vspace*{-10pt}
\begin{eqnarray}
	&v_i=0,\quad \cos(\theta_k-\theta_j)=0,\nonumber\\
	&m_{jj}=\frac{1}{2} v_k^2 \left(\lambda _{jk}^{\prime}+\lambda _{jk}-2 \lambda _i\right)+\lambda _{jj} v_j^2,\quad
	m_{kk}=\frac{1}{2} v_j^2 \left(\lambda _{jk}^{\prime}+\lambda _{jk}-2\lambda _i \right)+\lambda _{kk} v_k^2.
\end{eqnarray}

\noindent
{\it Solution 5 (CPC):}
\vspace*{-10pt}
\begin{eqnarray}
	&\sin (\theta_3-\theta_1)=\sin (\theta_2-\theta_1)=0,\nonumber\\
	&m_{11}=\frac{1}{2} \left(v_2^2 \left(\lambda _{12}^{\prime}+2 \lambda _3 +\lambda _{12}\right)+v_3^2 \left(\lambda _{13}^{\prime}+2 \lambda _2 +\lambda _{13}\right)+2 \lambda _{11} v_1^2\right),\nonumber\\
	&m_{22}=\frac{1}{2} \left(v_1^2 \left(\lambda _{12}^{\prime}+2 \lambda _3 +\lambda _{12}\right)+v_3^2 \left(\lambda _{23}^{\prime}+2 \lambda _1 +\lambda _{23}\right)+2 \lambda _{22} v_2^2\right),\nonumber\\
	&m_{33}=\frac{1}{2} \left(v_1^2 \left(\lambda _{13}^{\prime}+2 \lambda _2 +\lambda _{13}\right)+v_2^2 \left(\lambda _{23}^{\prime}+2 \lambda _1 +\lambda _{23}\right)+2 \lambda _{33} v_3^2\right).
\end{eqnarray}

\noindent
{\it Solution 6 (CPC):}
\vspace*{-10pt}
\begin{eqnarray}
	&\cos (\theta_3-\theta_1)=\sin (\theta_2-\theta_1)=0,\nonumber\\
	&m_{11}=\frac{1}{2} \left(v_2^2 \left(\lambda _{12}^{\prime}+2 \lambda _3 +\lambda _{12}\right)+v_3^2 \left(\lambda _{13}^{\prime}-2 \lambda _2 +\lambda _{13}\right)+2 \lambda _{11} v_1^2\right),\nonumber\\
	&m_{22}=\frac{1}{2} \left(v_1^2 \left(\lambda _{12}^{\prime}+2 \lambda _3 +\lambda _{12}\right)+v_3^2 \left(\lambda _{23}^{\prime}-2 \lambda _1 +\lambda _{23}\right)+2 \lambda _{22} v_2^2\right),\nonumber\\
	&m_{33}=\frac{1}{2} \left(v_1^2 \left(\lambda _{13}^{\prime}-2 \lambda _2 +\lambda _{13}\right)+v_2^2 \left(\lambda _{23}^{\prime}-2 \lambda _1 +\lambda _{23}\right)+2 \lambda _{33} v_3^2\right).
\end{eqnarray}

\noindent
{\it Solution 7 (CPC):}
\vspace*{-10pt}
\begin{eqnarray}
	&\sin (\theta_3-\theta_1)=\cos (\theta_2-\theta_1)=0,\nonumber\\
	&m_{11}=\frac{1}{2} \left(v_2^2 \left(\lambda _{12}^{\prime}-2 \lambda _3 +\lambda _{12}\right)+v_3^2 \left(\lambda _{13}^{\prime}+2 \lambda _2 +\lambda _{13}\right)+2 \lambda _{11} v_1^2\right),\nonumber\\
	&m_{22}=\frac{1}{2} \left(v_1^2 \left(\lambda _{12}^{\prime}-2 \lambda _3 +\lambda _{12}\right)+v_3^2 \left(\lambda _{23}^{\prime}-2 \lambda _1 +\lambda _{23}\right)+2 \lambda _{22} v_2^2\right),\nonumber\\
	&m_{33}=\frac{1}{2} \left(v_1^2 \left(\lambda _{13}^{\prime}+2 \lambda _2 +\lambda _{13}\right)+v_2^2 \left(\lambda _{23}^{\prime}-2 \lambda _1 +\lambda _{23}\right)+2 \lambda _{33} v_3^2\right).
\end{eqnarray}

\noindent
{\it Solution 8 (CPC):}
\vspace*{-10pt}
\begin{eqnarray}
	&\cos (\theta_3-\theta_1)=\cos (\theta_2-\theta_1)=0,\nonumber\\
	&m_{11}=\frac{1}{2} \left(v_2^2 \left(\lambda _{12}^{\prime}-2 \lambda _3 +\lambda _{12}\right)+v_3^2 \left(\lambda _{13}^{\prime}-2 \lambda _2 +\lambda _{13}\right)+2 \lambda _{11} v_1^2\right),\nonumber\\
	&m_{22}=\frac{1}{2} \left(v_1^2 \left(\lambda _{12}^{\prime}-2 \lambda _3 +\lambda _{12}\right)+v_3^2 \left(\lambda _{23}^{\prime}+2 \lambda _1 +\lambda _{23}\right)+2 \lambda _{22} v_2^2\right),\nonumber\\
	&m_{33}=\frac{1}{2} \left(v_1^2 \left(\lambda _{13}^{\prime}-2 \lambda _2 +\lambda _{13}\right)+v_2^2 \left(\lambda _{23}^{\prime}+2 \lambda _1 +\lambda _{23}\right)+2 \lambda _{33} v_3^2\right).
\end{eqnarray}

\noindent
{\it Solution 9 (CPC):}
\vspace*{-10pt}
\begin{eqnarray}
	&\sin(\theta_j-\theta_i)=0,\quad \lambda_i=\lambda_j=0,\nonumber\\
	&	m_{ii}=\frac{1}{2} \left(v_j^2 \left(\lambda _{ij}^\prime+\lambda _{ij}+2 \lambda _k\right)+v_k^2 \left(\lambda _{ik}^\prime+\lambda _{ik}\right)+2 \lambda _{ii} v_i^2\right),\nonumber\\
	&m_{jj}=\frac{1}{2} \left(v_i^2 \left(\lambda _{ij}^\prime+\lambda _{ij}+2 \lambda _k\right)+v_k^2 \left(\lambda _{jk}^\prime+\lambda _{jk}\right)+2 \lambda _{jj} v_j^2\right),\nonumber\\
	&m_{kk}=\frac{1}{2} \left(v_i^2 \left(\lambda _{ik}^\prime+\lambda _{ik}\right)+v_j^2 \left(\lambda _{jk}^\prime+\lambda _{jk}\right)+2 \lambda _{kk} v_k^2\right).
\end{eqnarray}

\noindent
{\it Solution 10 (CPC):}
\vspace*{-10pt}
\begin{eqnarray}
	&\cos(\theta_j-\theta_i)=0,\quad \lambda_i=\lambda_j=0,\nonumber\\
	&	m_{ii}=\frac{1}{2} \left(v_j^2 \left(\lambda _{ij}^\prime+\lambda _{ij}-2 \lambda _k\right)+v_k^2 \left(\lambda _{ik}^\prime+\lambda _{ik}\right)+2 \lambda _{ii} v_i^2\right),\nonumber\\
	&m_{jj}=\frac{1}{2} \left(v_i^2 \left(\lambda _{ij}^\prime+\lambda _{ij}-2 \lambda _k\right)+v_k^2 \left(\lambda _{jk}^\prime+\lambda _{jk}\right)+2 \lambda _{jj} v_j^2\right),\nonumber\\
	&m_{kk}=\frac{1}{2} \left(v_i^2 \left(\lambda _{ik}^\prime+\lambda _{ik}\right)+v_j^2 \left(\lambda _{jk}^\prime+\lambda _{jk}\right)+2 \lambda _{kk} v_k^2\right).
\end{eqnarray}

\noindent
{\it Solution 11 (CPV):}
\vspace*{-10pt}
\begin{eqnarray} \label{Eq:sol-12}
	&m_{11}=\frac{1}{2} \left(v_2^2 \left(\lambda _{12}^\prime+\lambda _{12}\right)+v_3^2 \left(\lambda _{13}^\prime+\lambda _{13}\right)+\frac{2 \lambda _1 v_2^2 v_3^2 \sin ^2 2 \left(\theta _2-\theta _3\right) }{v_1^2\sin 2\left( \theta _1-\theta _2\right) \sin 2 \left(\theta _1-\theta _3\right)}+2 \lambda _{11} v_1^2\right),\nonumber\\
	&m_{22}=\frac{1}{2} \left(v_1^2 \left(\lambda _{12}^\prime+\lambda _{12}\right)+v_3^2 \left(\lambda _{23}^\prime+\lambda _{23}\right)+2 \lambda _{22} v_2^2\right)+\lambda _1 v_3^2 \frac{\sin 2 \left(\theta _1-\theta _3\right) }{\sin 2 \left(\theta _1-\theta _2\right)},\nonumber\\
	&m_{33}=\frac{1}{2} \left(v_1^2 \left(\lambda _{13}^\prime+\lambda _{13}\right)+v_2^2 \left(\lambda _{23}^\prime+\lambda _{23}\right)+2 \lambda _{33} v_3^2\right)+\lambda _1 v_2^2 \frac{\sin 2 \left(\theta _1-\theta _2\right) }{\sin 2 \left(\theta _1-\theta _3\right)},\nonumber\\
	&\lambda _3=\frac{\lambda _1 v_3^2 \sin 2 \left(\theta _2-\theta _3\right) }{v_1^2\sin 2 \left(\theta _1-\theta _2\right)},\quad
	\lambda _2=-\frac{\lambda _1 v_2^2 \sin 2 \left(\theta _2-\theta _3\right) }{v_1^2\sin 2 \left(\theta _1-\theta _3\right)}.
\end{eqnarray}
The latter is the solution identified by Branco \cite{Branco:1980sz} and studied in our earlier work \cite{Plantey:2022jdg,Plantey:2022gwj}, as well as in the bulk of the present paper.

One can understand the existence of the two types of solutions that allow for spontaneous CP violation by looking at the scalar potential given by Eq.~(\ref{Eq:potential}). Spontaneous CP violation can only occur if there are, for a real potential, complex vevs. In the case of $\lambda_i = 0$ and $v_i$ = 0 (for a single $i$) there is only one relative phase in the vacuum which gives rise to a complex $\lambda_j$ in the basis were all vevs are real. The minimisation conditions for the vacuum phases in this case are trivially satisfied and therefore do not impose any additional constraints on the $\lambda$s, this is Solution~2. The most general case with all $\lambda_k \neq 0$ and a general complex vacuum corresponds to Solution~11 where the minimisation conditions for the two independent vacuum phases lead to the possibility of expressing $\lambda_i$ and $\lambda_j$ in terms of $\lambda_k$ and vevs, i.e., with $v_i\neq0$. We shall mainly work in this framework.

Solution 2 requires one $\lambda_i\neq0$. At first signt this would indicate an enlargement of the symmetry of the potential. By comparing to refs.~\cite{Ivanov:2011ae,deMedeirosVarzielas:2019rrp,Darvishi:2019dbh}, however, we were unable to find anything similar. If, by putting $\lambda_i=0$, there is an enhancement of the symmetry of the potential, this will have physical implications. If not, this condition is expected to go away at one-loop level.

\section{Conditions for CP conservation}
\label{sect-CP-invariants}

With complex vevs, the potential will in general violate CP. However, there are special cases in which CP is conserved, for example when the phases satisfy certain relations. These special cases can be identified from the study of CP-odd invariants. When they all vanish, CP is conserved.
The invariants are expressed in terms of the tensors $Z$, $Y$ and $\hat V$, where $Z$ and $Y$ are defined by the expansion 
\cite{Branco:1999fs}  (see also Refs.~\cite{Mendez:1991gp,Lavoura:1994fv,Botella:1994cs,Branco:2005em,Gunion:2005ja,Davidson:2005cw,Haber:2006ue})
\begin{equation}
V=Y_{ab}(\phi_a^\dagger\phi_b)+\half Z_{abcd}(\phi_a^\dagger\phi_b)(\phi_c^\dagger\phi_d),
\end{equation}
whereas the vacuum is represented by $\hat V$,
\begin{equation}
\hat V_{ab}=\frac{v_ae^{i\theta_a}}{v}\frac{v_be^{-i\theta_b}}{v}.
\end{equation}
Here, the indices $a$, $b$, $c, \ldots$ can take the values 1,2,3, identifying the three fields.
In terms of this notation, we can state the following \\

\noindent
{\bf Theorem:}\\
Whenever the stationary-point equations are satisfied, the real\footnote{By ``real'' we mean that all coefficients are real.} $\ztwo\times\ztwo$-symmetric three-Higgs-doublet potential conserves CP if and only if the following 15 CP-odd invariants all vanish:
\begin{subequations} \label{Eq:Weinberg-invariants}
\begin{align}
J_1&=\Im\{\hat V_{{ac}} \hat V_{{be}} Z_{{cadf}} Z_{{edfg}} Z_{{gbhh}}\},\\
J_2&=\Im\{\hat V_{{ac}} \hat V_{{be}} Z_{{cadf}} Z_{{edfg}} Z_{{ghhb}}\},\\
J_3&=\Im\{\hat V_{{ac}} \hat V_{{be}} Z_{{cadf}} Z_{{egfd}} Z_{{gbhh}}\},\\
J_4&=\Im\{\hat V_{{ac}} \hat V_{{bd}} Z_{{cedg}} Z_{{eafh}} Z_{{gbhf}}\},\\
J_5&=\Im\{\hat V_{{ac}} \hat V_{{bd}} Z_{{cedg}} Z_{{ehfa}} Z_{{gfhb}}\},\\
J_6&=\Im\{\hat V_{{ac}} \hat V_{{bd}} Z_{{cedf}} Z_{{eafg}} Z_{{gbhh}}\},\\
J_7&=\Im\{\hat V_{{ad}} \hat V_{{be}} \hat V_{{cf}} Z_{{daeh}} Z_{{fbgi}} Z_{{hcig}}\},\\
J_8&=\Im\{\hat V_{{ad}} \hat V_{{be}} \hat V_{{cf}} Z_{{daeh}} Z_{{figb}} Z_{{hgic}}\},\\
J_9&=\Im\{\hat V_{{ad}} \hat V_{{be}} \hat V_{{cf}} Z_{{daeg}} Z_{{fbgh}} Z_{{hcii}}\},\\
J_{10}&=\Im\{\hat V_{{ad}} \hat V_{{be}} \hat V_{{cf}} Z_{{daeg}} Z_{{fhgi}} Z_{{hbic}}\},\\
J_{11}&=\Im\{\hat V_{{ac}} \hat V_{{be}} Z_{{cadg}} Z_{{edff}} Z_{{gihh}} Z_{{ibjj}}\},\\
J_{12}&=\Im\{\hat V_{{ac}} \hat V_{{be}} Z_{{cadg}} Z_{{effd}} Z_{{ghhi}} Z_{{ijjb}}\},\\
J_{13}&=\Im\{\hat V_{{ac}} \hat V_{{be}} Z_{{cadf}} Z_{{edfg}} Z_{{gihj}} Z_{{ibjh}}\},\\
J_{14}&=\Im\{\hat V_{{ac}} \hat V_{{bd}} Z_{{cedf}} Z_{{eafg}} Z_{{gihj}} Z_{{ibjh}}\},\\
v^6J_{15}&=\Im\{\hat V_{{ac}} \hat V_{{bd}} Y_{{cf}} Y_{{dg}} Y_{{ea}} Z_{{fbge}}\}.
\end{align}
\end{subequations}

The following remarks are important:
\begin{enumerate}
\item
This result relies on the fact that the stationary-point equations are all satisfied (in other words: after electroweak symmetry breaking has taken place). There are several ways to solve the stationary-point equations (see section~\ref{sect:stationary-point}). The above result is valid for all possible ways they can be solved.

\item
There are no redundant invariants in the set of 15 invariants listed. For each invariant $J_i$, we have located a point in parameter-space where $J_i\neq0$ while the remaining 14 invariants vanish.

\item
One could imagine picking a set of CP-odd invariants different from the one presented here, whose simultaneous vanishing would be equivalent to the model being CP-conserving.
We do not know if there could exist such a set where the total number of invariants is less than 15.

\item
Although we have proven that the vanishing of the 15 CP-odd invariants $J_i$ is equivalent to having a CP conserving model, that does not mean that any CP-odd quantity can be written as a linear combination of the 15  $J_i$. This is also known from the 2HDM, where $\Im J_1=\Im J_2=\Im J_{30}=0$ is equivalent to having a CP conserving 2HDM \cite{Grzadkowski:2014ada}. Yet, in the 2HDM, it is not possible to write all CP-odd quantities as linear combinations of those three CP-odd invariants. Therefore, in \cite{Grzadkowski:2014ada}, the authors introduced additional CP-odd invariants in their eqs.~(4.6) and (4.14). Those additional invariants vanish simultaneously with the three CP-odd invariants that guarantee a CP-conserving model, yet they are redundant in order to guarantee CP conservation, but they were introduced in order to express the CP-odd quantities studied there in terms of CP-odd invariants. In a later work \cite{Grzadkowski:2016szj}, the same authors were able to express all CP-odd quantities as linear combinations just by adding one extra invariant $\Im J_{11}$ to the set of the original three CP-odd invariants.
\\ \hspace*{6mm}
This can be understood as follows. The vanishing of \{$\Im J_1, \Im J_2, \Im J_{30}$\} guarantees a CP-conserving model, but not all CP-odd quantities can be written as linear combinations of these three. The set \{$\Im J_1, \Im J_2, \Im J_{30}, \Im J_{11}$\} constitutes what may be called a ``linear algebraic basis" for CP-odd quantities in the 2HDM, making it possible to write all CP-odd quantities as linear combinations of these four CP-odd invariants. The same thing happens for the model we study here. While the simultaneous vanishing of the 15 $J_i$ is equivalent to CP conservation, we would need additional CP-odd invariants in order to express any CP-odd quantity of the model as a linear combination of CP-odd invariants. We would have to extend the set of the 15 $J_i$ with additional $J_i$ (number unknown) in order to get a ``linear algebraic basis" for CP-odd invariants of the model. This is, however, beyond the scope of the present work.
\end{enumerate}

In order to arrive at the result presented in this theorem, we constructed a large amount of CP-odd invariants by contracting indices among $\hat{V}$-, $Y$- and $Z$-tensors. We started by first carefully picking some invariants that were algebraically simple enough so that the equations that resulted from demanding the vanishing of the invariants were possible to solve algebraically. Through a process of trial and error, we carefully added more invariants to the set, until we finally arrived at the set of the 15 above invariants. Demanding the simultaneous vanishing of these 15 invariants, gives us a set of 15 equations,
\begin{equation} \label{Eq:the_equations}
J_i=0, \quad i=1,\ldots 15.
\end{equation}
It turns out that it is possible to solve this set of equations algebraically. All possible solutions of this set of equations are presented in appendix~\ref{app:CP-conserved}. We ended up with a total of 80 different solutions (counting all possible permutations) to the set of equations. For each of these solutions, we were able to show that the solution implies a CP-conserving model. This was done by explicit construction of a basis change that renders both the parameters of the potential and the vacuum real, implying CP conservation. These basis changes are also presented in appendix~\ref{app:CP-conserved} along with the solutions.

In summary, using Mathematica \cite{Mathematica}, we have shown that the fifteen conditions of Eq.~(\ref{Eq:the_equations}) are enough to force the potential to be real in a basis were the vevs are also real.

We note that all invariants involve two or three factors $\hat V$, each being a product of a vev with another that is complex conjugated.
These invariants also involve coefficients of the potential. As examples, for $\{i,j,k\}\in\{1,2,3\}$,
\begin{subequations} \label{Eq:J_i-explicit}
\begin{align}
J_1&=\frac{-2 v_i^2 v_j^2}{v^4} \lambda_k\sin2(\theta_j-\theta_i)(\lambda_{ij}+\lambda_{ij}^\prime)
[2(\lambda_{ii}-\lambda_{jj})+\lambda_{ik}-\lambda_{jk}] + \text{permutations}, \\
J_2&=\frac{-2 v_i^2 v_j^2}{v^4} \lambda_k\sin2(\theta_j-\theta_i)(\lambda_{ij}+\lambda_{ij}^\prime)
[2(\lambda_{ii}-\lambda_{jj})+\lambda_{ik}^\prime-\lambda_{jk}^\prime] + \text{permutations}, \\
J_3&=\frac{-4 v_i^2 v_j^2}{v^4} \lambda_k\sin2(\theta_j-\theta_i)\,\lambda_{ij}^\prime\,
[2(\lambda_{ii}-\lambda_{jj})+\lambda_{ik}-\lambda_{jk}] + \text{permutations}.
\end{align}
\end{subequations}

For the $\ztwo\times\ztwo$-symmetric real potential there are no CP-odd invariants with only {\it one} $\hat V_{ac}$. 
This is due to the high degree of symmetry of the potential, as commented on in appendix~\ref{app:need-two-Vhat}.

Whenever there is a CP-violating quantity, such as for instance an asymmetry of the form
\begin{equation} \label{Eq:asymmetry}
{\cal A}_\text{ch}=\frac{\sigma(h_i^+h_i^-\to h_2^+h_1^-)-\sigma(h_i^+h_i^-\to h_2^-h_1^+)}{\sigma(h_i^+h_i^-\to h_2^+h_1^-)+\sigma(h_i^+h_i^-\to h_2^-h_1^+)},
\end{equation}
it will be possible to express it in terms of non-vanishing CP-odd invariants like the ones in Eq.~(\ref{Eq:Weinberg-invariants}). Even though vanishing of the latter is sufficient to guarantee that CP is conserved, a given CP violating quantity may be directly related to a CP-odd invariant not contained in this set.  In the CP-conserving limit all possible CP-odd invariants will vanish simultaneously.

Spontaneous CP violation of the type discussed here could show up in processes like this, via interference involving trilinear couplings of charged and neutral scalars, each trilinear vertex proportional to a complex vev.

As global measures of CP violation, we shall explore two quantities based on these invariants,
\begin{equation} \label{Eq:CP-inv-measure}
A_\text{sum}=\log_{10}\sum_{i=1}^{15} J_i^2, \quad
A_\text{max}=\log_{10}(\max_i J_i^2).
\end{equation}

CP violation is also possible with only two vevs being non-zero. An example is provided by Solution~2 of section~\ref{sect:stationary-point},
\begin{subequations}
\begin{align}
v_1 &=0, \quad \lambda_1=0, \\
m_{22}&=v_2^2\lambda_{22}+\half v_3^2(\lambda_{23}+\lambda_{23}^\prime),\\
m_{33}&=v_3^2\lambda_{33}+\half v_2^2(\lambda_{23}+\lambda_{23}^\prime).
\end{align}
\end{subequations}
In this case, we find five non-vanishing invariants. All of these contain the factor
\begin{equation}
J_0\equiv v_2 v_3 \lambda_2 \lambda_3 \sin2(\theta_2-\theta_3),
\end{equation}
 which would have to be non-zero for CP to be spontaneously violated. In particular, we note that CP violation requires $2(\theta_2-\theta_3)\neq n\pi$ for $n$ integer.
Including non-trivial factors, the non-vanishing invariants are proportional to
\begin{subequations} 
\begin{align}
J_4 &\propto J_0[\lambda_{12}^\prime-\lambda_{13}^\prime], \\
J_5 &\propto J_0[\lambda_{12}-\lambda_{13}], \\
J_6 &\propto J_0[\lambda_{12}-\lambda_{13}+2\lambda_{22}-2\lambda_{33}], \\
J_{10}&\propto J_0[2(\lambda_{22}v_2^2-\lambda_{33}v_3^2)+(\lambda_{23}+\lambda_{23}^\prime)(v_3^2-v_2^2)], \\
J_{14}&\propto J_0[\lambda_{12}^2-\lambda_{13}^2+\lambda_{12}^{\prime\,2}-\lambda_{13}^{\prime\,2}
-4(\lambda_2^2-\lambda_3^2)+4(\lambda_{22}^2-\lambda_{33}^2)].
\end{align}
\end{subequations}

Actually, this Solution~2 yields a massless state, but that can be avoided by adding a soft symmetry-breaking term and dropping the constraint $\lambda_1=0$.
\section{Yukawa sector}
\label{sect:yukawa}

With three scalar doublets, there are more possible Yukawa structures than in the familiar 2HDM.
Table~\ref{table:yukawa-types} lists all possible Yukawa structures of the $\ztwo \times \ztwo$-symmetric 3HDMs that respect Natural Flavour Conservation (NFC). In addition to the $\mathbb{Z}_2$-symmetric 2HDM-like structures, any 3HDM allows for a ``democratic" structure (often referred to as ``Type~Z'') where each kind of fermion species ($u$, $d$, $e$) couples to a different doublet \cite{Akeroyd:2016ssd,Logan:2020mdz,Boto:2021qgu}. The table does not take into account permutations of indices for different pairs of fields. This permutation symmetry may be broken by the vacuum.
With one more doublet, there are also more possibilities for inert doublets. 

\begin{table}[!htbp]
\centering
\begin{tabular}{|c|c|c|c|c|}
\hline
 & $u$ & $d$ & $e$ & Inert doublets \\
\hline
Type I-like & $\phi_1$ & $\phi_1$ & $\phi_1$ & at most 2 \\
\hline
Type II-like & $\phi_1$ & $\phi_2$ & $\phi_2$ & at most 1 \\
\hline
Lepton specific-like & $\phi_1$ & $\phi_1$ & $\phi_2$ & at most 1 \\
\hline
Flipped-like & $\phi_1$ & $\phi_2$ & $\phi_1$ & at most 1 \\
\hline
Type~Z & $\phi_1$ & $\phi_2$ & $\phi_3$ & none \\
\hline
\end{tabular}
\caption{The different NFC-respecting Yukawa structures for the $\ztwo \times \ztwo$-symmetric 3HDM.}
\label{table:yukawa-types}
\end{table}

In democratic models, at tree level\footnote{Scalars which are massless at tree level could acquire mass at loop level, unless protected by a symmetry\cite{Hernandez-Sanchez:2022dnn}.}, all three vevs must be non-zero in order for fermions to be massive whereas in other types one or two doublets may have a vanishing vev and thus lead to dark-matter candidates \cite{Hernandez-Sanchez:2020aop,Hernandez-Sanchez:2022dnn,Boto:2024tzp}.

Consider the generic Yukawa Lagrangian (the indices $a$, $b$ and $c$ do not need to be different)
\begin{equation} \label{Eq:yukawa}
- \mathcal L_Y =  \bar Q^0_L Y^u \tilde\phi_a u^0_R +  \bar Q^0_L Y^d \phi_b d^0_R 
+  \bar E^0_L Y^e \phi_c e^0_R + \hc,
\end{equation}
where the superscript ``$0$'' on fermion fields is a reminder that these are not necessarily mass eigenfields. Even though the vevs may be complex, the complex phases may be absorbed by a redefinition of the right-handed fermion fields and hence play no role in the Yukawa interactions. This is a generic feature of models with natural flavour conservation (NFC) \cite{Glashow:1976nt,Paschos:1976ay}. 

In addition to providing the fermion mass matrices
\begin{equation}
\frac{v_a}{\sqrt{2}} Y^f = m_f,
\end{equation}
this Lagrangian will determine the interactions between the fermions and the physical scalar fields of Eq.~(\ref{Eq:physicalfields}). Below, we write these interactions in the fermion physical basis where $m_f=\text{diag}(m_{f_1}, m_{f_2}, m_{f_3})$, with $f\in\{u, d, e\}$.

We will parametrize the interaction of fermions with neutral scalars as
\begin{equation} \label{Eq:Yuk-kappa}
-\mathcal L_Y  \supset \bar f\frac{m_f}{v}(\kappa^S_{h_iff} \pm i\kappa^P_{h_iff}\gamma_5)fh_i \quad i=1,...,5,
\end{equation}
with
\begin{equation} \label{Eq:Yuk-kappa-S-P}
\kappa^S_{h_iff} = \frac{v}{v_a}\mathcal O_{ia}, \quad \kappa^P_{h_iff} = \frac{v}{v_a}\mathcal O_{i\,a+3}.
\end{equation}
The ``$\pm$'' sign in Eq.~(\ref{Eq:Yuk-kappa}) should be interpreted as ``plus'' for $d$-type quarks (and charged leptons) and ``minus'' for $u$-type quarks, and stems from the complex conjugation associated with the $\tilde \phi$ factor in Eq.~(\ref{Eq:yukawa}).
Furthermore, ${\cal O}$ is the diagonalization matrix, defined in Eq.~(\ref{Eq:physicalfields}).

For the purpose of studying the $b\to s\gamma$ constraint, it is convenient to write the interaction of the quarks with 
the charged scalars in the notation of Borzumati and Greub \cite{Borzumati:1998tg},
\begin{equation} \label{Eq:Yuk-BG}
- \mathcal L_Y  \supset \frac{g}{\sqrt2\, m_W} \, 
\sum_{i,j,k}V_{ij}\Bigl\{\bar u_i [Y_k\, m_{u_i}P_L+X_k\, m_{d_j}P_R]d_j h_k^++\hc\Bigr\}.
\end{equation}
The notation of Ciuchini, Degrassi, Gambino and Giudice \cite{Ciuchini:1997xe,Hermann:2012fc} is also often used:
\begin{equation} \label{Eq:Yuk-A_L-A_R}
- \mathcal L_Y  \supset (2\sqrt{2}G_\text{F})^{1/2} \, 
\sum_{i,j,k}V_{ij}\Bigl\{\bar u_i [A_u^k\, m_{u_i}P_L-A_d^k\, m_{d_j}P_R]d_j h_k^++\hc\Bigr\}.
\end{equation}
Here, 
\begin{equation}
\frac{g}{\sqrt2\, m_W}=\frac{\sqrt2}{v}=(2\sqrt{2}G_\text{F})^{1/2},
\end{equation}
$G_\text{F}$ is the Fermi constant, and $V_{ij}$ the appropriate element of the CKM matrix. The coefficients $X_k$ and $Y_k$ do not depend either on quark mass or on the CKM matrix element. However, they will depend on which type of Yukawa couplings we consider, as specified in the following. The subscript $k$ refers to the charged scalar, $k=1,2$.

The two notations (\ref{Eq:Yuk-BG}) and (\ref{Eq:Yuk-A_L-A_R}) are related by
\begin{equation}
X_k=-A_d^k, \qquad Y_k=A_u^k.
\end{equation}
\subsection{Type~I Yukawa couplings}
With both $u$- and $d$-quarks (as well as charged leptons) coupling to $\phi_a$, we find
\begin{equation}
Y_k=\frac{v}{v_a}\mathcal U_{ka}^\ast, \quad
X_k=-\frac{v}{v_a}\mathcal U_{ka}, \quad\text{no sum over }a.
\end{equation}
For comparison, in the familiar 2HDM with Type~I Yukawa couplings we have
\begin{equation}
Y=\frac{1}{\tan\beta}, \quad X=-Y.
\end{equation}
The 2HDM values are obtained if we consider $v_a\to v_2 = v\sin\beta$, and $\mathcal U_{ka}\to\cos\beta$, with $\beta$ the rotation angle identifying the Goldstone boson in the 2HDM.
\subsection{Type~II Yukawa couplings}
With $u$- and $d$-quarks (as well as charged leptons) coupling to $\phi_a$ and $\phi_b$, respectively, we find
\begin{equation}
Y_k=\frac{v}{v_a}\mathcal U_{ka}^\ast, \quad 
X_k=\frac{v}{v_b}\mathcal U_{kb}, \quad \text{no sum over $a$, with }b\neq a.
\end{equation}
For fixed $v_a$ and $v_b$, $|Y_k|$ and $|X_k|$ will be proportional.
In the familiar 2HDM we have
\begin{equation}
Y=\frac{1}{X}=\frac{1}{\tan\beta}.
\end{equation}

\subsection{Type~Z Yukawa couplings}
With $u$- and $d$-quarks coupling to $\phi_a$ and $\phi_b$, respectively, and charged leptons coupling to $\phi_c$, we find $Y_k$ and $X_k$ like for Type~II, and for charged leptons
\begin{align} \label{Eq:Z_k}
Z_k=\frac{v}{v_c}\mathcal U_{kc}, \quad \text{no sum over }c\neq a,b.
\end{align}

\begin{figure}[htb]
\begin{center}
\includegraphics[scale=0.30]{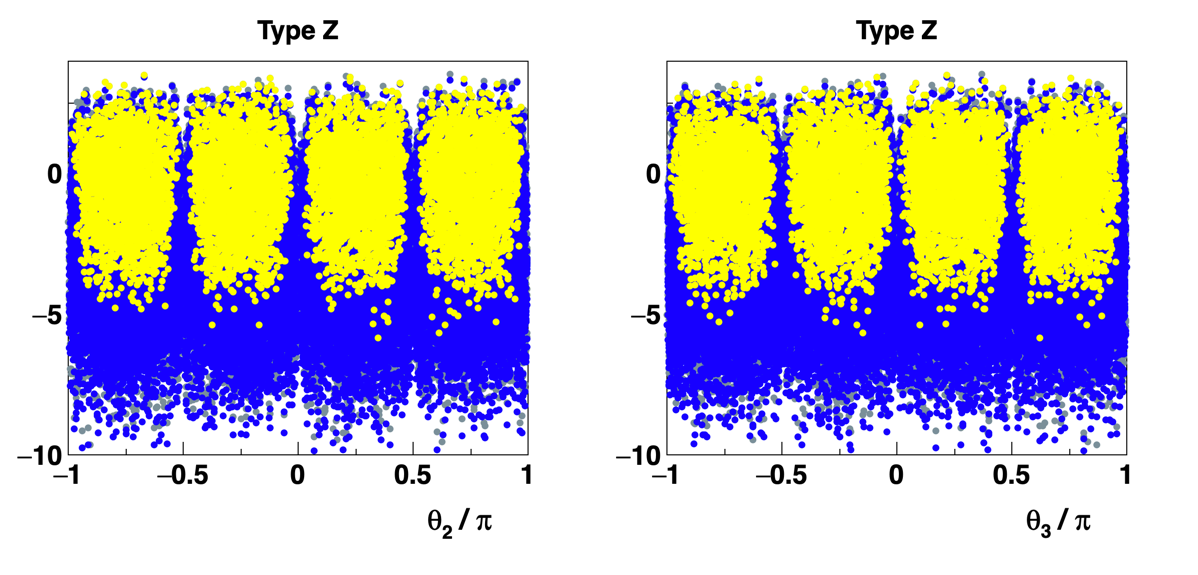}
\end{center}
\vspace*{-8mm}
\caption{Scatter plots of logarithms of the sum (and maxima) of the squares of the 15 invariants of Eq.~(\ref{Eq:Weinberg-invariants}) vs $\theta_2$ and $\theta_3$. 
If low masses are allowed (blue) the invariants can become quite small.
Note that the two bulk measures $A_\text{sum}$ (grey) and $A_\text{max}$ (blue, over grey) basically populate the same regions.
See the text after Eq.~(\ref{Eq:Z_k}) for further comments.}
\label{Fig:invar-raw}
\end{figure}

We show, in Fig.~\ref{Fig:invar-raw}, scatter plots of the measures of CP violation, $A_\text{sum}$ (grey, bottom layer, mostly covered), $A_\text{max}$ (blue, over grey), defined by Eq.~(\ref{Eq:CP-inv-measure}), vs $\theta_2/\pi$ and $\theta_3/\pi$, based on the scan performed in Ref.~\cite{Plantey:2022jdg}. The cut-off at high values is obviously caused by the upper bounds on the $\lambda$s (from perturbativity, since the invariants are polynomials in the $\lambda$s). Likewise, there is a thinning out at low values due to the finite resolution in the scan sample. Inspection shows that the lower points are associated with low masses of the lightest neutral state. This is reminiscent of the 2HDM case \cite{Grzadkowski:2014ada} where the corresponding invariants are proportional to masses squared or even differences of masses squared. 

Actually, the original scan \cite{Plantey:2022jdg} does not have any lower bound imposed on the masses of the neutral scalars. In order to explore this connection, we imposed a lower bound of 45~GeV on the lightest neutral scalar, and obtained the subset of points shown in yellow (on top of the blue). Yellow points correspond to parameter sets
where the lightest neutral scalar has mass $> 45~\text{GeV}$; these appear to
avoid regions with very small CP-violating invariants, suggesting that
suppressed CP violation may require one or more light scalars.

Our focus will be on Type~Z Yukawa couplings, since these produce a less constrained model, with independent couplings for quarks and leptons.

\section{Experimental constraints}
\label{sect:constraints}

We shall here impose further experimental constraints on the parameter points which already survived theoretical constraints (perturbativity, unitarity and boundedness from below) as well as the following experimental constraints  \cite{Plantey:2022jdg}: compatibility with the measured $WWh_\text{SM}$ coupling \cite{ParticleDataGroup:2022pth} and the CP constraint on the $h_\text{SM}\to \bar\tau\tau$ coupling \cite{CMS:2021sdq}. Now, we also impose the constraints from the electroweak precision observables, $S$, $T$ and $U$, from the digamma signal strength ($h_\text{SM}\to\gamma\gamma$), from $\bar B\to X_s\gamma$ and from the electron EDM. A more detailed analysis should also take into account other observables, like the neutron EDM \cite{Engel:2013lsa,Chupp:2017rkp}.

\subsection{Electroweak precision observables}
\label{sect:STU}

The oblique parameters $S,T,U$ parametrize possible BSM deviations in the electroweak precision observables~\cite{Peskin:1990zt,Peskin:1991sw,Maksymyk:1993zm} and have been calculated for a general multi-Higgs model consisting of SU(2) doublets and singlets~\cite{Grimus:2007if,Grimus:2008nb}. We take into account the correlation between $S$ and $T$, and adopt the 2-sigma contours from~\cite{Asadi:2022xiy} which are based on a $\chi^2$ fit of the electroweak precision observables with respect to their PDG values \cite{ParticleDataGroup:2022pth}.

\subsection{$h_i\rightarrow \gamma\gamma$}
\label{sect:h-gaga}

For the SM Higgs particle, the one-loop diphoton decay width has long been known~\cite{Ellis:1975ap, Shifman:1979eb, Gunion:1989we} and played an important role in its discovery.
The $h_iWW$ vertex coupling will slightly differ from the SM value. Furthermore, there will be an additional contribution to $h_i \rightarrow \gamma\gamma$ coming from one-loop processes involving the charged scalars. This new contribution will interfere with the one-loop SM processes involving fermions and the $W$ boson and could enhance or reduce the total decay rate.

For the Weinberg potential the relevant part of the Lagrangian for this process is \cite{Plantey:2022jdg}
\begin{align} \label{Eq:h-gaga}
\mathcal L_{h_i\gamma\gamma} &= gm_W O_{i1} h_iW^+_\mu W^{\mu-} - \frac{m_f}{v}(\kappa^S_{h_iff}h_i\bar f f + i\kappa^P_{h_idd}h_i\bar d \gamma_5 d - i\kappa^P_{h_iuu}h_i\bar u \gamma_5 u) \nonumber \\
&- \sum_{k,k'=1,2} v g_{ikk'} h_i h_k^+h_{k'}^-.
\end{align}
Here, and in Eq.~(\ref{Eq:Wloop}) below, $O_{i1}$ is an element of the rotation matrix, given by Eq.~(\ref{Eq:rot-matrx-55}). Since the neutral scalars are mixed CP states, their scalar and pseudoscalar Yukawa couplings $\kappa^{S,P}_{h_iff}$ are in general both non-zero. Reading off the coupling modifiers, the decay rate is given by \cite{Gunion:1989we,Djouadi:2005gj}
\begin{equation}
\Gamma(h_i \rightarrow \gamma\gamma) = \frac{\alpha^2}{256\pi^3}\frac{m_i^3}{v^2} 
\big\{\big |I_1 + I^S_{1/2} + I_0 \big|^2+\big| I^P_{1/2} \big |\big\},
\end{equation}
where the $W$-loop contribution is given by
\begin{equation}
I_1= O_{i1} F_1(\tau_W), \label{Eq:Wloop}
\end{equation}
the fermion loop contributions are given by\footnote{There is a sign ambiguity associated with this contribution. We allow for either sign. For a given parameter set, only one sign admits agreement with observations, usually the SM sign.}
\begin{equation}
I^r_{1/2}= \sum_f N_c^f e^2_f \kappa^r_{h_iff} F_{1/2}^r(\tau_f), \quad r=S,P,
\end{equation}
and the contributions of the charged scalars by
\begin{equation}
I_0= \sum_{k=1,2} \frac{v^2}{2m_{h_k^\pm}^2} g_{ikk} F_0(\tau_{h_k^\pm}).
\end{equation}
Here, $N_c^f$ and $e_f$ refer to the color multiplicity and electric charge of fermion $f$, respectively, $\tau_j = 4m_j^2/m_i^2$ and the functions $F$ are the one-loop functions given in \cite{Gunion:1989we}.
We adopt values for $\mu_{\gamma\gamma}=\Gamma(h_i \rightarrow \gamma\gamma)/\Gamma(h_\text{SM} \rightarrow \gamma\gamma)$ within $3\sigma$ of the PDG result $\mu_{\gamma\gamma}=1.10\pm0.07$ \cite{ParticleDataGroup:2022pth} for the $h_i$ that is interpreted as $h_\text{SM}$ ($h_2$ or $h_3$).
The choice of which is considered to be the $h_\text{SM}$-like state will be denoted by $h_2 = h_\text{SM}$ or $h_3 = h_\text{SM}$.

\subsection{$\bar B\rightarrow X_s\gamma$}
\label{sect:Bsga}

The importance of charged scalar exchange for the $\bar B\to X_s\gamma$ rate has been known since the late 1980s \cite{Grinstein:1987pu,Hou:1988gv,Grinstein:1990tj}. The rate is determined from an expansion of the relevant Wilson coefficients in powers of $\tilde\alpha_s\equiv\alpha_s/(4\pi)$, starting with (1) the matching of these coefficients to the full theory at a high scale ($\mu_0\sim m_W$ or $m_t$), then (2) evolve them down to the low scale $\mu_b\sim m_b$ (taking into account the mixing of operators in this process), and (3) determine the matrix elements at the low scale \cite{Buras:1993xp,Ciafaloni:1997un,Ciuchini:1997xe,Borzumati:1998tg,Bobeth:1999ww,Bobeth:1999mk,Gambino:2001ew,Cheung:2003pw,Misiak:2004ew,Czakon:2006ss,Hermann:2012fc,Misiak:2015xwa,Misiak:2017bgg,Misiak:2020vlo}.
 
The Weinberg potential, with the general vacuum we are considering, has {\it two} pairs of charged scalars. This means that there are two charged-scalar contributions to the $\bar B\to X_s\gamma$ decay amplitude. For the 2HDM, with just one charged scalar\footnote{Or a 3HDM with only one active charged scalar, i.e., a charged scalar that couples to fermions.}, the calculation of this branching ratio is already rather complicated, involving a BSM contribution to the amplitude that contains a $bH^\pm t$ and an $sH^\pm t$ (or a $bH^\pm c$ and an $sH^\pm c$) coupling, subject to large QCD corrections \cite{Hermann:2012fc}, plus a non-perturbative part.

Since the experimental data are in good agreement with the most precise SM calculations, the BSM contribution has to be small. This imposes constraints on the above-mentioned couplings (in the 2HDM, these are often expressed in terms of $\tan\beta$) and the charged-Higgs mass (a high mass will suppress this loop contribution).

The minimisation conditions of the potential, Eq.~(\ref{Eq:sol-12}), force all mass terms in the potential to be fully written in terms of the quartic couplings \cite{Plantey:2022jdg}, and since the latter are constrained by perturbativity, the masses of the charged scalars are at the order of the electroweak scale. This will have an impact on the $\bar B\to X_s\gamma$ constraint.

\begin{figure}[htb]
\begin{center}
\includegraphics[scale=0.30]{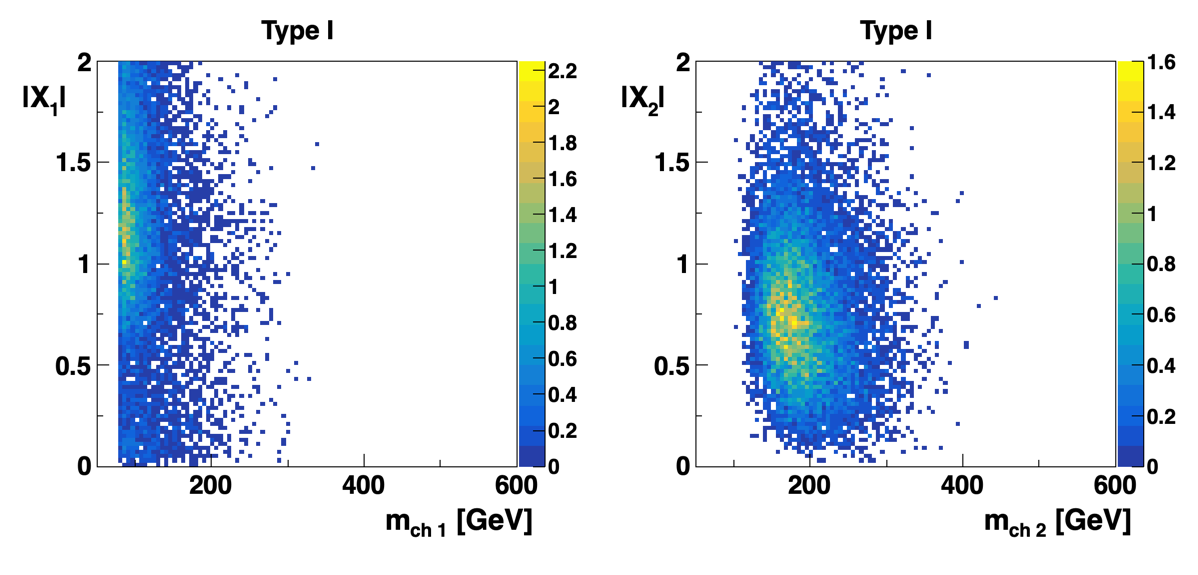}
\end{center}
\vspace*{-8mm}
\caption{Scatter plots (often referred to as a ``temperature'' plot, regions shown at a higher ``temperature'' are more populated) of the moduli $|X_k|=|Y_k|$ (see Eq.~(\ref{Eq:Yuk-BG})) vs the charged-scalar masses for Type~I Yukawa couplings. These points, from the initial scan of Ref.~\cite{Plantey:2022jdg}, also satisfy the constraints discussed in the previous subsections, i.e., the electroweak precision observables and $h_i\to\gamma\gamma$. In order to avoid a cumbersome notation, we refer to the masses of the charged states $h^\pm_{1.2}$ as $m_\text{ch\! 1}$ and $m_\text{ch\! 2}$.}
\label{Fig:bsga-type-I}
\end{figure}

\subsubsection{The $b\bar t h_k^+$ couplings}

As a first step in acquiring some intuition for this constraint based on the considered potential, we show  in Fig.~\ref{Fig:bsga-type-I} scatter plots of the Type~I vs the charged-scalar mass.  By definition, for Type~I we have $|X_k| =  |Y_k|$, see Eq.~(\ref{Eq:Yuk-BG}).  The points shown satisfy the theoretical constraints, as well as the electroweak $S$, $T$, $U$ constraints, and the digamma rate.
Colours reflect the density of points in the scan that satisfy the theoretical and experimental constraints listed at the beginning of section~\ref{sect:constraints}, as well as those addressed in subsections~\ref{sect:STU} and \ref{sect:h-gaga}.

A new feature (as compared with the familiar 2HDMs) is that there are {\it two} charged states. There could thus be regions of parameter space in which the contributions of the two charged scalars $h_1^+$ and $h_2^+$ have opposite signs, and partially cancel, allowing for rather light charged-scalar masses without violating the $\bar B\rightarrow X_s\gamma$ constraint \cite{Akeroyd:2016ssd}.

Also, we should stress that the individual (due to a particular charged state $h_k^\pm$) contributions to the $\bar B\to X_s\gamma$ amplitude are complex. First of all, the basic interaction, Eq.~(\ref{Eq:Yuk-kappa}) is complex, due to the non-zero phase of some VEV. Secondly, there are complex coefficients involved in the high-scale matching \cite{Misiak:2004ew,Czakon:2006ss,Hermann:2012fc,Misiak:2015xwa,Misiak:2017bgg,Misiak:2020vlo}. As a result, there is a non-trivial interference between the contributions involving the $h_1^\pm$ and $h_2^\pm$ intermediate states.

\begin{figure}[htb]
\begin{center}
\includegraphics[scale=0.25]{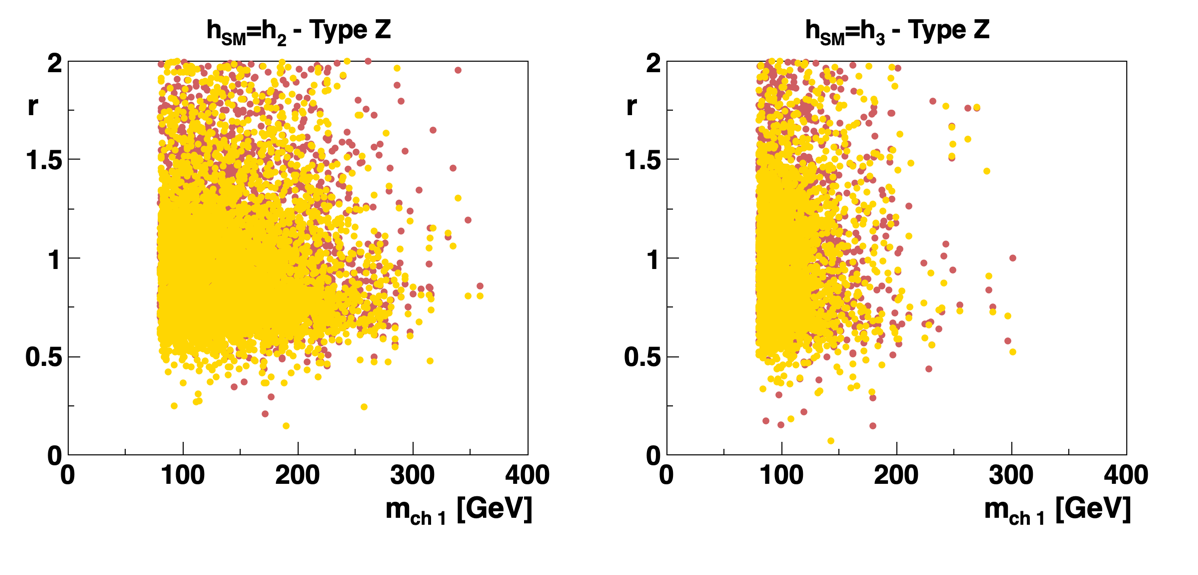}
\end{center}
\vspace*{-8mm}
\caption{Scatter plots of relative sums $r$ defined as the ratio between the absolute value of (\ref{Eq:two-charged-scalars}) and the absolute value of the first term ($k=1$). Reddish: ratios $r$ for $C_7$, yellowish (on top): for $C_8$. Left: $h_2=h_\text{SM}$, right: $h_3=h_\text{SM}$.  See the text for more comments.}
\label{Fig:ratio-s}
\end{figure}

\subsubsection{Treatment of the $\bar B\to X_s\gamma$ constraint}
As mentioned above, compared to the familiar $\bar B\to X_s\gamma$ constraint for the 2HDM, we here face two complications:
\begin{itemize}
\item
The couplings $X_k$ and $Y_k$ are complex. Thus, some Wilson coefficients will become complex.
\item
There are two charged scalars, thus two non-SM contributions to the {\it amplitude} determining the decay rate.
\end{itemize}

The basic BSM contribution to the amplitude for $\bar B\to X_s\gamma$ involves the couplings $X$ and $Y$ in the following combination for both Wilson coefficients $C_7$ and $C_8$ \cite{Borzumati:1998tg},
\begin{equation} \label{Eq:one-charged-scalar}
f(y)|Y|^2+g(y)(X^\ast Y),
\end{equation}
where for the lowest-order (in $\tilde\alpha_s$) amplitude $f(y)$ and $g(y)$ are just Inami-Lim functions \cite{Inami:1980fz} of the ratio $y=(m_t/m_{h^+})^2$, whereas at the NLO level they are more complicated functions. Loop contributions involving both charged scalars will obviously interfere at the amplitude level. Thus, in order to account for both charged scalars, we replace the expression (\ref{Eq:one-charged-scalar}) (at the high scale, often referred to as $\mu_0$) by
\begin{equation} \label{Eq:two-charged-scalars}
\sum_{k=1}^2\big[f(y_k)|Y_k|^2+g(y_k)(X_k^\ast Y_k)\bigr],
\end{equation}
both at the LO and the NLO level.

\begin{figure}[htb]
\begin{center}
\includegraphics[scale=0.19]{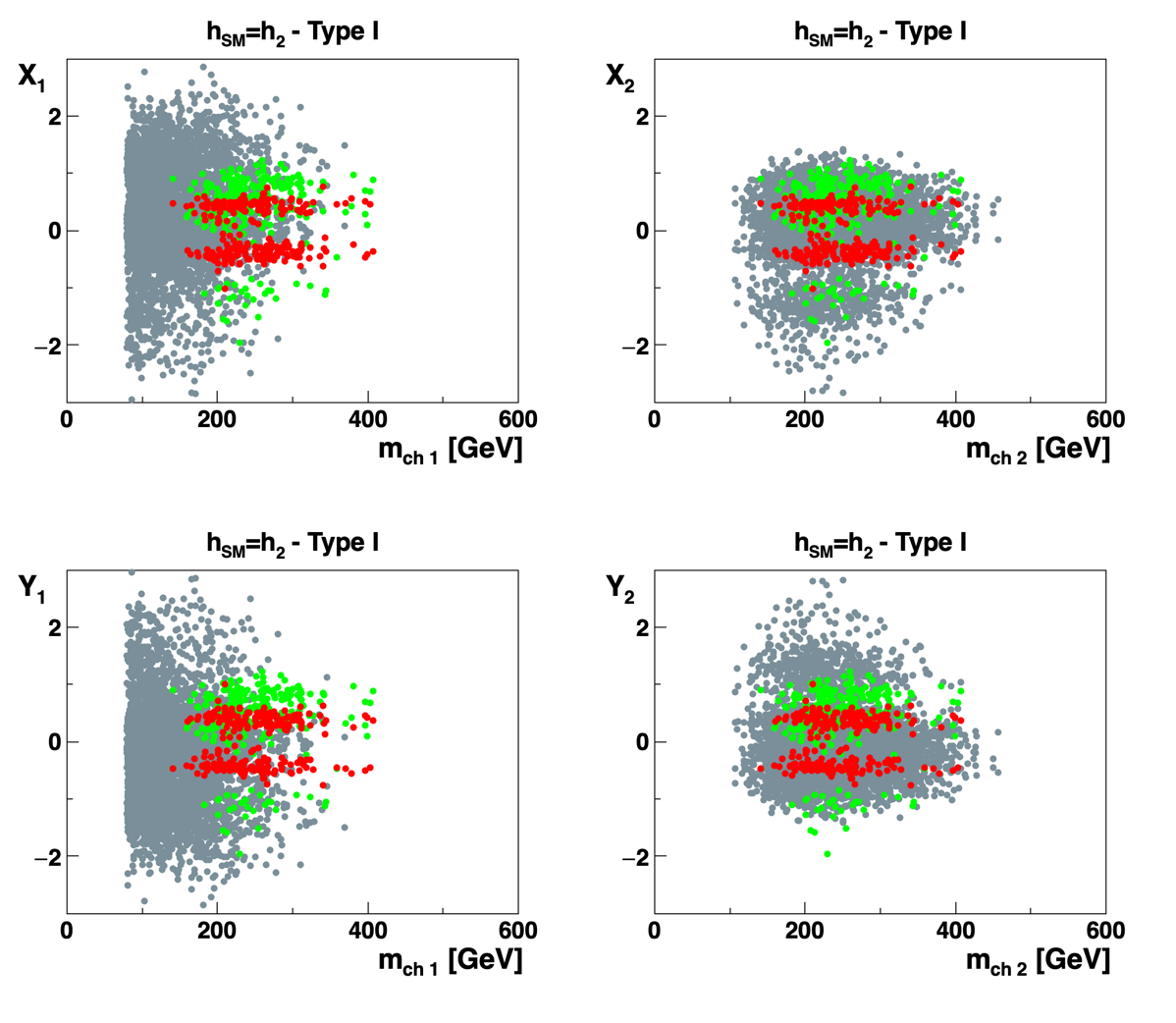}
\includegraphics[scale=0.19]{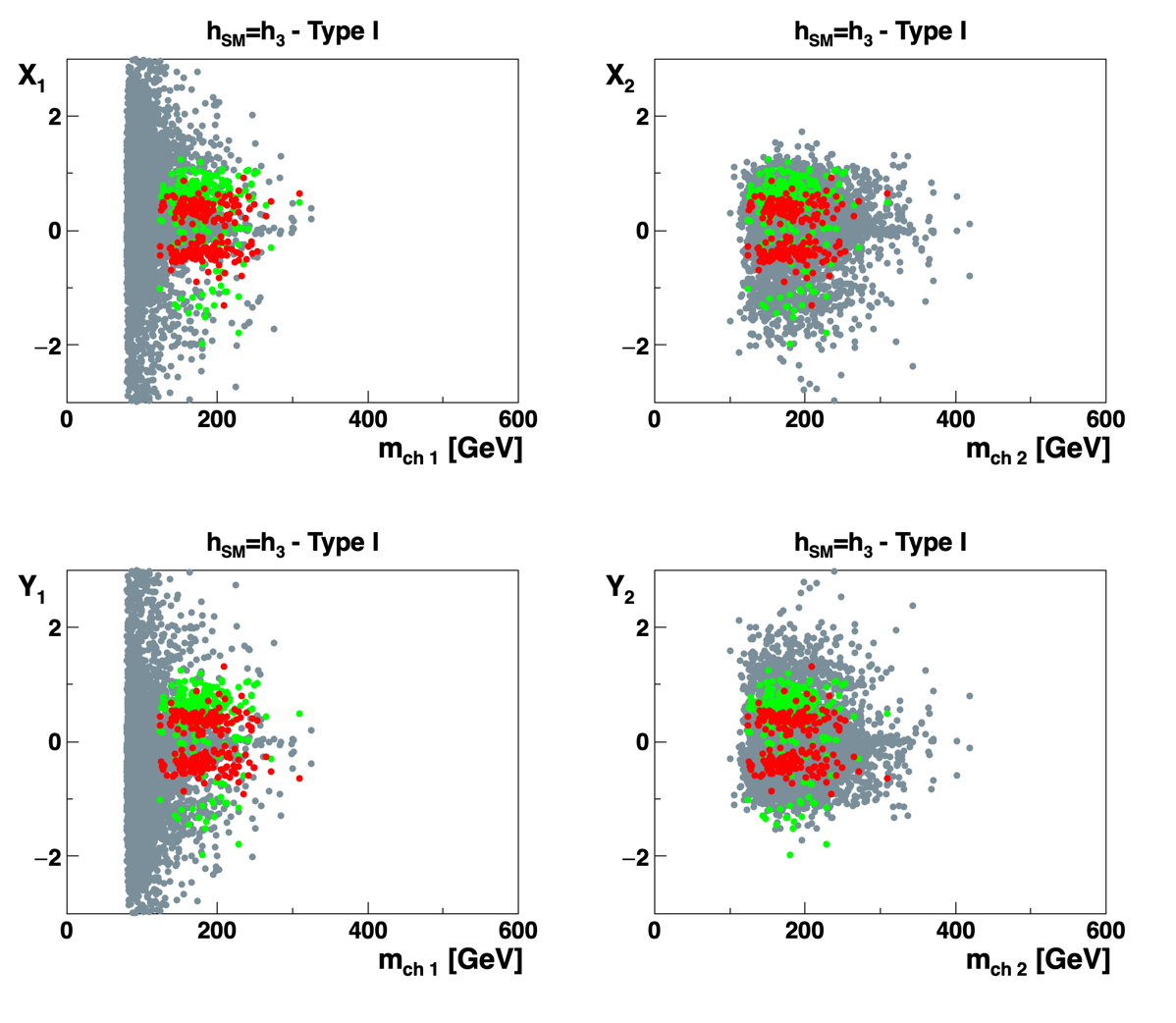}
\end{center}
\vspace*{-8mm}
\caption{Scatter plots of real (green) and imaginary (red, on top of green) parts of $X_k$ and $Y_k$ vs the charged-scalar masses for Type~I Yukawa couplings, with $h_2$ (left) or $h_3$ (right) identified as $h_\text{SM}$. The constraints imposed are those of Fig.~\ref{Fig:bsga-type-I}, plus $\bar B\to X_s\gamma$. See text for more details.}
\label{Fig:bsga-type-I-2-3-ok}
\end{figure}

In order to explore the possible cancellations, in Fig.~\ref{Fig:ratio-s} we show the ratios
\begin{equation} \label{Eq:ratio:bsga}
r=\frac{|\sum_{k=1}^2\bigl[f(y_k)|Y_k|^2+g(y_k)(X_k^\ast Y_k)\bigr]|}
{|f(y_1)|Y_1|^2+g(y_1)(X_1^\ast Y_1)|}
\end{equation}
for contributions to $C_7$ and to $C_8$, for both $h_2=h_\text{SM}$ and $h_3=h_\text{SM}$. Only parameter points surviving the previous constraints including $\bar B\rightarrow X_s\gamma$ are shown. A considerable fraction of the surviving scan points lie below unity, i.e., representing cases of destructive interference between the contributions of $h_1^\pm$ and $h_2^\pm$.

\begin{figure}[htb]
\begin{center}
\includegraphics[scale=0.19]{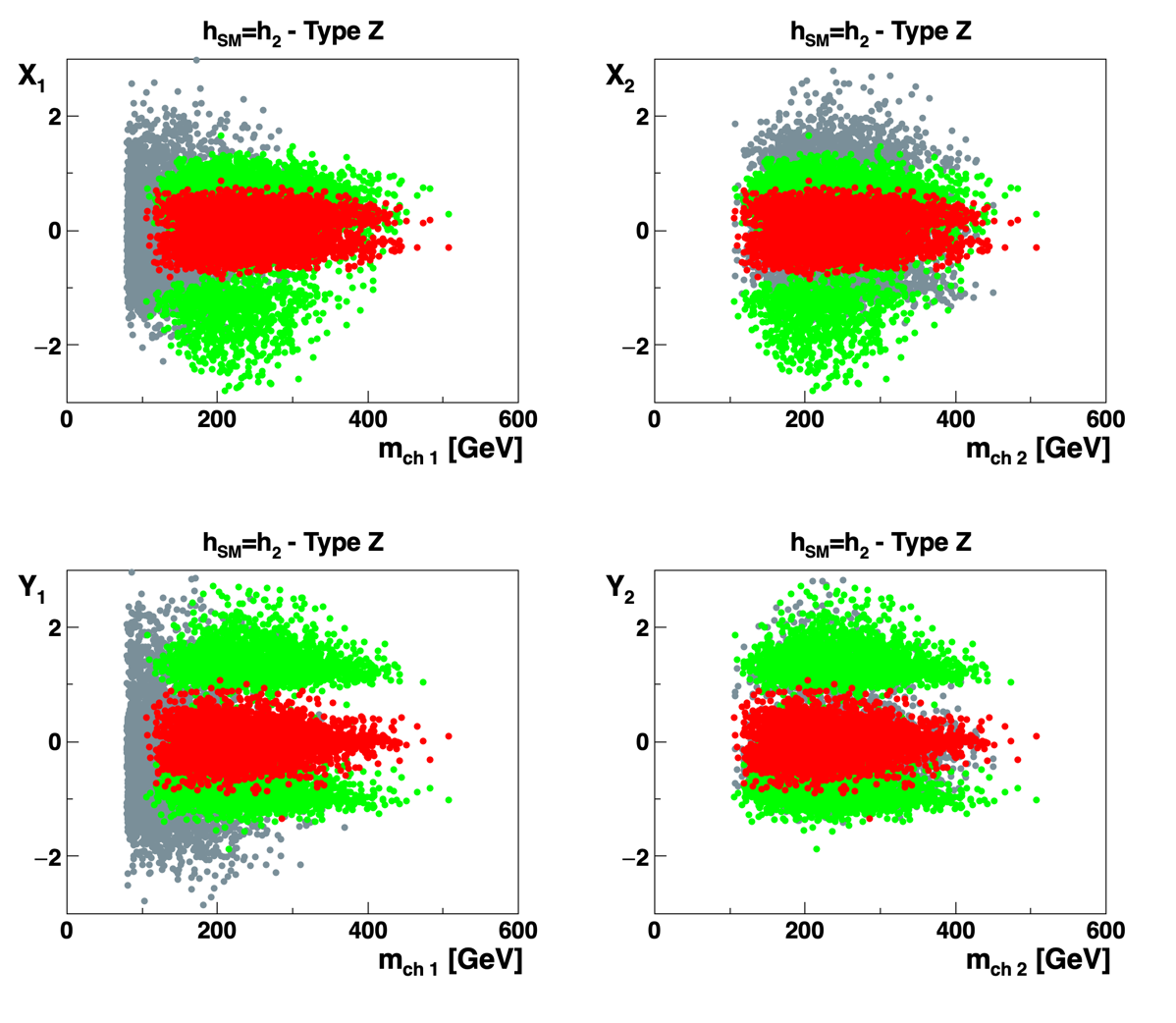}
\includegraphics[scale=0.19]{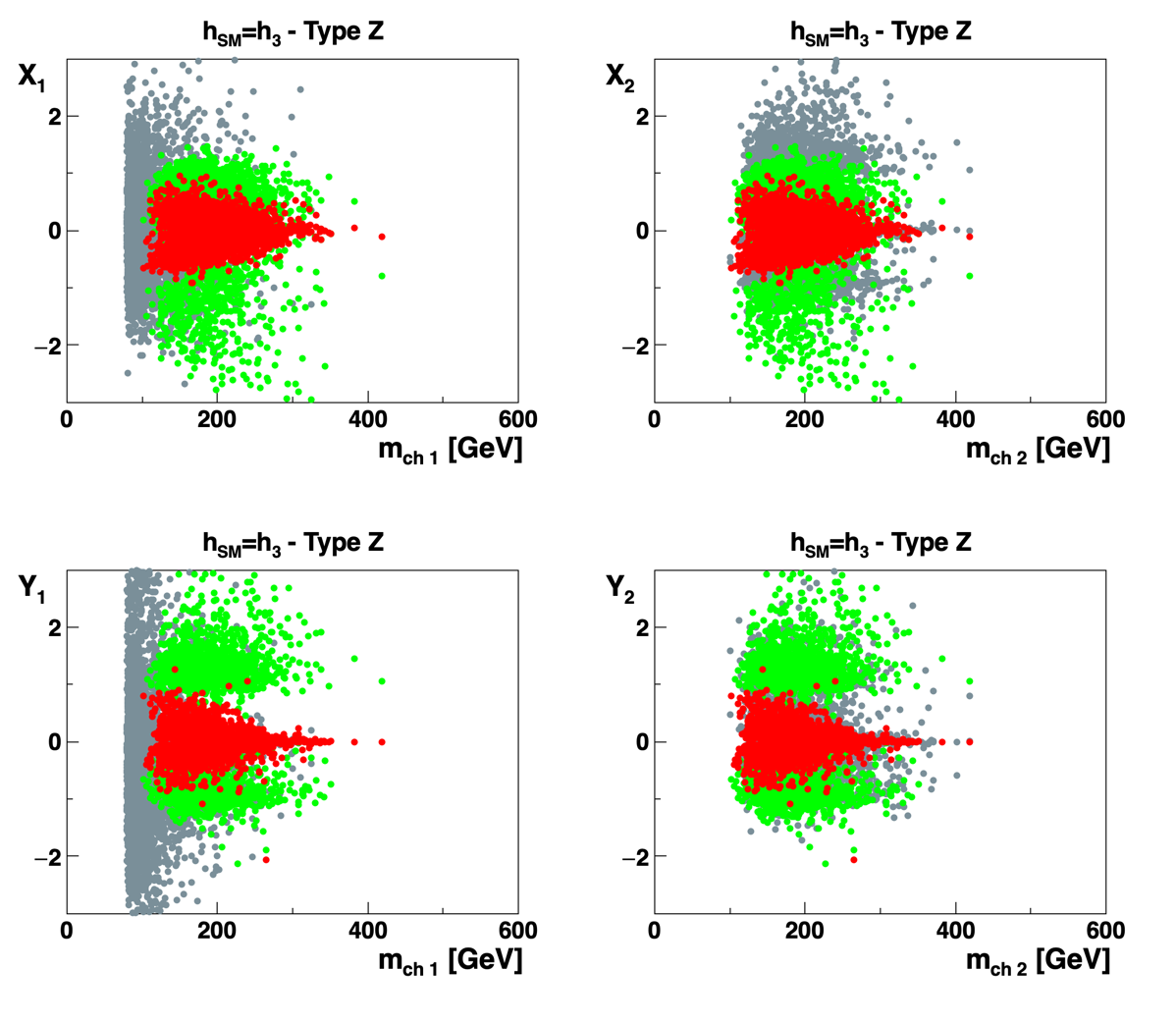}
\end{center}
\vspace*{-8mm}
\caption{Scatter plots of real (green) and imaginary (red, on top of green) parts of $X_k$ and $Y_k$ vs the charged-scalar masses for Type~Z (or Type~II) Yukawa couplings, with $h_2$ (left) or $h_3$ (right) identified as $h_\text{SM}$. The constraints imposed are those of Fig.~\ref{Fig:bsga-type-I}, plus $\bar B\to X_s\gamma$. See text for more details.}
\label{Fig:bsga-type-Z-2-3-ok}
\end{figure}

Next, we show in Figs.~\ref{Fig:bsga-type-I-2-3-ok} and \ref{Fig:bsga-type-Z-2-3-ok} scatter plots of real and imaginary parts of $X_k$ and $Y_k$ vs $m_\text{ch $k$}$, for Type~I and Type~Z Yukawa couplings. In grey, as an underlying background, we show real and imaginary parts of these couplings for parameter points that survive all previous constraints, not considering $\bar B\to X_s\gamma$. Superimposed on this background, we show separately real (green) and imaginary (red) parts of $X_k$ and $Y_k$ that in addition survive an approximate implementation of the $\bar B\to X_s\gamma$ constraint $\text{Br}(\bar B\to X_s\gamma)\times10^4=3.32\pm0.15$ \cite{ParticleDataGroup:2022pth} at the 3-$\sigma$ level. The points are layered in the following sequence. Bottom, grey: no $\bar B\to X_s\gamma$ constraints;  intermediate, green: real part; top, red: imaginary part. Grey points come in pairs. For each green point there is also a red point. 

The cases $h_2=h_\text{SM}$ (left, $k=1$ and 2) and $h_3=h_\text{SM}$ (right, $k=1$ and 2) are considered. Figure~\ref{Fig:bsga-type-Z-2-3-ok}, labelled ``Type~Z'', applies equally to Type~II Yukawa couplings since the Yukawa couplings to leptons have no bearing on the $\bar B\to X_s\gamma$ constraint. 
The same would apply to other types with similar property.
In evaluating the constraints, we include interference with the SM amplitude, up to the second order in $\tilde\alpha_S$ \cite{Misiak:2006ab}.

The red bands in Fig.~\ref{Fig:bsga-type-I-2-3-ok} (Type~I) show that non-zero imaginary parts of $X_k$ and $Y_k$ are required for the $\bar B\to X_s\gamma$ constraint to be satisfied. This is no longer the case in Fig.~\ref{Fig:bsga-type-Z-2-3-ok} (Type~Z).
For Type~I Yukawa couplings we observe both lower and upper bounds on $|\Im X_k|$ and $|\Im Y_k|$, whereas for Type~Z we observe only upper bounds.
Type~Z populates a somewhat different region than Type~I, since the quark Yukawa couplings are different.

\subsection{The electron EDM}
The mixing of CP-even and odd fields will induce a contribution to the electron electric dipole moment via the Barr--Zee effect \cite{Barr:1990vd}, which is experimentally rather constrained. Following Pilaftsis \cite{Pilaftsis:2002fe}, the result can be expressed in terms of the couplings $\kappa^S$ and $\kappa^P$ of Eqs.~(\ref{Eq:Yuk-kappa}) and (\ref{Eq:Yuk-kappa-S-P}) as
\begin{equation}
\frac{d_e}{e}=-\frac{3\alpha_\text{e.m.}}{8\pi^2\sin^2\theta_\text{W}}\frac{m_e}{m_W^2}
\sum_{i=1}^5\sum_{q=t,b}Q_q^2
\left[\kappa_{h_iee}^P\kappa_{h_iqq}^S\,f\left(\frac{m_q^2}{m_{h_i}^2}\right)
+\kappa_{h_iee}^S\kappa_{h_iqq}^P\,g\left(\frac{m_q^2}{m_{h_i}^2}\right)
\right].
\end{equation}
Here, $f$ and $g$ are two-loop functions \cite{Barr:1990vd}, not to be confused with the $f$ and $g$ of Eqs.~(\ref{Eq:one-charged-scalar})--(\ref{Eq:ratio:bsga}).
The experimental upper bound has recently been tightened from $1.1\times10^{-29}~e\cdot\text{cm}$ \cite{ACME:2018yjb} to  $4.1\times10^{-30}~e\cdot\text{cm}$ \cite{Roussy:2022cmp}.

The SM contribution to this process, induced by a complex CKM matrix is of the order $10^{-38}~e\cdot\text{cm}$ which is much smaller than the scalar contribution of the present model and therefore does not play any relevant role here.  We do not consider at this stage what mechanism would generate a complex CKM matrix.

\begin{figure}[htb]
\begin{center}
\includegraphics[scale=0.30]{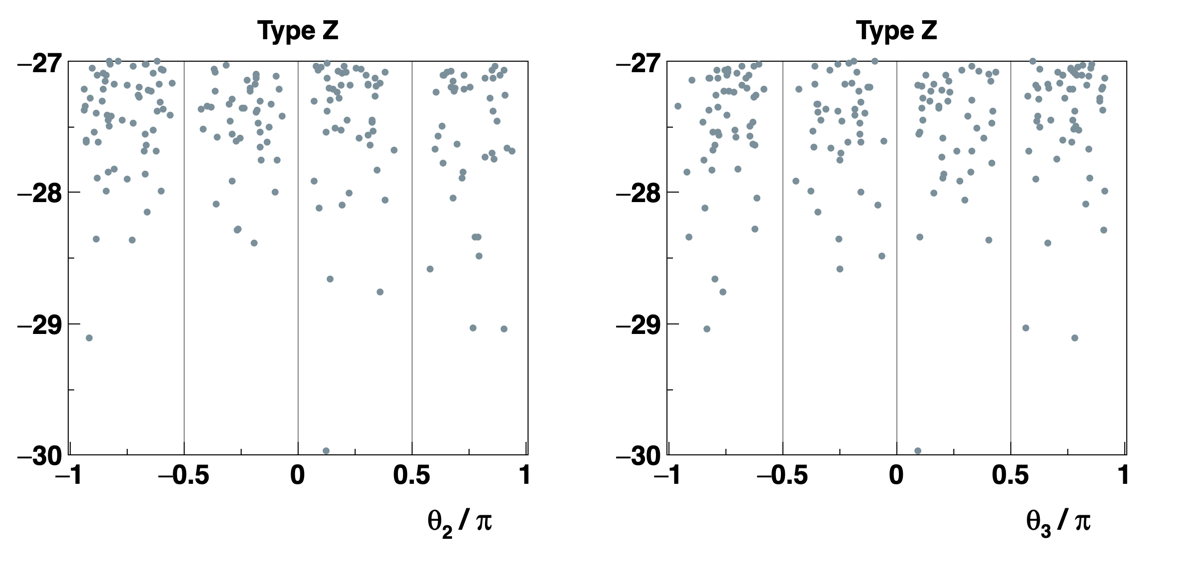}
\end{center}
\vspace*{-8mm}
\caption{Scatter plots of logarithms of the electron EDM vs the angles $\theta_2$ and $\theta_3$. 
These points correspond to random scans over the scalar potential parameters
subject to theoretical and collider bounds specified above.
Few parameter points that survive the previous constraints survive the experimental EDM bound allowing only the lower region of $|d_e/e|$.}
\label{Fig:edm_vs_theta}
\end{figure}

Since the complex vevs induce a mixing of the CP even and odd fields, it is interesting to see how $|d_e/e|$ varies with the phases of the vevs. We show in Fig.~\ref{Fig:edm_vs_theta} parameter points for which $|d_e/e|<10^{-27}~e\cdot\text{cm}$ vs $\theta_2$ and $\theta_3$. These points satisfy the constraints listed at the beginning of this section, plus the electroweak, the diphoton signal strength, and the $\bar B\to X_s\gamma$ ones. While the distribution is thinning out at low values of $|d_e|$, there are parameter points for which $|d_e/e|<10^{-29}~e\cdot\text{cm}$, as required by experiment.
Clearly, a larger sample of scan points would be required in order to explore this region of the parameter space.

Since a non-zero $d_e$ is a sign of CP violation, and CP is conserved in the limits $\theta_i=n\pi/2$, one would expect parameter points of small $|d_2|$ to be found near these limits. This is however not the case, due to other constraints, as discussed in section~\ref{sect:theta}.

\section{CP violation---the invariants}
\label{sect:CP-violation-invariants}
The experimental upper bound on the electric dipole moment is an important constraint on CP-violating theories. We shall here explore how this constraint is reflected in the CP-violating invariants presented in section~\ref{sect-CP-invariants}. If CP is conserved, no scalar would have simultaneous non-zero $\kappa^S$ and $\kappa^P$ couplings. While we there (in section~\ref{sect-CP-invariants}) presented a global measure, involving all 15 invariants of Eq.~(\ref{Eq:Weinberg-invariants}), we shall here explore the individual invariants, and see how they correlate with the value of the electric dipole moment.

\begin{figure}[htb]
\begin{center}
\includegraphics[scale=0.30]{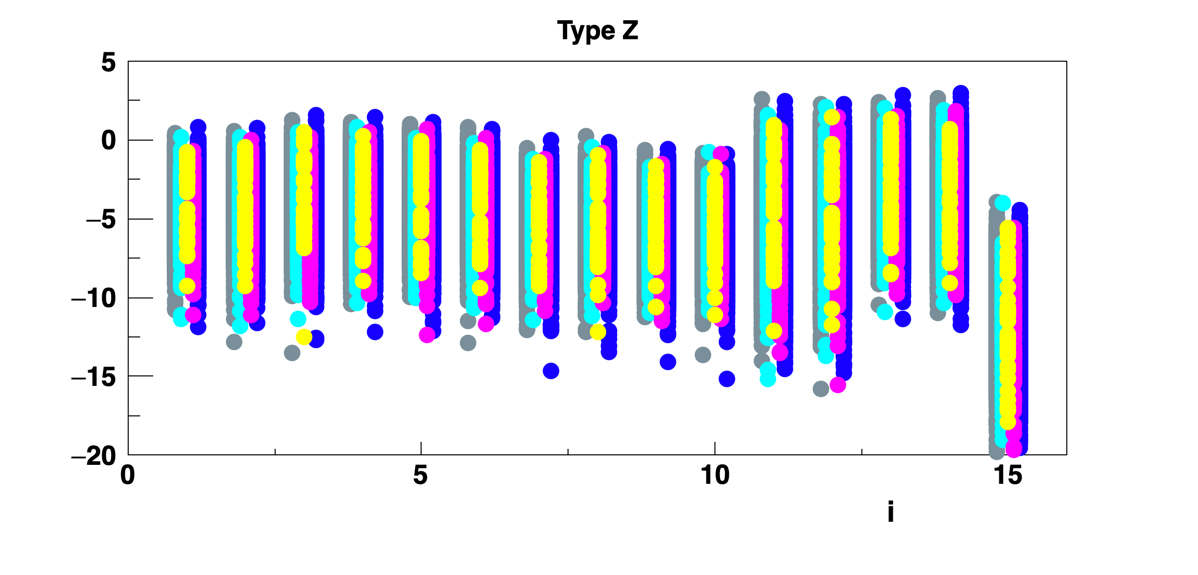}
\end{center}
\vspace*{-8mm}
\caption{Scatter plots of $\log_{10}(J_i^2)$ for the 15 invariants of Eq.~(\ref{Eq:Weinberg-invariants}). Different colours denote different ranges of $|d_e|$ as specified by the list~(\ref{Eq:inv_colors}).}
\label{Fig:J_i}
\end{figure}

In Fig.~\ref{Fig:J_i} we show scatter plots of $\log_{10}(J_i^2)$ vs $i$, for $i$ ranging from 1 to 15, where the points are colour coded according to the range of the electric dipole moment $d_e$ as follows (a small horizontal offset is introduced to improve readability, for the reason explained below):
\begin{alignat}{2} \label{Eq:inv_colors}
&\text{grey (bottom layer)} &\quad &|d_e|/e<10^{-27}~\text{cm}, \text{and }\left(\tan\alpha\right)_\tau<0.1 \text{\cite{Plantey:2022jdg}}, 
\nonumber \\
&\text{blue} &\quad &|d_e|/e<5\cdot10^{-28}~\text{cm}, \nonumber \\
&\text{green} &\quad &|d_e|/e<10^{-28}~\text{cm}, \\
&\text{purple} &\quad &|d_e|/e<5\cdot10^{-29}~\text{cm}, \nonumber \\
&\text{yellow (on top)} &\quad &|d_e|/e<10^{-29}~\text{cm}. \nonumber
\end{alignat}
In order to represent both positive and negative values of the invariants, and focus on orders of magnitude, we show the logarithms of their squares.

We recall that the initial scan points (or data set) had been screened for its CP-odd admixture in the coupling of the $\tau$ to the observed Higgs boson. This constraint is expressed in terms of the mixing angle $\alpha$, defined in ref.~\cite{Plantey:2022jdg}.

The invariant $J_{15}$, is the only one of the selected invariants that involves the bilinear terms of the potential. These couplings are not normalised in the same way as the quartic couplings, this invariant turns out to be overall smaller than the others. Its smallness compared to the other invariants is just an artefact and not physically meaningful. While these scatter plots reflect their underlying stochastic nature from the finite number of scan points, there is an interesting correlation of the {\it maximum} values with the range of $|d_e|$. In fact, as $|d_e|$ is constrained to successively smaller values, {\it all} $J_i$ invariant are also constrained in the sense that the maximum is lower.

\begin{figure}[htb]
\begin{center}
\includegraphics[scale=0.27]{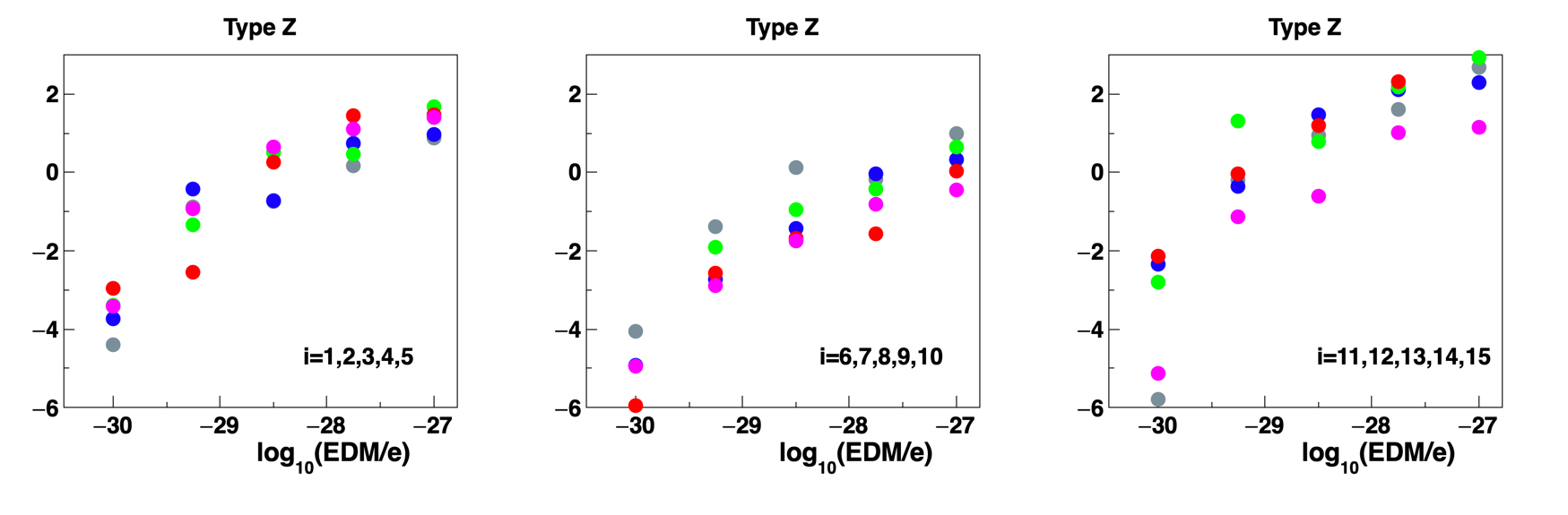}
\end{center}
\vspace*{-8mm}
\caption{Plots of $\max(\log_{10}(J_i^2))$ for the 15 invariants of Eq.~(\ref{Eq:Weinberg-invariants}). The maximum value of the invariant within each EDM bin is plotted. Different colours refer to different $J_i$. Each frame displays the result for invariants appearing in increasing order of $i$. Grey: $i=1$, 6, 11; blue: $i=2$, 7, 12; green: $i=3$, 8, 13; red: $i=4$, 9, 14; purple: $i=5$, 10, 15. For $i=15$, we have shifted the values upwards by 5 units (see Fig.~\ref{Fig:J_i}, i.e., we plot $\log_{10}J_{15}^2+5$.}
\label{Fig:max-J_i}
\end{figure}

We see that a constraint on {\it one} CP-violating observable (here: $|d_e|$) impacts {\it all} the CP-violating invariants. In order to illustrate this connection, we split the range of $d_e$ values into bins, analogous to those of Eq.~(\ref{Eq:inv_colors}), and plot for each of the 15 $J_i$ the maximum value of $\log_{10}(J_i^2)$ in each bin of $d_e$-values, using a different colour code. The result is shown in Fig.~\ref{Fig:max-J_i}.

Apart from some fluctuations due to the finite scan sample, there is a clear tendency for the maximum value of the invariants to get smaller in bins of smaller values of $|d_e|$. And, importantly, we see that the potential is able to produce models with $d_e$ below $10^{-29}e\cdot\text{cm}$ as required by recent data.

What this shows is that the simple basis of invariants chosen in Eq.~(\ref{Eq:Weinberg-invariants}) is not well ``aligned'' with this particular observable, an invariant describing only the EDM would be a linear combination of those given in Eq.~(\ref{Eq:Weinberg-invariants}).

\section{Two CP-violating processes}
\label{sect:CP-processes}
We shall next comment on two CP non-conserving processes,
\begin{align}
W^+W^-&\to Z, \\
h_i^+h_i^-&\to h_1^\pm h_2^\mp, \label{eq:processes}
\end{align}
the first of which could perhaps be experimentally accessible, whereas the second one is more of academic interest.\footnote{A relevant discussion of CP violation in the bosonic sector can be found in Ref.~\cite{Haber:2022gsn}.} However, it has the unique feature of involving only scalar particles in the initial and final states, no fermions and no gauge bosons.

\subsection{$W^+W^-\to Z$}
\label{sect:WWZ}

\begin{figure}[htb]
\centerline{
\includegraphics[scale=0.6]{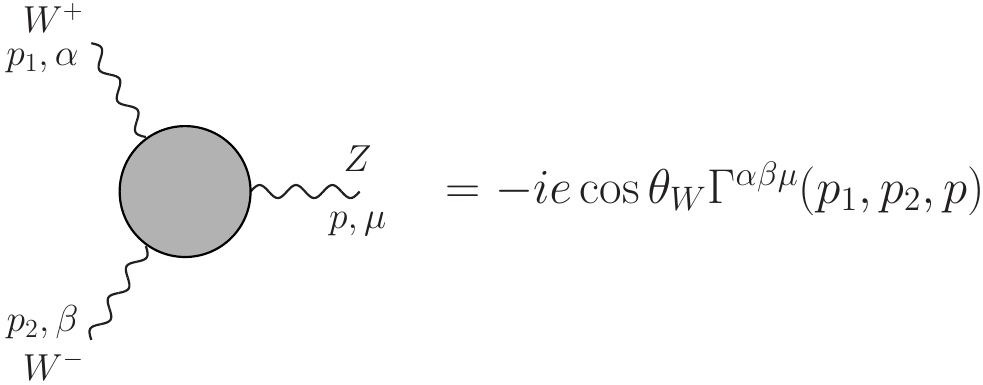}}
\caption{The general $WWZ$ vertex.}
\label{fig:blob-wwz}
\end{figure}

For the $WWZ$ vertex we adopt the notation given in Fig.~\ref{fig:blob-wwz}, with all momenta incoming.

The vertex is present at the tree level, with a well-known, CP-conserving structure:
\begin{align}
ig_{WWZ}\Gamma^{\mu_1\mu_2\mu}_\text{tree}
&=-ig\cos\theta_W[g^{\mu_1\mu_2}(p_2-p_1)^\mu + g^{\mu_2\mu}(p-p_2)^{\mu_1}
+ g^{\mu\mu_1}(p_1-p)^{\mu_2}] \nonumber \\
&=-ig\cos\theta_W[g^{\mu_1\mu_2}\ell^\mu + g^{\mu_2\mu}(\threehalf p-\half\ell)^{\mu_1}
- g^{\mu\mu_1}(\threehalf p+\half\ell)^{\mu_2}],
\end{align}
where $g_{WWZ}=-e\cot\theta_W$, and in the second line, we make use of the abbreviation $\ell=p_2-p_1$. 

Triangle diagrams discussed below contribute to CP-violation in the $W^+W^-Z$ vertex. In fact, in the Two-Higgs-Doublet model they give a contribution proportional to the 2HDM-invariant $\Im J_2$, which is one measure of CP violation \cite{Gunion:2005ja,Branco:2005em} (referred to as $J_1$ in earlier work by Lavoura, Silva and Botella \cite{Lavoura:1994fv,Botella:1994cs}), not to be confused with the $J_i$ of Eq.~(\ref{Eq:Weinberg-invariants}).

Phenomenological discussions \cite{Hagiwara:1986vm} of the $WWZ$ vertex have presented its most general Lorentz structure. We assume both $W^\pm$ to be on-shell and $Z$ off-shell, furthermore assuming that $Z$ couples to a pair of light leptons like $e^+e^-$ so that we may neglect terms proportional to the lepton mass. Then, according to \cite{Hagiwara:1986vm} the structure reads
\begin{align} \label{eq:hagiwara2}
\Gamma_{WWZ}^{\alpha\beta\mu}
&=f_1^Z\ell^{\mu}g^{\alpha\beta}
-\frac{f_2^Z}{m_W^2}\ell^{\mu}p^{\alpha}p^{\beta}
+f_3^Z(p^{\alpha}g^{\mu\beta}-p^{\beta}g^{\mu\alpha}) \nonumber \\
&+if_4^Z(p^{\alpha}g^{\mu\beta}+p^{\beta}g^{\mu\alpha})
+if_5^Z\epsilon^{\mu\alpha\beta\rho}\ell_\rho \nonumber \\
&-f_6^Z\epsilon^{\mu\alpha\beta\rho}p_\rho
-\frac{f_7^Z}{m_W^2}\ell^{\mu}\epsilon^{\alpha\beta\rho\sigma}p_\rho\ell_\sigma.
\end{align}
The tree-level vertex contributes to $f_1$ and $f_3$: 
\begin{equation}
f_1^\text{tree}=1, \quad
f_3^\text{tree}=2.
\end{equation}
The dimensionless form factors $f_4^Z$, $f_6^Z$ and $f_7^Z$ violate CP while the others conserve CP \cite{Hagiwara:1986vm}. LHC experiments \cite{ATLAS:2017pbb,CMS:2019ppl} have constrained the CP-conserving anomalous couplings $f_2^Z$ and $f_5^Z$, and even started to constrain the CP-violating $f_4^Z$ to below 0.0015 \cite{CMS:2020gtj}.

\begin{figure}[htb]
\begin{center}
\includegraphics[scale=0.6]{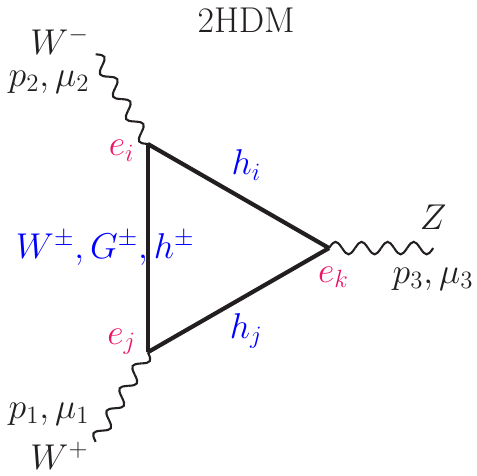}
\end{center}
\vspace*{-8mm}
\caption{Loop contributions to the CP-violating $W^+W^-\to Z$ amplitude in the 2HDM, Here, $i,j\in{\{1,2,3\}}$ label the neutral scalars.}
\label{Fig:Feynman-WWZ-2hdm}
\end{figure}

In the CP-violating 2HDM there are contributions to $f_4^Z$ that arise from triangle diagrams of the kind shown in Fig.~\ref{Fig:Feynman-WWZ-2hdm} \cite{Grzadkowski:2016lpv}. Like for the $ZZZ$ vertex, this contribution is proportional to the product of the three $VVh$ couplings, $e_1e_2e_3$ (see Table~\ref{table:couplings} and ref.~\cite{Grzadkowski:2014ada}), and it was shown that it is also proportional to the differences of masses squared that make up the invariant $\Im J_2$, resulting in $f_4^Z\propto \Im J_2$.

\begin{figure}[htb]
\begin{center}
\includegraphics[scale=0.6]{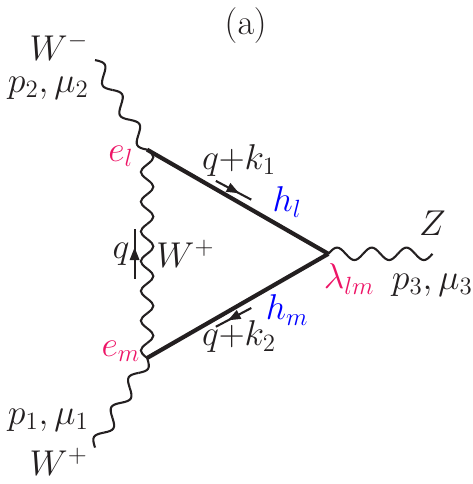}
\includegraphics[scale=0.6]{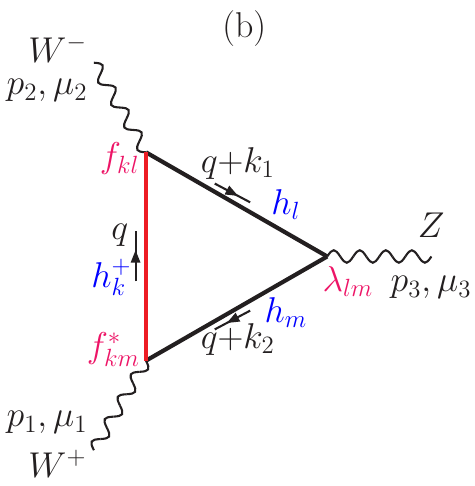}
\end{center}
\vspace*{-6mm}
\caption{Loop contributions to the CP-violating $W^+W^-\to Z$ amplitude. Here, $k,l,m\in{\{1,2,3,4,5\}}$ label the neutral scalars.}
\label{Fig:Feynman-WWZ}
\end{figure}

In the framework of the 3HDM, and in particular in the case of the Weinberg potential, the situation is more complicated for several reasons. Diagrams corresponding to the one given in Fig.~\ref{Fig:Feynman-WWZ-2hdm} for the 2HDM are shown in Fig.~\ref{Fig:Feynman-WWZ} for a 3HDM.

\begin{table}[!htbp]
\centering
\begin{tabular}{|c|c|c|}
\hline
 &  2HDM & 3HDM \\
\hline
$Zh_i h_j$ & $\epsilon_{ijk}\,e_k$ & $\tilde\lambda_{ij}$ \\
\hline
$ZZh_k$ & $e_k$ & $e_k$ \\
\hline
\end{tabular}
\caption{Trilinear vector-scalar couplings (the $Zh_i h_j$ coupling was referred to as $P_{ij}$ in \cite{Plantey:2022jdg,Plantey:2022gwj}), modulo normalization and Lorentz structure.}
\label{table:couplings}
\end{table}

In the 2HDM, the relevant part of the product $f_i f_j^\ast$ can be expressed in terms of $e_i e_j$ and the $Zh_i h_j$ couplings\footnote{The 2HDM coupling $e_i$ is real, whereas $f_i$ in general is complex \cite{Grzadkowski:2014ada}. Here, and in Fig.~\ref{Fig:Feynman-WWZ} and Table~\ref{table:couplings}, the two-index $\tilde\lambda$ refers to the coupling of a vector boson with two neutral scalars. It should not be confused with any $\lambda$ appearing in the potential.}  $\tilde\lambda_{ij}$ can be expressed in terms of  $\epsilon_{ijk}\,e_k$, i.e., the $VVh_k$-couplings. This is not the case in a 3HDM. 
In a 3HDM, we have five couplings $e_i$ and ten couplings $\tilde\lambda_{ij}$. However, there exist relations between the $e_i$ and the $\tilde\lambda_{ij}$ that can be used to show that the number of independent couplings are not 5+10=15, but in fact the number of independent neutral gauge couplings is only 7, as discussed in appendix~\ref{app:3HDM-gauge}. One can for instance pick the seven independent couplings to consist of the five $e_i$ along with two of the $\tilde\lambda_{ij}$. Then, the remaining eight $\tilde\lambda_{ij}$ can always be expressed in terms of the seven independent couplings that we pick. Since these relations between the different gauge couplings are not as transparent in a 3HDM as in a 2HDM, we will just express the results in terms of the five $e_i$ and the ten $\tilde\lambda_{ij}$. 

In the 2HDM, the CP-odd contribution to the $ZWW$ vertex was directly proportional to the CP-odd invariant $\Im J_2$ (expressed in terms of masses and gauge couplings). 
In the 3HDM, one can also express the CP-odd contribution to the $ZWW$-vertex in terms of similar CP-odd invariants translated into a sum involving a combination of masses and the couplings appearing in the triangle diagrams. We did not calculate the contribution to the CP-odd form factor for our model, but we expect it to be a linear combination of CP-odd invariants $J_i$, where we include not only the 15 invariants of the theorem, but also those extra ones that are needed in order to form a ``linear algebraic basis" for all the CP-odd invariants of the model (see Section~\ref{sect-CP-invariants}).  CP-odd invariants that appear in those contributions are not necessarily the same as the set of invariants $J_1,\ldots, J_{15}$ whose simultaneous vanishing implies CP-conservation, but they will all vanish simultaneously with $J_1,\ldots, J_{15}$. 

\subsection{A Gedanken Experiment: $h_i^+h_i^-\to h_1^\pm h_2^\mp$}
In the SM, CP violation involves fermions. In the 2HDM, CP violation can take place in the non-fermionic sector, and will often involve the couplings to gauge bosons \cite{Mendez:1991gp,Lavoura:1994fv} (see also \cite{Grzadkowski:2014ada}).
In a recent study \cite{Haber:2022gsn}, it was pointed out that the simultaneous observation of three non-vanishing processes constitutes evidence for P-even CP-violating scalar exchange. But in that study, couplings to gauge bosons are essential. We shall here present a charge asymmetry that would constitute a P-even CP violation, not involving gauge bosons.

The Weinberg potential allows for CP violation in the scalar sector, without involvement of either fermions or gauge bosons. 
The simultaneous observation of several processes involving only scalars can reflect the existence of CP violation, even if any one of them could conserve CP (see an interesting discussion of CP violation in the bosonic sector in Ref~\cite{Haber:2022gsn}). However, such conclusion would only be valid if a large enough set of neutral states of the theory are known.\footnote{ For the 2HDM one would need all 3, for the 3HDM one would need at least 4.} Below, we will sketch a process that only involves scalars, but when comparing only with its charge-conjugated process, may display CP violation, without reference to other processes.
This phenomenon may be common to 3HDMs that violate CP.

The CP violation induced on the scalar sector by the complex vevs can be explored by comparing the two processes
\begin{subequations} \label{eq:chargedprocesses}
\begin{align}
h_i^+h_i^-&\to h_1^+h_2^-, \quad\text{and}\\
h_i^+h_i^-&\to h_1^-h_2^+.
\end{align}
\end{subequations}
Importantly, the initial state is even under CP, whereas the final state is not an eigenstate of CP since $h_1^\pm$ and $h_2^\pm$ have different masses.
While it may be hard to imagine actually measuring this asymmetry, it is interesting to see how the different couplings can conspire to produce a non-zero result.

\begin{figure}[htb]
\begin{center}
\includegraphics[scale=0.8]{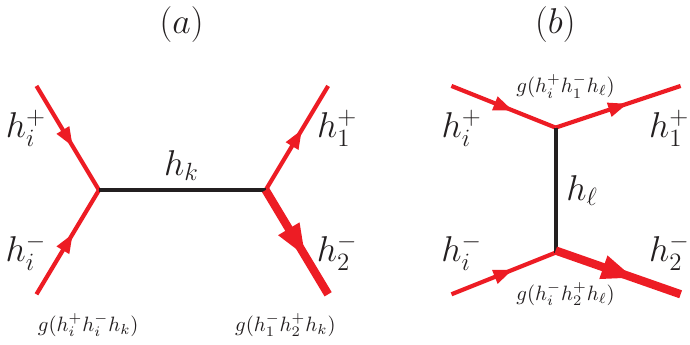}
\end{center}
\vspace*{-8mm}
\caption{Feynman diagrams contributing to $h_i^+h_i^-\to h_1^+h_2^-$, with $i\in{\{1,2\}}$ and 
$k,\ell\in{\{1,2,3,4,5\}}$.
The couplings $g$ refer to the related process where all fields are incoming.}
\label{Fig:feyn-a}
\end{figure}

At tree level, the process may proceed via the diagrams shown in Fig.~\ref{Fig:feyn-a}
with the amplitudes
\begin{subequations} \label{Eq:M_and_M}
\begin{align}
{\cal M}_\text{tree}(h_i^+h_i^-\to h_1^+h_2^-)&=-ig(h_i^+ h_i^-  h_1^-h_2^+)+{\cal M}_b^{+-}+{\cal M}_c^{+-}, \\
{\cal M}_\text{tree}(h_i^+h_i^-\to h_1^-h_2^+)&=-ig(h_i^+ h_i^-  h_1^+h_2^-)+{\cal M}_b^{-+}+{\cal M}_c^{-+},
\end{align}
\end{subequations}
where $i=1$ or 2 refers to the initial state and the subscripts ``b'' and ``c'' refer to the diagrams in Fig.~\ref{Fig:feyn-a}.
Furthermore,
\begin{subequations} 
\begin{align}
{\cal M}_b^{+-}&=\sum_{k=1}^5 [-ig(h_i^+h_i^-h_k)]\frac{i}{s-m_k^2}[-ig(h_2^+h_1^-h_k)], \\
{\cal M}_c^{+-}&=\sum_{\ell=1}^5 [-ig(h_i^+h_1^-h_\ell)]\frac{i}{t-m_\ell^2}[-ig(h_2^+h_i^-h_\ell)],
\end{align}
\end{subequations} 
where $g(xyz)$ refers to the trilinear coupling of fields $x$, $y$ and $z$.
While the Feynman diagrams are drawn in terms of outgoing particles in the final state, we define the vertices $g(xyz)$ in terms of all fields being incoming.

Replacing fields by their charge conjugated ones, individual vertex couplings will be replaced by their complex conjugated ones,\begin{equation}
g(h_i^+ h_i^-  h_1^+h_2^-)=[g(h_i^+ h_i^-  h_1^-h_2^+)]^*\quad \text{and}\quad
g(h_2^+h_1^-h_k)=[g(h_2^-h_1^+h_k)]^*.
\end{equation}
Charge-symmetric vertices are real.
In a CP-conserving theory, in diagram (b), only a CP-even $h_k$ would couple to the initial state, and thus only couple to the CP-even part of $h_1^\pm h_2^\mp$. Here, however, since $h_k$ has indefinite CP, it can couple to both the CP-even and the CP-odd part of the final state. The same is obviously true for the $h_\ell$ in the $t$-channel. 

Thus, the two expressions in (\ref{Eq:M_and_M}) are related by complex conjugation as follows (the kinematical variable $t$ will not change):
\begin{equation} \label{Eq:tree-equality}
i{\cal M}_\text{tree}(h_i^+h_i^-\to h_1^-h_2^+)=[i{\cal M}_\text{tree}(h_i^+h_i^-\to h_1^+h_2^-)]^*,
\end{equation}
and the two rates (\ref{eq:chargedprocesses}) will be the same at tree level, since the two couplings only differ by a phase.

In fact, the amplitudes can be expanded as follows,
\begin{subequations} \label{Eq:full-ampl}
\begin{align}
{\cal M}(h_i^+h_i^-\to h_1^+h_2^-)&=-i\sum_k A_k g_k(h_1^+h_2^-), \\
{\cal M}(h_i^+h_i^-\to h_1^-h_2^+)&=-i\sum_k A_k g_k(h_1^-h_2^+),
\end{align}
\end{subequations} 
where
\begin{equation}
g_k(h_1^-h_2^+)=g_k(h_1^+h_2^-)^*\quad\text{for all }k.
\end{equation}
At tree level the $A_k$ are all real, composed of real charge-symmetric couplings and propagators, leading to Eq.~(\ref{Eq:tree-equality}) and the equality of the two rates. 

However, at loop level, some $A_k$ will involve a loop integral, and thus in general be complex, given by the {\it same} complex function for both charge configurations. Thus, the two amplitudes ${\cal M}(h_i^+h_i^-\to h_1^+h_2^-)$ and ${\cal M}(h_i^+h_i^-\to h_1^-h_2^+)$  of Eq.~(\ref{Eq:full-ampl}) will {\it not} be related by complex conjugation, the two rates of Eq.~(\ref{eq:processes}) will differ. 

It should be noted that this CP asymmetry originates in the scalar sector, as a C asymmetry. No parity violating observable can be constructed involving only scalars. In reality, however, also other exchanges would contribute to the asymmetry, such as gauge bosons and fermions.

\begin{figure}[htb]
\begin{center}
\includegraphics[scale=0.8]{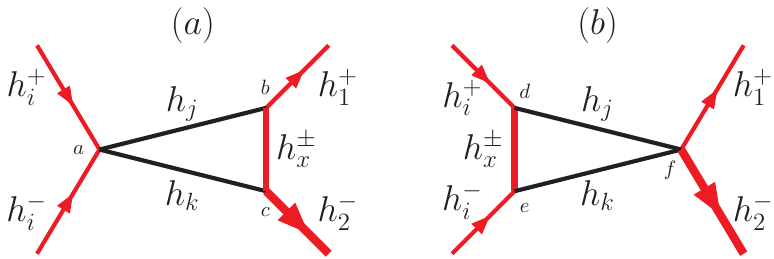}
\end{center}
\caption{Loop diagrams contributing to $h_i^+h_i^-\to h_1^+h_2^-$. 
Here, $i,x\in{\{1,2\}}$, and $j,k\in{\{1,2,3,4,5\}}$.
The couplings $a$...$f$ refer to the related process where all fields are incoming.}
\label{Fig:feyn-loop}
\end{figure}

For the CP-odd contribution to the $ZWW$ vertex, we argued that the amplitude itself would contain linear combinations of CP-odd invariants $J_i$. For the $h_i^+h_i^-\to h_1^\pm h_2^\mp$, the situation is somewhat more complicated. In order to extract the CP-odd parts of such processes, we need to form an asymmetry involving the two processes of eq.~(\ref{eq:chargedprocesses}), and it will contain differences of the squared amplitudes of those two processes. The contributions from the tree-level diagrams alone will vanish, so to lowest order we expect to get contributions from the interference between the tree-level diagrams and one-loop diagrams (there are more one-loop diagrams than is shown in Fig.~\ref{Fig:feyn-loop}). We expect that this lowest-order contribution to the asymmetry can be written as a linear combination of the elements of the ``linear algebraic basis" of CP-odd invariants $J_i$ (more than 15).

\section{Scalar phenomenology}
\label{sect:light-states}

While details of the Yukawa sector would be relevant for a detailed model, there are several aspects of the scalar sector that are generic.
We summarise here the most important features of the parameter points that are compatible with the imposed constraints. 

\subsection{Mass distributions}
\label{sect:mass-distributions}

\begin{figure}[htb]
\begin{center}
\includegraphics[scale=0.30]{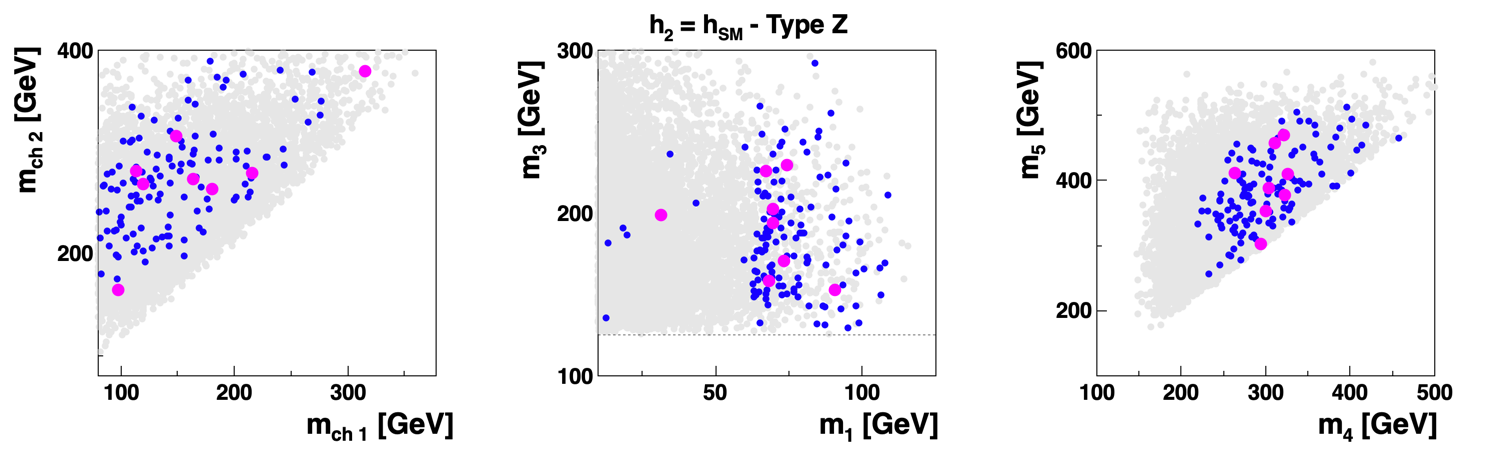}
\includegraphics[scale=0.30]{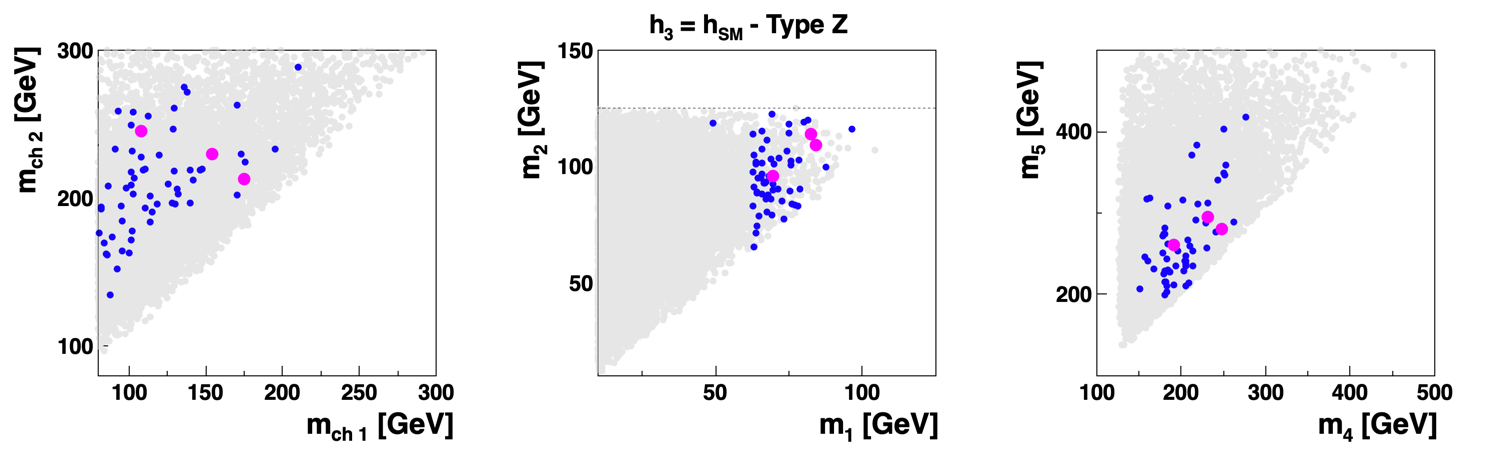}
\end{center}
\vspace*{-8mm}
\caption{Scatter plots of masses allowed by the initial scan in Ref.~\cite{Plantey:2022jdg}, as well as those discussed in section~\ref{sect:constraints}.  We recall that the masses are ordered, $m_1<m_2<m_3<m_4<m_5$. Grey background: parameter points in the initial scan \cite{Plantey:2022jdg}; blue: parameter points satisfying all constraints for $h_2$ (top) or $h_3$ (bottom) identified as $h_\text{SM}$ at 125~GeV. The (larger) purple points satisfy a stronger EDM constraint, $|d_e|<5\cdot10^{-29}e\cdot\text{cm}$.
Parts of these mass--mass planes are empty due to our convention for the labelling of states.}
\label{Fig:masses-jM2-3}
\end{figure}

The different constraints remove the majority of the 2.8 million ``raw'' scan points of Refs.~\cite{Plantey:2022jdg,Plantey:2022gwj}, the remaining points are shown in blue and purple in fig.~\ref{Fig:masses-jM2-3}. Since the quartic couplings ($\lambda$s) are subject to the perturbativity constraints, and the coefficients of the bilinear terms are related to the $\lambda$s by the minimisation conditions, the masses are subject to an upper bound at the electroweak scale. Furthermore, the vicinity of the U(1)$\times$U(1) symmetry favours low masses \cite{Plantey:2022gwj}.

In this figure, the blue points satisfy a rather loose constraint on the electron EDM, $|d_e|<10^{-27}e\cdot\text{cm}$, whereas the purple ones satisfy $|d_e|<5\cdot10^{-29}e\cdot\text{cm}$, There is no obvious systematic difference between the two distributions.
The mass ranges for the different states may be characterised as follows.
\smallskip

\noindent
$h_2=h_\text{SM}$:
\begin{subequations}
\begin{alignat}{2}
m_{\text{ch}\,1}&\in\{80~\text{GeV},\sim 300~\text{GeV}\},&\quad
m_{\text{ch}\,2}&\in\{150~\text{GeV},\sim400~\text{GeV}\}, \\
m_1&\in\{\sim 10~\text{GeV},m_{h_\text{SM}}\}, &\quad 
m_3&\in\{m_{h_\text{SM}},\sim300~\text{GeV}\}, \\
m_4&\in\{\sim250~\text{GeV},\sim450~\text{GeV}\}, &\quad
m_5&\in\{\sim 250~\text{GeV}, \sim 500~\text{GeV}\}
\end{alignat}
\end{subequations}
\smallskip

\noindent
$h_3=h_\text{SM}$:
\begin{subequations}
\begin{alignat}{2}
m_{\text{ch}\,1}&\in\{80~\text{GeV},\sim 200~\text{GeV}\},&\quad
m_{\text{ch}\,2}&\in\{150~\text{GeV},\sim300~\text{GeV}\}, \\
m_1&\in\{\half m_{h_\text{SM}},\sim90~\text{GeV}\}, &\quad 
m_2&\in\{\half m_{h_\text{SM}},m_{h_\text{SM}}\}, \\
m_4&\in\{\sim150~\text{GeV},\sim300~\text{GeV}\}, &\quad
m_5&\in\{\sim 200~\text{GeV}, \sim 400~\text{GeV}\}
\end{alignat}
\end{subequations}

As a general feature we note that there are very few points with $m_1<m_{h_\text{SM}}/2$. This holds both for $h_2=h_\text{SM}$ and for $h_3=h_\text{SM}$.
Also, the very low values of the EDM (purple) are seemingly ``randomly'' distributed over these ranges.

Here, $m_{\text{ch}\,1}>80~\text{GeV}$ is a ``hard'' cut inherited with the original data set \cite{Plantey:2022jdg}. We did not impose a similar lower cut on the scalar masses. One might worry that the invisible width of the $Z$ would impose a lower bound. While there is no $Zh_ih_i$ coupling, there is a $Zh_ih_j$ coupling, proportional to the $\tilde\lambda_{ij}$ of Table~\ref{table:couplings}. This coupling (referred to as $P_{ij}$ in ref.~\cite{Plantey:2022jdg}) can be expressed in terms of the Higgs basis rotation matrix $O$ of Eq.~(\ref{Eq:rot-matrx-55}) as
\begin{equation}
\tilde\lambda_{ij}=v[O_{i2}O_{j4}+O_{i3}O_{j5}-(i\leftrightarrow j)].
\end{equation}
The coupling is also proportional to the c.m.\ momentum in the final state, given by the K\"allen function 
$\lambda_K=m_Z^4+m_1^4+m_2^4-2m_Z^2(m_1^2+m_2^2)-2m_1^2m_2^2$ as $p_\text{c.m.}^2=\lambda_K/(4m_Z^2)$.

In terms of this notation, the $Z\to h_1h_2$ decay width is given by
\begin{equation}
\Delta\Gamma_Z=\frac{g^2}{24\pi\,\cos^2\theta_W}
\tilde\lambda_{12}^2 
\frac{p_\text{c.m.}^3}{v^2m_Z^2},
\end{equation}
where $g$ is the SU(2) coupling.
The largest value encountered among the surviving scan points is 68~MeV, whereas the average is 5~MeV. These numbers should be compared to the $Z$ invisible width of 499~MeV \cite{ParticleDataGroup:2022pth}, from which the number of active neutrinos is determined. It is clear that the $Z\to h_1 h_2$ decays do not further constrain the parameters of the potential.
There are two reasons this decay rate is small: (i) the amplitude contains a momentum factor, and (ii) the smallness of $\tilde\lambda_{12}$, which effectively projects out the CP-odd component of the $h_1h_2$ state.

\begin{figure}[htb]
\begin{center}
\includegraphics[scale=0.30]{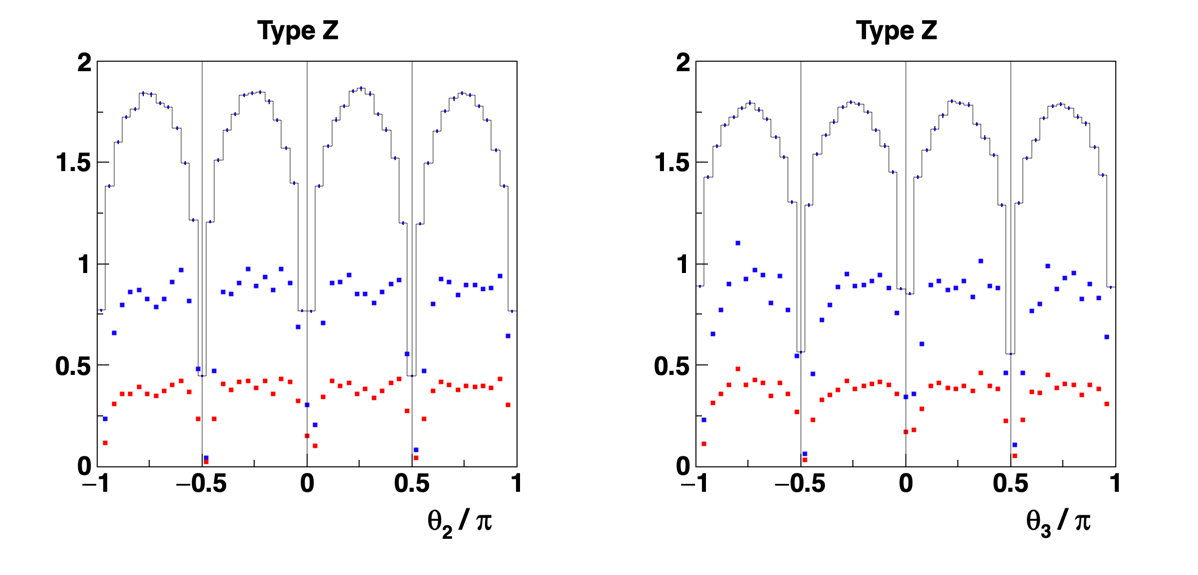}
\end{center}
\vspace*{-8mm}
\caption{Scatter point distributions vs $\theta_2$ and $\theta_3$. The (black) histograms at the top show the ``input'' distributions of scan points from Ref.~\cite{Plantey:2022jdg}. There are fewer points at $\theta_i=n\pi/2$, since the minimisation conditions would force $\lambda_2$ and $\lambda_3$ to exceed the perturbativity bound in these limits. The blue points represent the ratios of points surviving the constraints of section~\ref{sect:constraints}, except for the EDM, normalized to the initial distribution. The red points show this same ratio, but restricted to parameter points for which $m_1<50~\text{GeV}.$ The absolute normalizations contain no information.}
\label{Fig:dist_vs_theta}
\end{figure}

\subsection{Some $\theta$ distributions}
\label{sect:theta}

We show in Fig.~\ref{Fig:dist_vs_theta} scatter plots of scan points vs $\theta_2$ and $\theta_3$ we adopt the convention $\theta_1=0$, such that these become abbreviations for $\theta_2-\theta_1$ and $\theta_3-\theta_1$). As a reference, we display (black) histograms at the top showing the ``input'' distributions of scan points from Ref.~\cite{Plantey:2022jdg}. Shown here are points for which $h_2$ or $h_3$ can be identified as the SM Higgs boson at 125~GeV. There are fewer points around $\theta_i=n\pi/2$. This is due to a depletion of these regions caused by the perturbativity constraints imposed on $\lambda_2$ and $\lambda_3$. The way we solve the minimisation conditions, is to express $\lambda_2$ and $\lambda_3$ in terms of $\lambda_1$, $\theta_2$ and $\theta_3$. These constraints make $\lambda_2$ and $\lambda_3$ diverge for $\theta_i\to n\pi/2$.

However, there seems to be a further depletion of these limits, which corresponds to CP conservation (see Appendix~\ref{app:CP-conserved}), as illustrated by the blue points. These blue points show the ratio of two distributions: the parameter points surviving the constraints of (1) $S$, $T$, $U$ (section~\ref{sect:STU}), (2) $h_\text{SM}\to\gamma\gamma$ (section~\ref{sect:h-gaga}), (3) $\bar B\to X_s\gamma$ (section~\ref{sect:Bsga}), divided by the input distribution shown in black at the top.
While neither of these constraints refers to CP violation, there is a relative loss of points in the regions of CP conservation, as illustrated by the dips in the blue-point distributions. 

\begin{figure}[htb]
\begin{center}
\includegraphics[scale=0.30]{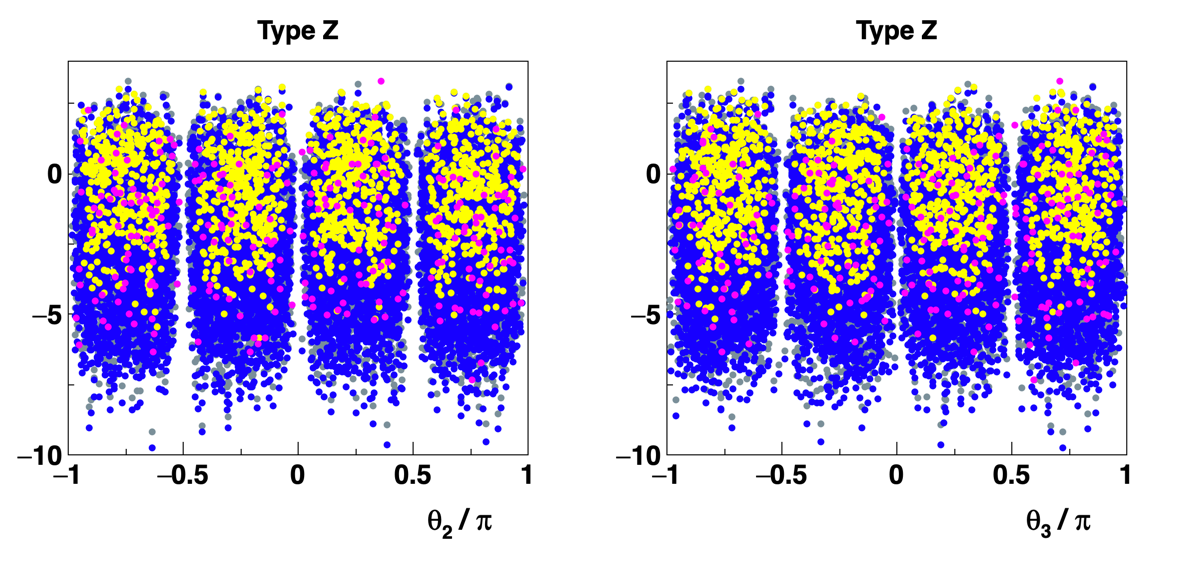}
\end{center}
\vspace*{-8mm}
\caption{Scatter plots of logarithms of the sum and maxima of the squares of the 15 invariants. Grey: $\log_{10}(\sum_i|J_i^2|)$ (``sums''), blue: $\log_{10}(\max|J_i^2|)$ vs $\theta_2/\pi$ and $\theta_3/\pi$. Yellow: ``sums'' for which $m_1>50~\text{GeV}$, purple: ``sums'' for which $|d_e|<5\cdot 10^{-29}e\cdot\text{cm}$.}
\label{Fig:invar}
\end{figure}

This ``loss of points'' for $\theta_i$ near $n\pi/2$ is also seen in Fig.~\ref{Fig:invar}, where we show logarithms of the squared invariants vs $\theta_2$ and $\theta_3$. This figure has some similarity to Fig.~\ref{Fig:invar-raw}, except that we now have imposed all the discussed constraints. The grey points (``sums'') have all the discussed constraints imposed, except for the electron electric dipole moment. Likewise, the blue and yellow points refer to the ``max'' and ``sum'', the latter excluding low masses. If we drop the constraint on $m_1$, but require the electron EDM to be below $5\cdot10^{-29}e\cdot\text{cm}$ we obtain the purple points. Some values below $10^{-30}e\cdot\text{cm}$ were also encountered in the scan, as shown in Fig.~\ref{Fig:max-J_i}.

In order to illustrate how low scalar masses are correlated with less CP violation, we illustrate in Fig.~\ref {Fig:dist_vs_theta} the distribution that survives if we impose an upper bound of 50~GeV on $m_1$. Also, in Fig.~\ref{Fig:invar} we show in yellow the distribution that emerges if we impose a lower bound of 50~GeV on $m_1$.

It may be surprising that values of $\theta_2$ and $\theta_3$ well away from $n\pi/2$ are compatible with a small value of the electron EDM. A possible explanation is as follows. Suppose the EDM can be expanded in the CP-violating invariants,
\begin{equation}
d_e = \sum_{i=1}^{15} c_i\, J_i.
\end{equation}
Then, without requiring the individual $J_i$ to all be small, one could imagine cancellations among the 15 terms. See Eq.~(\ref{Eq:J_i-explicit}).

\subsection{Properties of the light states}

A very important aspect of the Weinberg 3HDM potential is that, when requiring that some $h_i$ has SM-like couplings to $WW$ (and thus referred to as $h_\text{SM}$), then the potential tends to lead to one or two neutral states lighter than the $h_\text{SM}$ \cite{Plantey:2022jdg,Plantey:2022gwj}.

The light neutral scalars can be produced directly in $pp$ collisions at the LHC (dominantly via gluon-gluon fusion), and also, if $m_1<(m_j=m_\text{SM})/2$, by decay of the SM Higgs, $h_\text{SM}\to h_1 h_1$, at a rate given by
\begin{equation}
\Gamma=\frac{v^2}{32\pi\, m_j}\left|\hat g_{h_jh_1h_1}\right|^2\sqrt{1-\frac{4\,m_1^2}{m_j^2}},
\end{equation}
where $h_j=h_\text{SM}$ and a factor of $v$ has been extracted from the $h_j h_1 h_1$ coupling, $g=v\hat g$.

While it may be kinematically allowed for the SM Higgs candidate to decay to the lighter states, we impose the condition that the rate for such decays does not exceed the invisible width, $\Gamma_\text{inv}$ of the observed SM-like Higgs boson, since any subsequent decay to a low-energy fermion-antifermion pair might be hard to detect.
According to the PDG \cite{ParticleDataGroup:2022pth}, $\Gamma_\text{inv}<0.19\cdot\Gamma_\text{tot}$. The latter is however poorly determined, to $3.2\mycom{+2.8}{-2.2}\text{ MeV}$. We adopt the generous constraint $\Gamma_\text{inv}=0.19\times 6~\text{MeV}$, and require
\begin{alignat}{2}
h_3&=h_\text{SM}: &\quad \Gamma(h_3\to h_1h_1)&+\Gamma(h_3\to h_2h_1)<\Gamma_\text{inv}, \label{Eq:Gamma3}\\
h_2&=h_\text{SM}: &\quad \Gamma(h_2\to h_1h_1)&<\Gamma_\text{inv}.  \label{Eq:Gamma2}
\end{alignat}
Light neutral scalars can not a priori be excluded, provided they only couple weakly \cite{Feng:2017uoz,Drechsel:2018mgd,Kling:2022uzy,Robens:2023bzp}.

Since the coupling to $WW$ (and $ZZ$) is strongly suppressed, the $h_i$ (lighter than $h_\text{SM}$) would dominantly decay to $b\bar b$ and $\tau\bar\tau$ with partial decay rates given by
\begin{subequations}
\begin{align}
\Gamma(h_i\to b\bar b)&=\frac{3}{8\pi} m_i\left(\frac{m_b}{v}\right)^2 \beta
\left\{\beta^2|\kappa_{h_ib\bar b}^S|^2+|\kappa_{h_ib\bar b}^P|^2\right\}, \\
\Gamma(h_i\to \tau\bar \tau)&=\frac{1}{8\pi} m_i\left(\frac{m_\tau}{v}\right)^2 \beta
\left\{\beta^2|\kappa_{h_i\tau\bar \tau}^S|^2+|\kappa_{h_i\tau\bar \tau}^P|^2\right\},
\end{align}
\end{subequations}
where the couplings $\kappa^S$ and $\kappa^P$ are defined in Eq.~(\ref{Eq:h-gaga}).
For Type~Z Yukawa interactions the coupling modifiers $\kappa$ for $b$-quarks and for $\tau$-leptons will be independent.

\begin{figure}[htb]
\begin{center}
\includegraphics[scale=0.30]{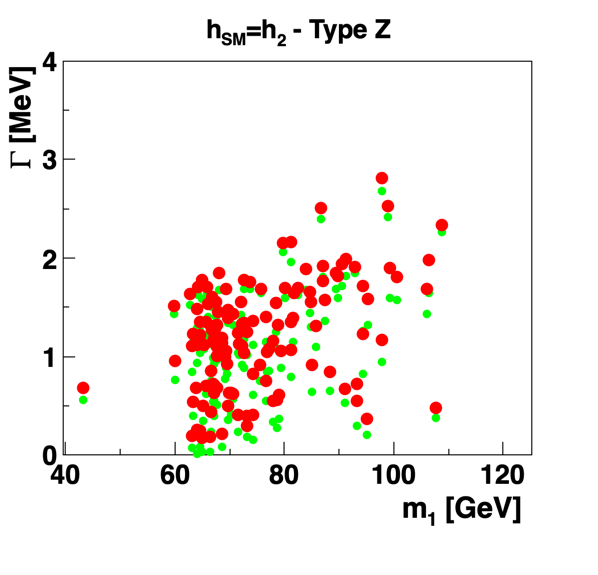}
\end{center}
\vspace*{-8mm}
\caption{Scatter plots of decay widths for $h_1$: $\Gamma(h_1\to b\bar b)$ (green) and $\Gamma(h_1\to b\bar b)+\Gamma(h_1\to \tau\bar \tau)$ (red) vs mass, when $h_2=h_\text{SM}$.}
\label{Fig:widths-jM2}
\end{figure}

In Fig.~\ref{Fig:widths-jM2} we show the dominant decay width $\Gamma(h_i\to b\bar b)$ as well as the approximate total width, $\Gamma(h_i\to b\bar b)+\Gamma(h_i\to \tau\bar \tau)$ vs $m_1$ for the case when $h_2=h_\text{SM}$. The total width is of the order 1--2~MeV, making a discovery of such decays very challenging.

The $Z$ boson cannot decay to a pair of identical spinless particles, whatever CP properties they might have. However, if also $m_2<\half m_Z$, then this mass region would be constrained by the $Z$ width.\footnote{See, however, the discussion at the end of section~\ref{sect:mass-distributions}.}
Such decays would lead to final states like $h_1h_2\to b\bar b b\bar b$ and $h_1h_2\to b\bar b\tau\bar\tau$ with $m(f\bar f)<\half m_Z$.

\begin{figure}[htb]
\begin{center}
\includegraphics[scale=0.30]{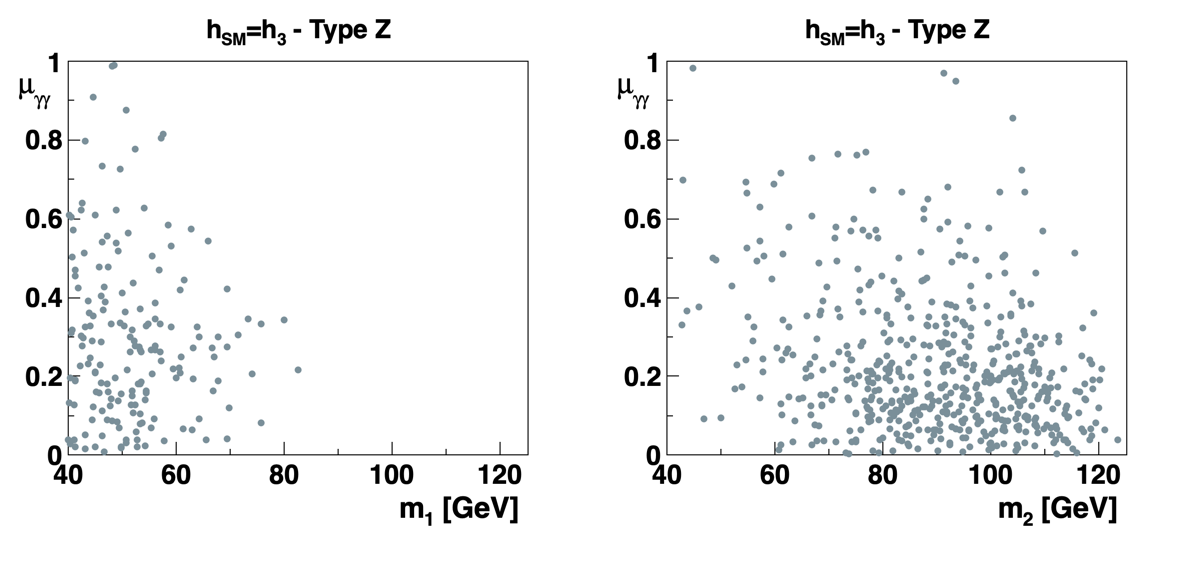}
\end{center}
\vspace*{-8mm}
\caption{Scatter plots of relative digamma rate $\mu_{\gamma\gamma}$ vs mass, for the two lightest states, when $h_3=h_\text{SM}$.
Left: $\mu_{\gamma\gamma}$ for $h_1\to\gamma\gamma$; right: for $h_2\to\gamma\gamma$. In the latter case, $m_1<m_2$.}
\label{Fig:digamma-jM3}
\end{figure}

\subsubsection{The 95~GeV mass region}

CMS has noted the possibility of an excess in the digamma rates at an invariant mass of around 95~GeV \cite{CMS:2018cyk}. This has recently been vigorously explored, in a 2HDM plus a real or complex singlet \cite{Biekotter:2019kde,Heinemeyer:2021mnz,Heinemeyer:2021msz,Biekotter:2022jyr,Biekotter:2022abc,Biekotter:2023jld,Biekotter:2023oen}, in the related NMSSM \cite{Cao:2019ofo}, as well as in the 2HDM with unconventional Yukawa sectors \cite{Benbrik:2022azi,Azevedo:2023zkg}. Several of these studies have also addressed the LEP excess in $b\bar b$ around the same energy \cite{LEPHiggsWorkingGroupforHiggsbosonsearches:2001dnp,McNamara:2002nk} and found it to be compatible with the model considered. More recently, also some excess in the $\tau^+\tau^-$ channel has been observed around the same mass \cite{CMS:2022goy}
and analysed within the same framework \cite{Iguro:2022dok,Azevedo:2023zkg}. For an overview of these excesses, see Ref.~\cite{Crivellin:2023zui}, but it should also be noted that ATLAS does not see any excess \cite{ATLAS-CONF-2023-035,ATLAS:2024itc}.

\begin{figure}[htb]
\begin{center}
\includegraphics[scale=0.30]{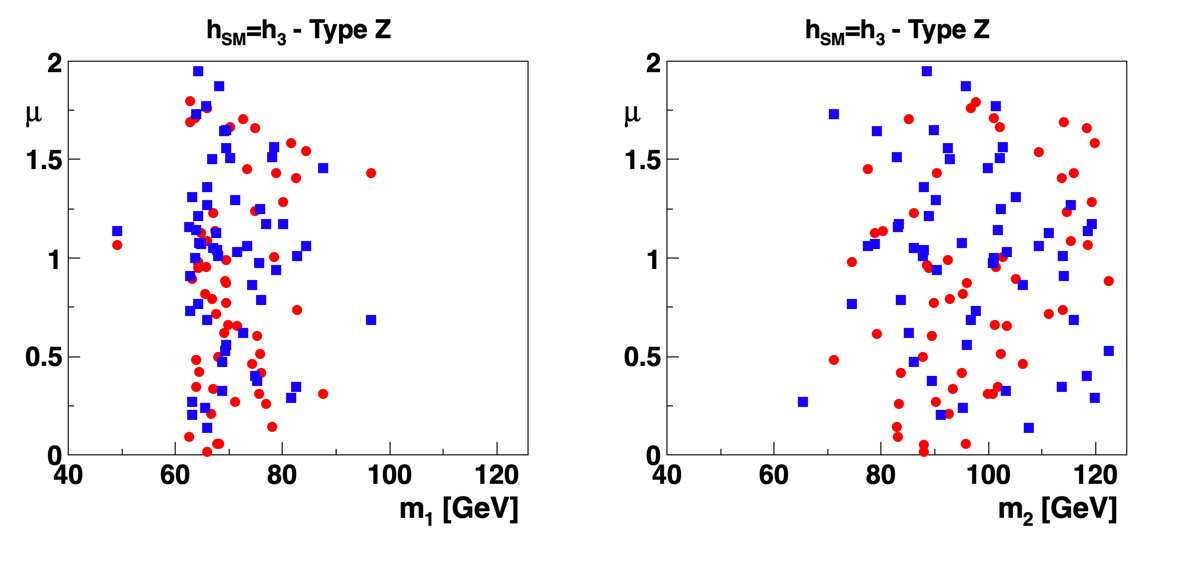}
\end{center}
\vspace*{-8mm}
\caption{Scatter plots of signal strengths $\mu_{bb}$ (red bullets) and $\mu_{\tau\tau}$ (blue squares) vs mass, for the two lightest states, $h_1$ and $h_2$, when $h_3=h_\text{SM}$.}
\label{Fig:signalstrength-jM3}
\end{figure}

It is obviously of interest to see how the light neutral states of the Weinberg potential (those with mass below 125~GeV) compare with the claimed excesses. We show in Fig.~\ref{Fig:digamma-jM3} the digamma rate for $h_1\to\gamma\gamma$ and $h_2\to\gamma\gamma$, relative to the rate for an SM scalar at the same mass decaying to two photons. Here, ``SM scalar'' means that we calculate the digamma rate for a scalar at the considered mass, with couplings to fermions and the $W$ the same as for the Standard Model.  Among the parameter points that
survive the applied constraints, some predict a di-photon rate near the
reported excess.

Next, we turn to fermion final states. The light states will decay to $b\bar b$ and $\tau \bar\tau$ according to the Yukawa couplings given by Eq.~(\ref{Eq:Yuk-kappa}). The relevant signal strengths will be given by
\begin{equation}
\mu_{h_i b\bar b}=(\kappa_{h_ib\bar b}^S)^2+\frac{m_i^2}{m_i^2-4m_b^2}(\kappa_{h_ib\bar b}^P)^2,
\end{equation}
and similarly for $\mu_{h_i \tau\bar \tau}$.
We also show, in Fig.~\ref{Fig:signalstrength-jM3}, the signal strength for $h_i\to b\bar b$ and $h_i\to \tau\bar \tau$, for $h_1$  and $h_2$, assuming $h_3=h_\text{SM}$. A wide range of values is populated, basically from 0 up to 2, for the parameter points that are compatible with the constraints. The case $h_2=h_\text{SM}$ is similar, but then obviously only $h_1$ is available as a ``light'' state.

It should also be stressed that, as compared with an SM-like scalar, the light scalars of the Weinberg potential have a much reduced decay width to $\gamma\gamma$, because of the reduced $h_1WW$ coupling, illustrated in fig.~4 of Ref.~\cite{Plantey:2022jdg}. The dominant decay would be to $b\bar b$.

\section{Concluding remarks}
\label{sect:conclusions}

We have provided a full set of CP-odd invariants for the real Weinberg potential with complex vevs. Complex vevs normally signal CP violation. However, these invariants would all have to vanish for CP not to be violated. Thus, a full set of solutions is given, identifying limits in which they all vanish and CP is conserved.

Furthermore, we have illustrated some aspects of the phenomenology of this potential, which tends to yield two light CP-mixed states. Even with sizeable CP-violating phases in the scalar sector, the model can remain consistent with the current electron EDM bound.
This shows that it is not the presence of complex vevs {\it per se}, but the size of the induced CP-violating effects that matter for the EDM constraints. Notice that the structure of the vevs is basis dependent.

While our main focus has been on exhibiting mathematical features of the Weinberg 3HDM potential, it is clearly also of interest to ask how experimental data could support it or reject it. In particular, if one discovers an additional scalar, can one argue for the Weinberg 3HDM potential rather than some version of the 2HDM? In the parameter space explored, light scalars and sizeable $Zh_\text{SM}h_i$ couplings appear frequently. Also, it should be noted that the Weinberg potential can lead to {\it two} light neutral scalars, as well as at least one pair of light charged scalars.

\section*{Acknowledgements}
We thank Robin Plantey and Marius A. Solberg for contributions to the early stages of this work.
We also thank the referee for prompting us to clarify several issues.
PO is supported in part by the Research Council of Norway.
The work of MNR was partially supported by Funda\c c\~ ao 
para  a  Ci\^ encia e a Tecnologia  (FCT, Portugal)  through  the  projects  
CFTP-FCT Unit UIDB/00777/2020 and UIDP/00777/2020, CERN/FIS-PAR/0002/2021, 
2024.02004.CERN, which are  partially  funded  through
POCTI  (FEDER),  COMPETE,  QREN  and  EU. We also thank 
CFTP/IST/University of Lisbon and the University of Bergen, where collaboration visits took place.

It is also a pleasure to thank Thomas Hahn for advice on the diagonalization of the mass matrices \cite{Hahn:2006hr}, as well as Mihail Misiak, Francesca Borzumati and Andreas Crivellin for discussions of the $\bar B \to X_s\gamma$ constraints.

\appendix

\section{CP-conserving limits}
\label{app:CP-conserved}
Let $\{i,j,k\}$ be any permutation of $\{1,2,3\}$. We shall in the following interpret $\lambda_{ij}=\lambda_{ji}$  and $\lambda_{ij}'=\lambda_{ji}'$ whenever $i>j$. We assume the most general form of the vevs, given by Eq.~(\ref{Eq:vacuum-general}).

Below, we list all cases where CP is conserved, i.e., when Eq.~(\ref{Eq:the_equations}) is satisfied, together with the solutions of the minimisation conditions. They will be ordered according to the number of vanishing vevs.

Furthermore, for all the presented cases, we shall also give the explicit form of the basis transformation that takes us to a real basis, i.e., a potential with only real parameters as well as real vevs. The transformation is of the form
\begin{eqnarray}
\begin{pmatrix}
	\bar{\phi}_1 \\
	\bar{\phi}_2 \\
	\bar{\phi}_3
\end{pmatrix}
=U \begin{pmatrix}
	{\phi}_1 \\
	{\phi}_2 \\
	{\phi}_3
\end{pmatrix},
\end{eqnarray}
and we present the matrix $U$ for each case.

All the cases listed below are particular cases of the CP-conserving solutions presented in section~\ref{sect:stationary-point}.

\subsection{Solutions with two vanishing vevs}
\begin{equation}
	v_i=0,\quad v_j=0,\quad m_{kk}=\lambda _{kk} v_k^2.
\end{equation}
\begin{eqnarray}
U=
\left(
\begin{array}{ccc}
	e^{-i \theta _k} & 0 & 0 \\
	0 & e^{-i \theta _k} & 0 \\
	0 & 0 & e^{-i \theta _k} \\
\end{array}
\right),\quad \text{for }\{i,j,k\}=\{1,2,3\}.	
\end{eqnarray}
\subsection{Solutions with one vanishing vev}
There are seven cases. In addition to $v_i=0$ they require

\noindent
{\it Case~1-1:}
\vspace*{-10pt}
\begin{subequations}
	\begin{align}
		m_{jj}&=\lambda _{jj} v_j^2+\frac{1}{2} \left(\lambda _{jk}'+\lambda _{jk}\right) v_k^2,\\
		m_{kk}&=\lambda _{kk} v_k^2+\frac{1}{2} \left(\lambda _{jk}'+\lambda _{jk} \right)v_j^2,\quad
		\lambda _i=\lambda _j=0.
	\end{align}
\end{subequations}
\begin{eqnarray}
U=
\left(
\begin{array}{ccc}
	e^{-i \theta _j} & 0 & 0 \\
	0 & e^{-i \theta _j} & 0 \\
	0 & 0 & e^{-i \theta _k} \\
\end{array}
\right)
,\quad \text{for }\{i,j,k\}=\{1,2,3\}.	
\end{eqnarray}
\noindent
{\it Case~1-2:}
\vspace*{-10pt}
\begin{subequations}
	\begin{align}
		m_{jj}&=m_{kk}=\lambda _{jj} \left(v_j^2+v_k^2\right),\quad\lambda _i=0,\quad\lambda _{ik}'=\lambda _{ij}',\quad
		\lambda _{ik}=\lambda _{ij},\quad \lambda _{kk}=\lambda _{jj},\\
		\lambda _k&=\lambda _j,\quad
		\lambda _{jk}'=2 \lambda _{jj}-\lambda _{jk}.
	\end{align}
\end{subequations}
\begin{eqnarray}
	U=
	\left(
	\begin{array}{ccc}
		e^{i \alpha} & 0 & 0 \\
		0 & \frac{ ie^{i \left(\alpha-\beta\right)}}{\sqrt{2}} & \frac{e^{i \left(\alpha-\beta \right)}}{\sqrt{2}} \\
		0 & -\frac{ ie^{i \left(\alpha+\beta  \right)}}{\sqrt{2}} & \frac{e^{i \left(\alpha+\beta \right)}}{\sqrt{2}} \\
	\end{array}
	\right),
	\quad \text{for }\{i,j,k\}=\{1,2,3\}.	
\end{eqnarray}
where
\begin{eqnarray}
	\alpha&=& \arctan\left(\frac{v_j^2 \cos 2 \theta _j+ v_k^2 \cos 2 \theta _k-\sqrt{v_j^4+v_k^4+ 2 v_j^2 v_k^2 \cos 2 \left(\theta _j-\theta _k\right)}}{v_j^2 \sin 2 \theta _j+v_k^2 \sin 2 \theta _k}\right),\\
	\beta&=& \arctan\left(\frac{v_j^2-v_k^2+ \sqrt{v_j^4+v_k^4+ 2 v_j^2 v_k^2 \cos 2 \left(\theta _j-\theta _k\right)}}{2 v_j v_k\cos \left(\theta _j-\theta _k\right)}\right).
\end{eqnarray}

\noindent
{\it Case~1-3:}
\vspace*{-10pt}
\begin{subequations}
	\begin{align}
		m_{jj}&=m_{kk}=\lambda _{jj} \left(v_j^2+v_k^2\right),\quad \lambda _i=0,\quad\lambda _{ik}'=\lambda _{ij}',\quad
		\lambda _{ik}=\lambda _{ij},\quad\lambda _{kk}=\lambda _{jj},\\
		\lambda _k&=-\lambda _j,\quad
		\lambda _{jk}'=2 \lambda _{jj}-\lambda _{jk}.
	\end{align}
\end{subequations}
\begin{eqnarray}
	U=
	\left(
	\begin{array}{ccc}
		e^{i \alpha} & 0 & 0 \\
		0 & \frac{ e^{i \left(\alpha-\beta\right)}}{\sqrt{2}} & \frac{e^{i \left(\alpha-\beta \right)}}{\sqrt{2}} \\
		0 & -\frac{ e^{i \left(\alpha+\beta  \right)}}{\sqrt{2}} & \frac{e^{i \left(\alpha+\beta \right)}}{\sqrt{2}} \\
	\end{array}
	\right),
	\quad \text{for }\{i,j,k\}=\{1,2,3\}.	
\end{eqnarray}
where
\begin{eqnarray}
	\alpha&=& \arctan\left(\frac{v_j^2 \cos 2 \theta _j- v_k^2 \cos 2 \theta _k-\sqrt{v_j^4+v_k^4- 2 v_j^2 v_k^2 \cos 2 \left(\theta _j-\theta _k\right)}}{v_j^2 \sin 2 \theta _j-v_k^2 \sin 2 \theta _k}\right),\\
	\beta&=& \arctan\left(\frac{v_j^2-v_k^2- \sqrt{v_j^4+v_k^4- 2 v_j^2 v_k^2 \cos 2 \left(\theta _j-\theta _k\right)}}{2 v_j v_k\sin \left(\theta _j-\theta _k\right)}\right).
\end{eqnarray}

\noindent
{\it Case~1-4:}
\vspace*{-10pt}
\begin{subequations}
	\begin{align}
		m_{jj}&=m_{kk}=\frac{1}{2} \left(\lambda _{jk}'+2 \lambda _{jj}+\lambda _{jk}\right)v_j^2 ,\quad\lambda _i=0,\quad
		\lambda _{ik}'=\lambda _{ij}',\quad\lambda _{ik}=\lambda _{ij},\\
		\lambda _{kk}&=\lambda _{jj}, \quad \lambda _k=\lambda _j,\quad v_k=v_j.
	\end{align}
\end{subequations}
\begin{eqnarray}
	U=
	\left(
	\begin{array}{ccc}
		e^{-\frac{1}{2} i \left(\theta _j+\theta _k\right)} & 0 & 0 \\
		0 & \frac{1}{\sqrt{2}} & \frac{e^{-i \left(\theta _j+\theta _k\right)}}{\sqrt{2}} \\
		0 & -\frac{i}{\sqrt{2}} & \frac{i e^{-i \left(\theta _j+\theta _k\right)}}{\sqrt{2}} \\
	\end{array}
	\right)
	,\quad \text{for }\{i,j,k\}=\{1,2,3\}.	
\end{eqnarray}

\noindent
{\it Case~1-5:}
\vspace*{-10pt}
\begin{subequations}
	\begin{align}
		m_{jj}&=m_{kk}=\frac{1}{2} \left(\lambda _{jk}'+2 \lambda _{jj}+\lambda _{jk}\right)v_j^2 ,\quad \lambda _i=0,\quad
		\lambda _{ik}'=\lambda _{ij}',\quad \lambda _{ik}=\lambda _{ij},\\
		\lambda _{kk}&=\lambda _{jj},\quad \lambda _k=-\lambda _j,\quad v_k=v_j.
	\end{align}
\end{subequations}
\begin{eqnarray}
	U=
	\left(
	\begin{array}{ccc}
		e^{-\frac{1}{2} i \left(\theta _j+\theta _k-\frac{\pi}{2}\right)} & 0 & 0 \\
		0 & \frac{e^{-\frac{1}{2} i \left(\theta _j+\theta _k\right)}}{\sqrt{2}} & \frac{e^{-\frac{1}{2} i \left(\theta _j+\theta _k\right)}}{\sqrt{2}} \\
		0 & -\frac{i e^{-\frac{1}{2} i \left(\theta _j+\theta _k\right)}}{\sqrt{2}} & \frac{i e^{-\frac{1}{2} i \left(\theta _j+\theta _k\right)}}{\sqrt{2}} \\
	\end{array}
	\right)
	,\quad \text{for }\{i,j,k\}=\{1,2,3\}.	
\end{eqnarray}

\noindent
{\it Case~1-6:}
\vspace*{-10pt}
\begin{subequations}
	\begin{align}
		m_{jj}&=\lambda _{jj} v_j^2+\frac{1}{2} \left(2 \lambda _i+\lambda _{jk}'+\lambda _{jk}\right)v_k^2 ,\\
		\,m_{kk}&= \lambda _{kk} v_k^2+\frac{1}{2} \left(2 \lambda _i +\lambda _{jk}' +\lambda _{jk} \right)v_j^2,\quad
		\sin(\theta _k-\theta _j)=0.
	\end{align}
\end{subequations}
\begin{eqnarray}
	U=
\left(
\begin{array}{ccc}
	e^{-i \theta _j} & 0 & 0 \\
	0 & e^{-i \theta _j} & 0 \\
	0 & 0 & e^{-i \theta _j} \\
\end{array}
\right)
	,\quad \text{for }\{i,j,k\}=\{1,2,3\}.	
\end{eqnarray}
\noindent
{\it Case~1-7:}
\vspace*{-10pt}
\begin{subequations}
	\begin{align}
		m_{jj}&=\lambda _{jj} v_j^2+\frac{1}{2} \left(-2 \lambda _i+\lambda _{jk}'+\lambda _{jk}\right) v_k^2,\\
		m_{kk}&=\lambda _{kk} v_k^2+\frac{1}{2} \left(-2 \lambda _i+\lambda _{jk}' +\lambda _{jk} \right)v_j^2,\quad
		\cos(\theta _k-\theta _j)=0.
	\end{align}
\end{subequations}
\begin{eqnarray}
	U=
\left(
\begin{array}{ccc}
	e^{-i \theta _j} & 0 & 0 \\
	0 & e^{-i \theta _j} & 0 \\
	0 & 0 & i e^{-i \theta _j} \\
\end{array}
\right)
	,\quad \text{for }\{i,j,k\}=\{1,2,3\}.	
\end{eqnarray}
\subsection{Solutions with no vanishing vev}
With all three vevs non-zero, there are 11 cases of CP conservation. They all require special values of the angles, or the vanishing of couplings sensitive to the angles:

\noindent
{\it Case~0-1:}
\vspace*{-10pt}
\begin{subequations}
	\begin{align}
		m_{11}&= \lambda _{11} v_1^2+\frac{1}{2} \left( 2 \lambda _3+\lambda _{12}' +\lambda _{12} \right)v_2^2+\frac{1}{2} \left(2 \lambda _2+ \lambda _{13}' +\lambda _{13} \right)v_3^2,\\
		m_{22}&= \lambda _{22} v_2^2+\frac{1}{2} \left(2 \lambda _3 +\lambda _{12}'+\lambda _{12} \right)v_1^2 +\frac{1}{2} \left(2 \lambda _1+ \lambda _{23}' +\lambda _{23} \right)v_3^2,\\
		m_{33}&= \lambda _{33} v_3^2+\frac{1}{2} \left( 2 \lambda _2+\lambda _{13}' +\lambda _{13} \right)v_1^2+\frac{1}{2} \left( 2 \lambda _1+\lambda _{23}' +\lambda _{23} \right)v_2^2,\\
		\sin(\theta _2-\theta_1)&=\sin(\theta _3-\theta_1)=0.
	\end{align}
\end{subequations}
\begin{eqnarray}
	U=
	\left(
	\begin{array}{ccc}
		e^{-i \theta _1} & 0 & 0 \\
		0 & e^{-i \theta _1} & 0 \\
		0 & 0 & e^{-i \theta _1} \\
	\end{array}
	\right)
\end{eqnarray}
\noindent
{\it Case~0-2:}
\vspace*{-10pt}
\begin{subequations}
	\begin{align}
		m_{11}&= \lambda _{11} v_1^2+\frac{1}{2} \left( 2 \lambda _3+\lambda _{12}' +\lambda _{12} \right)v_2^2+\frac{1}{2} \left(-2 \lambda _2+ \lambda _{13}' +\lambda _{13} \right)v_3^2,\\
		m_{22}&= \lambda _{22} v_2^2+\frac{1}{2} \left(2 \lambda _3 +\lambda _{12}'+\lambda _{12} \right)v_1^2 +\frac{1}{2} \left(-2 \lambda _1+ \lambda _{23}' +\lambda _{23} \right)v_3^2,\\
		m_{33}&= \lambda _{33} v_3^2+\frac{1}{2} \left( -2 \lambda _2+\lambda _{13}' +\lambda _{13} \right)v_1^2+\frac{1}{2} \left( -2 \lambda _1+\lambda _{23}' +\lambda _{23} \right)v_2^2,\\
		\sin(\theta _2-\theta_1)&=\cos(\theta _3-\theta_1)=0.
	\end{align}
\end{subequations}
\begin{eqnarray}
	U=
\left(
\begin{array}{ccc}
	e^{-i \theta _1} & 0 & 0 \\
	0 & e^{-i \theta _1} & 0 \\
	0 & 0 & i e^{-i \theta _1} \\
\end{array}
\right)
\end{eqnarray}
\noindent
{\it Case~0-3:}
\vspace*{-10pt}
\begin{subequations}
	\begin{align}
		m_{11}&= \lambda _{11} v_1^2+\frac{1}{2} \left( -2 \lambda _3+\lambda _{12}' +\lambda _{12} \right)v_2^2+\frac{1}{2} \left(2 \lambda _2+ \lambda _{13}' +\lambda _{13} \right)v_3^2,\\
		m_{22}&= \lambda _{22} v_2^2+\frac{1}{2} \left(-2 \lambda _3 +\lambda _{12}'+\lambda _{12} \right)v_1^2 +\frac{1}{2} \left(-2 \lambda _1+ \lambda _{23}' +\lambda _{23} \right)v_3^2,\\
		m_{33}&= \lambda _{33} v_3^2+\frac{1}{2} \left( 2 \lambda _2+\lambda _{13}' +\lambda _{13} \right)v_1^2+\frac{1}{2} \left( -2 \lambda _1+\lambda _{23}' +\lambda _{23} \right)v_2^2,\\
		\cos(\theta _2-\theta_1)&=\sin(\theta _3-\theta_1)=0.
	\end{align}
\end{subequations}
\begin{eqnarray}
	U=
\left(
\begin{array}{ccc}
	e^{-i \theta _1} & 0 & 0 \\
	0 & i e^{-i \theta _1} & 0 \\
	0 & 0 & e^{-i \theta _1} \\
\end{array}
\right)
\end{eqnarray}
\noindent
{\it Case~0-4:}
\vspace*{-10pt}
\begin{subequations}
	\begin{align}
		m_{11}&= \lambda _{11} v_1^2+\frac{1}{2} \left( -2 \lambda _3+\lambda _{12}' +\lambda _{12} \right)v_2^2+\frac{1}{2} \left(-2 \lambda _2+ \lambda _{13}' +\lambda _{13} \right)v_3^2,\\
		m_{22}&= \lambda _{22} v_2^2+\frac{1}{2} \left(-2\lambda _3 +\lambda _{12}'+\lambda _{12} \right)v_1^2 +\frac{1}{2} \left(2 \lambda _1+ \lambda _{23}' +\lambda _{23} \right)v_3^2,\\
		m_{33}&= \lambda _{33} v_3^2+\frac{1}{2} \left(- 2 \lambda _2+\lambda _{13}' +\lambda _{13} \right)v_1^2+\frac{1}{2} \left( 2 \lambda _1+\lambda _{23}' +\lambda _{23} \right)v_2^2,\\
		\cos(\theta _2-\theta_1)&=\cos(\theta _3-\theta_1)=0.
	\end{align}
\end{subequations}
\begin{eqnarray}
	U=
\left(
\begin{array}{ccc}
	e^{-i \theta _1} & 0 & 0 \\
	0 & i e^{-i \theta _1} & 0 \\
	0 & 0 & i e^{-i \theta _1} \\
\end{array}
\right)
\end{eqnarray}
\noindent
{\it Case~0-5:}
\vspace*{-10pt}
\begin{subequations}
	\begin{align}
		m_{ii}&= \lambda _{ii} v_i^2+\frac{1}{2} \left( \lambda _{ij}'+\lambda _{ij} \right)v_j^2+\frac{1}{2} \left( \lambda _{ik}'+\lambda _{ik}\right) v_k^2,\\
		m_{jj}&= \lambda _{jj} v_j^2+\frac{1}{2} \left( \lambda _{ij}'+\lambda _{ij} \right)v_i^2+\frac{1}{2} \left(2 \lambda _i+ \lambda _{jk}' +\lambda _{jk} \right)v_k^2,\\
		m_{kk}&=\lambda _{kk} v_k^2+\frac{1}{2} \left( \lambda _{ik}'+\lambda _{ik} \right)v_i^2+\frac{1}{2} \left(2 \lambda _i+ \lambda _{jk}' +\lambda _{jk} \right)v_j^2,\\
		\lambda _k&=\lambda _j=0,\quad\sin(\theta _k-\theta _j)=0.
	\end{align}
\end{subequations}
\begin{eqnarray}
	U=
\left(
\begin{array}{ccc}
	e^{-i \theta _i} & 0 & 0 \\
	0 & e^{-i \theta _j} & 0 \\
	0 & 0 & e^{-i \theta _j} \\
\end{array}
\right)
	,\quad \text{for }\{i,j,k\}=\{1,2,3\}.	
\end{eqnarray}
\noindent
{\it Case~0-6:}
\vspace*{-10pt}
\begin{subequations}
	\begin{align}
		m_{ii}&= \lambda _{ii} v_i^2+\frac{1}{2} \left( \lambda _{ij}'+\lambda _{ij} \right)v_j^2+\frac{1}{2} \left( \lambda _{ik}'+\lambda _{ik}\right) v_k^2,\\
		m_{jj}&= \lambda _{jj} v_j^2+\frac{1}{2} \left( \lambda _{ij}'+\lambda _{ij} \right)v_i^2+\frac{1}{2} \left(-2 \lambda _i+ \lambda _{jk}' +\lambda _{jk} \right)v_k^2,\\
		m_{kk}&=\lambda _{kk} v_k^2+\frac{1}{2} \left( \lambda _{ik}'+\lambda _{ik} \right)v_i^2+\frac{1}{2} \left(-2 \lambda _i+ \lambda _{jk}' +\lambda _{jk} \right)v_j^2,\\
		\lambda _k&=\lambda _j=0,\quad \cos(\theta _k-\theta _j)=0.
	\end{align}
\end{subequations}
\begin{eqnarray}
	U=
\left(
\begin{array}{ccc}
	e^{-i \theta _i} & 0 & 0 \\
	0 & e^{-i \theta _j} & 0 \\
	0 & 0 & i e^{-i \theta _j} \\
\end{array}
\right)
	,\quad \text{for }\{i,j,k\}=\{1,2,3\}.	
\end{eqnarray}
\noindent
{\it Case~0-7:}
\vspace*{-10pt}
\begin{subequations}
	\begin{align}
		m_{11}&=\lambda _{11} v_1^2+\frac{1}{2} \left( \lambda _{12}'+\lambda _{12} \right)v_2^2+\frac{1}{2} \left( \lambda _{13}'+\lambda _{13} \right)v_3^2,\\
		m_{22}&= \lambda _{22} v_2^2+\frac{1}{2} \left( \lambda _{12}'+\lambda _{12} \right)v_1^2+\frac{1}{2} \left( \lambda _{23}'+\lambda _{23} \right)v_3^2,\\
		m_{33}&= \lambda _{33} v_3^2+\frac{1}{2} \left( \lambda _{13}'+\lambda _{13} \right)v_1^2+\frac{1}{2} \left( \lambda _{23}'+\lambda _{23} \right)v_2^2,\\
		\lambda _3&=\lambda _2=\lambda _1=0.
	\end{align}
\end{subequations}
\begin{eqnarray}
	U=
\left(
\begin{array}{ccc}
	e^{-i \theta _1} & 0 & 0 \\
	0 & e^{-i \theta _2} & 0 \\
	0 & 0 & e^{-i \theta _3} \\
\end{array}
\right)
\end{eqnarray}

\noindent
{\it Case~0-8:}
\vspace*{-10pt}
\begin{subequations}
	\begin{align}
		m_{ii}&=\lambda _{ii} v_i^2+ \frac{1}{2}\left(\lambda _{ij}'+\lambda _{ij}\right)\left(1+\frac{\sin 2 \left(\theta _i-\theta _j\right)}{\sin 2 \left(\theta _i-\theta _k\right)}\right) v_j^2
		+ \frac{\lambda _i v_j^4 \sin ^2 2 \left(\theta _j-\theta _k\right) }{v_i^2\sin ^2 2 \left(\theta _i-\theta _k\right)},\\
		m_{jj}&=m_{kk}=(\lambda _{jj}+\lambda_i)\left(1+\frac{\sin 2 \left(\theta _i-\theta _j\right)}{\sin 2 \left(\theta _i-\theta _k\right)}\right) v_j^2
		+\frac{1}{2}(\lambda _{ij}'+ \lambda _{ij}) v_i^2,\\
		\lambda_k&=-\lambda _j=\frac{\lambda _i v_j^2 \sin 2 \left(\theta _j-\theta _k\right) }
		{v_i^2\sin 2 \left(\theta _i-\theta _k\right)}, \quad
		\lambda _{ik}'=\lambda _{ij}',\quad\lambda _{ik}=\lambda _{ij},\quad\lambda _{kk}=\lambda _{jj},\\
		v_k^2&=v_j^2 \frac{\sin 2 \left(\theta _i-\theta _j\right)}{\sin 2 \left(\theta _i-\theta _k\right)} ,\quad\lambda _{jk}'=2 \lambda _i+2 \lambda _{jj}-\lambda _{jk}.
	\end{align}
\end{subequations}
\begin{eqnarray}
	U=
\left(
\begin{array}{ccc}
	e^{-i \theta _i} & 0 & 0 \\
	0 & e^{-i \theta _j} \cos \alpha  & e^{-i \theta _k} \sin \alpha  \\
	0 & -e^{i (\theta _k-2  \theta _i)} \sin \alpha  & e^{i (\theta _j-2  \theta _i)} \cos \alpha  \\
\end{array}
\right)
	,\quad \text{for }\{i,j,k\}=\{1,2,3\},	
\end{eqnarray}
where
\begin{equation}
\alpha=\arctan\sqrt{\frac{\sin 2 \left(\theta _i-\theta _j\right)}{\sin 2 \left(\theta _i-\theta _k\right)}}=\arctan\left(\frac{v_k}{v_j}\right).
\end{equation}

\noindent
{\it Case~0-9:}
\vspace*{-10pt}
\begin{subequations}
	\begin{align}
		m_{ii}&=\lambda _{ii} v_i^2+ \frac{1}{2}\left(\lambda _{ij}'+\lambda _{ij}\right)\left(1-  \frac{\sin 2 \left(\theta _i-\theta _j\right)}{ \sin 2 \left(\theta _i-\theta _k\right)}\right) v_j^2
		-\frac{\lambda _i v_j^4 \sin ^2 2 \left(\theta _j-\theta _k\right) }{v_i^2\sin ^2 2 \left(\theta _i-\theta _k\right)},\\
		m_{jj}&=m_{kk}=\left(\lambda _{jj}-\lambda _i\right) \left(1-  \frac{\sin 2 \left(\theta _i-\theta _j\right)}{ \sin 2\left( \theta _i-\theta _k\right)}\right)v_j^2+\frac{1}{2}  \left(\lambda _{ij}'+\lambda _{ij}\right)v_i^2,\\
		\lambda_k&=\lambda _j=-\frac{\lambda _i v_j^2 \sin 2 \left(\theta _j-\theta _k\right) }
		{v_i^2\sin 2 \left(\theta _i-\theta _k\right)},\quad\lambda _{ik}'=\lambda _{ij}',\quad\lambda _{ik}=\lambda _{ij},\quad
		\lambda _{kk}=\lambda _{jj},\\
		v_k^2&=-v_j^2 \frac{\sin 2 \left(\theta _i-\theta _j\right)}{ \sin 2 \left(\theta _i-\theta _k\right)},\quad
		\lambda _{jk}'=-2 \lambda _i+2 \lambda _{jj}-\lambda _{jk}.
	\end{align}
\end{subequations}
\begin{eqnarray}
	U=
	\left(
	\begin{array}{ccc}
		e^{-i \theta _i} & 0 & 0 \\
		0 & e^{-i \theta _j} \cos \alpha  & e^{-i \theta _k} \sin \alpha  \\
		0 & -i e^{i (\theta _k-2  \theta _i)} \sin \alpha  & i e^{i (\theta _j-2  \theta _i)} \cos \alpha  \\
	\end{array}
	\right)
	,\quad \text{for }\{i,j,k\}=\{1,2,3\},	
\end{eqnarray}
where
\begin{equation}
\alpha =\arctan\sqrt{-\frac{\sin 2 \left(\theta _i-\theta _j\right)}{\sin 2 \left(\theta _i-\theta _k\right)}}=\arctan\left(\frac{v_k}{v_j}\right).
\end{equation}
\noindent
{\it Case~0-10:}
\vspace*{-10pt}
\begin{subequations}
	\begin{align}
		m_{ii}&=\lambda _{ii} v_i^2+ \left(\lambda _{ij}'+\lambda _{ij}\right)v_j^2
		+\frac{4 \lambda _i v_j^4 \cos ^2 2 \left(\theta _i-\theta _j\right)}{v_i^2},\\
		m_{jj}&=m_{kk}=(\lambda _{jj}+\lambda_i) v_j^2+\frac{1}{2} \left(\lambda _{ij}'+\lambda _{ij}\right)v_i^2
		+\frac{1}{2} \left(\lambda _{jk}'+\lambda _{jk}\right)v_j^2,\\
		\lambda_k&=-\lambda _j=\frac{2 \lambda _i v_j^2 \cos 2 \left(\theta _i-\theta _j\right)}{v_i^2},\\
		\lambda _{ik}'&=\lambda _{ij}',\quad\lambda _{ik}=\lambda _{ij},\quad\lambda _{kk}=\lambda _{jj},\quad v_k=v_j,\\
		\cos&(2 \theta _i-\theta _j-\theta _k) =0,\quad \sin(2 \theta _i-\theta _j-\theta _k)=\pm 1.
	\end{align}
\end{subequations}
\begin{eqnarray}
	U=e^{-i \theta _i}
\left(
\begin{array}{ccc}
	1 & 0 & 0 \\
	0 & \pm\frac{i }{\sqrt{2}} & \frac{1}{\sqrt{2}} \\
	0 & \pm\frac{1}{\sqrt{2}} & \frac{i }{\sqrt{2}} \\
\end{array}
\right)
	,\quad \text{for }\{i,j,k\}=\{1,2,3\}.	
\end{eqnarray}

\noindent
{\it Case~0-11:}
\vspace*{-10pt}
\begin{subequations}
	\begin{align}
		m_{ii}&=\lambda _{ii} v_i^2+ \left(\lambda _{ij}'+\lambda _{ij}\right)v_j^2-\frac{4 \lambda _i v_j^4 \cos ^2 2 \left(\theta _i-\theta _j\right)}{v_i^2},\\
		m_{jj}&=m_{kk}=(\lambda_{jj}-\lambda_i) v_j^2+\frac{1}{2} \left(\lambda _{ij}'+\lambda _{ij}\right)v_i^2+\frac{1}{2} 
		\left(\lambda _{jk}'+\lambda _{jk}\right)v_j^2,\\
		\lambda_k&=\lambda _j=-\frac{2 \lambda _i v_j^2 \cos 2 \left(\theta _i-\theta _j\right)}{v_i^2},\\
		\lambda _{ik}'&=\lambda _{ij}',\quad\lambda _{ik}=\lambda _{ij},\quad\lambda _{kk}=\lambda _{jj},\quad v_k=v_j,\\
		\sin&(2 \theta _i-\theta _j-\theta _k)=0,\quad \cos(2 \theta _i-\theta _j-\theta _k)=\pm 1.
	\end{align}
\end{subequations}
\begin{eqnarray}
	U=e^{-i \theta _i}
	\left(
	\begin{array}{ccc}
		1 & 0 & 0 \\
		0 & \pm\frac{1}{\sqrt{2}} & \frac{1}{\sqrt{2}} \\
		0 & \mp\frac{i }{\sqrt{2}} & \frac{i }{\sqrt{2}} \\
	\end{array}
	\right)
	,\quad \text{for }\{i,j,k\}=\{1,2,3\}.	
\end{eqnarray}
This last case corresponds to the ``simple model'' of ref.~\cite{Plantey:2022jdg}.
\section{Invariants require two $\hat V$ factors}
\label{app:need-two-Vhat}
The CP-odd invariants discussed in section~\ref{sect-CP-invariants} involve at least {\it two} factors $\hat V$. For a real potential invariants with just one such factor vanish.
The imaginary part of $\hat V_{ac}$ is antisymmetric in the indices $a$ and $c$. However, with potential coefficients that are subject to the $\ztwo\times\ztwo$ symmetry there is no way to construct an antisymmetric tensor $F_{ca}$ that could project out this imaginary part. 
While there could be many different $Z$ couplings, some are equal and some vanish. In the Weinberg potential, we have only four patterns of indices for which the $Z_{acbd}$ can be non-zero:
\begin{equation}
Z_{iiii}=\lambda_{ii}, \quad Z_{iijj}=\lambda_{ij}, \quad Z_{ijji}=\lambda_{ij}^\prime, \quad Z_{ijij}=\lambda_k.
\end{equation}
These all have pairs of repeated indices.

With two factors $\hat V_{ac}$ and $\hat V_{be}$ one can construct an imaginary quantity that is symmetric under two pairs of indices, $a\leftrightarrow b$ and $c\leftrightarrow e$, proportional to
\begin{equation}
\cos(\theta_a-\theta_c)\sin(\theta_b-\theta_e)+\cos(\theta_b-\theta_e)\sin(\theta_a-\theta_c).
\end{equation}

For $a=b$ and $c=e$, this is simply $\sin2(\theta_a-\theta_c)$.
This can for example be multiplied by a product of $Z$'s involving a factor $(Z_{aaaa}-Z_{cccc})=2(\lambda_{aa}-\lambda_{cc})$, 
or $(Z_{aadd}-Z_{ccdd})=\lambda_{ad}-\lambda_{cd}$, see Eq.~(\ref{Eq:J_i-explicit}).
The relative sign arises from rewriting
\begin{equation}
\sin2(\theta_c-\theta_a) = -\sin2(\theta_a-\theta_c).
\end{equation}
\section{3HDM gauge coupling relations}
\label{app:3HDM-gauge}
As mentioned in section~\ref{sect:WWZ}, there exist relations between the gauge-gauge-scalar couplings $e_i$ and the gauge-scalar-scalar couplings $\tilde\lambda_{ij}$. In the 2HDM they are very simple \cite{Grzadkowski:2014ada}, whereas in the 3HDM they are more involved.

In order to form a set of independent gauge couplings let us first pick the set of $e_1, e_2, e_3, e_4, e_5$ which are all independent.
Next, by applying identities derived from the orthogonality of the rotation matrix we find that
\begin{align}
\tilde\lambda _{12}&= \frac{-e_5 \tilde\lambda _{34}+e_4 \tilde\lambda _{35}-e_3 \tilde\lambda _{45}}{v},\\
\tilde\lambda _{13}&=\frac{ \left(e_1 e_4 v-e_2 e_3 e_5\right)\tilde\lambda _{34}+ \left(e_1 e_5 v+e_2 e_3 e_4\right)\tilde\lambda _{35}+e_2 \left(v^2-e_3^2\right) \tilde\lambda _{45}}{\left(e_1^2+e_2^2\right) v},\\
\tilde\lambda _{14}&=\frac{ -\left(e_1 e_3 v+e_2 e_4 e_5\right)\tilde\lambda _{34}-e_2 \left(v^2-e_4^2\right) \tilde\lambda _{35}+ \left(e_1 e_5 v-e_2 e_3 e_4\right)\tilde\lambda _{45}}{\left(e_1^2+e_2^2\right) v},\\
\tilde\lambda _{15}&=\frac{e_2 \left(v^2-e_5^2\right) \tilde\lambda _{34}+ \left(e_2 e_4 e_5-e_1 e_3 v\right)\tilde\lambda _{35}- \left(e_1 e_4 v+e_2 e_3 e_5\right)\tilde\lambda _{45}}{\left(e_1^2+e_2^2\right) v},\\
\tilde\lambda _{23}&=\frac{ \left(e_2 e_4 v+e_1 e_3 e_5\right)\tilde\lambda _{34}+ \left(e_2 e_5 v-e_1 e_3 e_4\right)\tilde\lambda _{35}-e_1 \left(v^2-e_3^2\right) \tilde\lambda _{45}}{\left(e_1^2+e_2^2\right) v},\\
\tilde\lambda _{24}&=\frac{ \left(e_1 e_4 e_5-e_2 e_3 v\right)\tilde\lambda _{34}+e_1 \left(v^2-e_4^2\right) \tilde\lambda _{35}+ \left(e_2 e_5 v+e_1 e_3 e_4\right)\tilde\lambda _{45}}{\left(e_1^2+e_2^2\right) v},\\
\tilde\lambda _{25}&=-\frac{e_1 \left(v^2-e_5^2\right) \tilde\lambda _{34}+ \left(e_2 e_3 v+e_1 e_4 e_5\right)\tilde\lambda _{35}+\left(e_2 e_4 v-e_1 e_3 e_5\right)\tilde\lambda _{45} }{\left(e_1^2+e_2^2\right) v}.
\end{align}
Hence, all gauge couplings involving neutral scalars can be expressed in terms of \\
$e_1, e_2, e_3, e_4, e_5, \tilde\lambda_{34},\tilde\lambda_{35}$ and $\tilde\lambda_{45}$, a total of eight couplings.

Finally, using
\begin{eqnarray}
&&(v^2-e_5^2)\tilde\lambda_{34}^2+(v^2-e_4^2)\tilde\lambda_{35}^2+(v^2-e_3^2)\tilde\lambda_{45}^2\nonumber\\
&&+2 e_4 e_5\tilde\lambda_{34}\tilde\lambda_{35}-2 e_3 e_5\tilde\lambda_{34}\tilde\lambda_{45}+2 e_3 e_4\tilde\lambda_{35}\tilde\lambda_{45}-\left(e_1^2+e_2^2\right) v^2=0.
\end{eqnarray}
we may bring the number of independent couplings down to seven by solving this quadratic equation for $\tilde\lambda_{34}$, $\tilde\lambda_{35}$ or $\tilde\lambda_{45}$.

\bibliographystyle{JHEP}

\bibliography{ref}

\providecommand{\href}[2]{#2}\begingroup\raggedright\begin{thebibliography}{10}

\bibitem{Weinberg:1976hu}
S.~Weinberg, \emph{{Gauge Theory of CP Violation}},
  \href{http://dx.doi.org/10.1103/PhysRevLett.37.657}{\emph{Phys. Rev. Lett.}
  {\bf 37} (1976) 657}.

\bibitem{Plantey:2022jdg}
R.~Plantey, O.~M. Ogreid, P.~Osland, M.~N. Rebelo and M.~A. Solberg,
  \emph{{Weinberg\textquoteright{}s 3HDM potential with spontaneous CP
  violation}}, \href{http://dx.doi.org/10.1103/PhysRevD.108.075029}{\emph{Phys.
  Rev. D} {\bf 108} (2023) 075029},
  [\href{http://arxiv.org/abs/2208.13594}{{\tt 2208.13594}}].

\bibitem{Plantey:2022gwj}
R.~Plantey, O.~M. Ogreid, P.~Osland, M.~N. Rebelo and M.~A. Solberg,
  \emph{{Light scalars in the Weinberg 3HDM potential with spontaneous CP
  violation}}, \href{http://dx.doi.org/10.22323/1.405.0064}{\emph{PoS} {\bf
  DISCRETE2020-2021} (2022) 064}, [\href{http://arxiv.org/abs/2209.06499}{{\tt
  2209.06499}}].

\bibitem{Chupp:2017rkp}
T.~Chupp, P.~Fierlinger, M.~Ramsey-Musolf and J.~Singh, \emph{{Electric dipole
  moments of atoms, molecules, nuclei, and particles}},
  \href{http://dx.doi.org/10.1103/RevModPhys.91.015001}{\emph{Rev. Mod. Phys.}
  {\bf 91} (2019) 015001}, [\href{http://arxiv.org/abs/1710.02504}{{\tt
  1710.02504}}].

\bibitem{Ivanov:2014doa}
I.~P. Ivanov and C.~C. Nishi, \emph{{Symmetry breaking patterns in 3HDM}},
  \href{http://dx.doi.org/10.1007/JHEP01(2015)021}{\emph{JHEP} {\bf 01} (2015)
  021}, [\href{http://arxiv.org/abs/1410.6139}{{\tt 1410.6139}}].

\bibitem{Branco:1980sz}
G.~C. Branco, \emph{{Spontaneous {CP} Nonconservation and Natural Flavor
  Conservation: A Minimal Model}},
  \href{http://dx.doi.org/10.1103/PhysRevD.22.2901}{\emph{Phys. Rev. D} {\bf
  22} (1980) 2901}.

\bibitem{Ivanov:2011ae}
I.~P. Ivanov, V.~Keus and E.~Vdovin, \emph{{Abelian symmetries in
  multi-Higgs-doublet models}},
  \href{http://dx.doi.org/10.1088/1751-8113/45/21/215201}{\emph{J. Phys. A}
  {\bf 45} (2012) 215201}, [\href{http://arxiv.org/abs/1112.1660}{{\tt
  1112.1660}}].

\bibitem{deMedeirosVarzielas:2019rrp}
I.~de~Medeiros~Varzielas and I.~P. Ivanov, \emph{{Recognizing symmetries in a
  3HDM in a basis-independent way}},
  \href{http://dx.doi.org/10.1103/PhysRevD.100.015008}{\emph{Phys. Rev. D} {\bf
  100} (2019) 015008}, [\href{http://arxiv.org/abs/1903.11110}{{\tt
  1903.11110}}].

\bibitem{Darvishi:2019dbh}
N.~Darvishi and A.~Pilaftsis, \emph{{Classifying Accidental Symmetries in
  Multi-Higgs Doublet Models}},
  \href{http://dx.doi.org/10.1103/PhysRevD.101.095008}{\emph{Phys. Rev. D} {\bf
  101} (2020) 095008}, [\href{http://arxiv.org/abs/1912.00887}{{\tt
  1912.00887}}].

\bibitem{Branco:1999fs}
G.~C. Branco, L.~Lavoura and J.~P. Silva, \emph{{CP Violation}}, vol.~103.
\newblock {Oxford University Press}, 1999.

\bibitem{Mendez:1991gp}
A.~Mendez and A.~Pomarol, \emph{{Signals of CP violation in the Higgs sector}},
  \href{http://dx.doi.org/10.1016/0370-2693(91)91836-K}{\emph{Phys. Lett. B}
  {\bf 272} (1991) 313--318}.

\bibitem{Lavoura:1994fv}
L.~Lavoura and J.~P. Silva, \emph{{Fundamental CP violating quantities in a
  SU(2) $\times$ U(1) model with many Higgs doublets}},
  \href{http://dx.doi.org/10.1103/PhysRevD.50.4619}{\emph{Phys. Rev. D} {\bf
  50} (1994) 4619--4624}, [\href{http://arxiv.org/abs/hep-ph/9404276}{{\tt
  hep-ph/9404276}}].

\bibitem{Botella:1994cs}
F.~J. Botella and J.~P. Silva, \emph{{Jarlskog - like invariants for theories
  with scalars and fermions}},
  \href{http://dx.doi.org/10.1103/PhysRevD.51.3870}{\emph{Phys. Rev. D} {\bf
  51} (1995) 3870--3875}, [\href{http://arxiv.org/abs/hep-ph/9411288}{{\tt
  hep-ph/9411288}}].

\bibitem{Branco:2005em}
G.~C. Branco, M.~N. Rebelo and J.~I. Silva-Marcos, \emph{{CP-odd invariants in
  models with several Higgs doublets}},
  \href{http://dx.doi.org/10.1016/j.physletb.2005.03.075}{\emph{Phys. Lett. B}
  {\bf 614} (2005) 187--194}, [\href{http://arxiv.org/abs/hep-ph/0502118}{{\tt
  hep-ph/0502118}}].

\bibitem{Gunion:2005ja}
J.~F. Gunion and H.~E. Haber, \emph{{Conditions for CP-violation in the general
  two-Higgs-doublet model}},
  \href{http://dx.doi.org/10.1103/PhysRevD.72.095002}{\emph{Phys. Rev. D} {\bf
  72} (2005) 095002}, [\href{http://arxiv.org/abs/hep-ph/0506227}{{\tt
  hep-ph/0506227}}].

\bibitem{Davidson:2005cw}
S.~Davidson and H.~E. Haber, \emph{{Basis-independent methods for the
  two-Higgs-doublet model}},
  \href{http://dx.doi.org/10.1103/PhysRevD.72.099902}{\emph{Phys. Rev. D} {\bf
  72} (2005) 035004}, [\href{http://arxiv.org/abs/hep-ph/0504050}{{\tt
  hep-ph/0504050}}].

\bibitem{Haber:2006ue}
H.~E. Haber and D.~O'Neil, \emph{{Basis-independent methods for the
  two-Higgs-doublet model. II. The Significance of tan$\beta$}},
  \href{http://dx.doi.org/10.1103/PhysRevD.74.015018}{\emph{Phys. Rev. D} {\bf
  74} (2006) 015018}, [\href{http://arxiv.org/abs/hep-ph/0602242}{{\tt
  hep-ph/0602242}}].

\bibitem{Grzadkowski:2014ada}
B.~Grzadkowski, O.~M. Ogreid and P.~Osland, \emph{{Measuring CP violation in
  Two-Higgs-Doublet models in light of the LHC Higgs data}},
  \href{http://dx.doi.org/10.1007/JHEP11(2014)084}{\emph{JHEP} {\bf 11} (2014)
  084}, [\href{http://arxiv.org/abs/1409.7265}{{\tt 1409.7265}}].

\bibitem{Grzadkowski:2016szj}
B.~Grzadkowski, O.~M. Ogreid and P.~Osland, \emph{{Spontaneous CP violation in
  the 2HDM: physical conditions and the alignment limit}},
  \href{http://dx.doi.org/10.1103/PhysRevD.94.115002}{\emph{Phys. Rev. D} {\bf
  94} (2016) 115002}, [\href{http://arxiv.org/abs/1609.04764}{{\tt
  1609.04764}}].

\bibitem{Mathematica}
W.~R. Inc., ``Mathematica.''

\bibitem{Akeroyd:2016ssd}
A.~G. Akeroyd, S.~Moretti, K.~Yagyu and E.~Yildirim, \emph{{Light charged Higgs
  boson scenario in 3-Higgs doublet models}},
  \href{http://dx.doi.org/10.1142/S0217751X17501457}{\emph{Int. J. Mod. Phys.
  A} {\bf 32} (2017) 1750145}, [\href{http://arxiv.org/abs/1605.05881}{{\tt
  1605.05881}}].

\bibitem{Logan:2020mdz}
H.~E. Logan, S.~Moretti, D.~Rojas-Ciofalo and M.~Song, \emph{{CP violation from
  charged Higgs bosons in the three Higgs doublet model}},
  \href{http://dx.doi.org/10.1007/JHEP07(2021)158}{\emph{JHEP} {\bf 07} (2021)
  158}, [\href{http://arxiv.org/abs/2012.08846}{{\tt 2012.08846}}].

\bibitem{Boto:2021qgu}
R.~Boto, J.~C. Rom\~ao and J.~a.~P. Silva, \emph{{Current bounds on the type-Z
  Z3 three-Higgs-doublet model}},
  \href{http://dx.doi.org/10.1103/PhysRevD.104.095006}{\emph{Phys. Rev. D} {\bf
  104} (2021) 095006}, [\href{http://arxiv.org/abs/2106.11977}{{\tt
  2106.11977}}].

\bibitem{Hernandez-Sanchez:2022dnn}
J.~Hernandez-Sanchez, V.~Keus, S.~Moretti and D.~Sokolowska,
  \emph{{Complementary collider and astrophysical probes of multi-component
  Dark Matter}}, \href{http://dx.doi.org/10.1007/JHEP03(2023)045}{\emph{JHEP}
  {\bf 03} (2023) 045}, [\href{http://arxiv.org/abs/2202.10514}{{\tt
  2202.10514}}].

\bibitem{Hernandez-Sanchez:2020aop}
J.~Hernandez-Sanchez, V.~Keus, S.~Moretti, D.~Rojas-Ciofalo and D.~Sokolowska,
  \emph{{Complementary Probes of Two-component Dark Matter}},
  \href{http://arxiv.org/abs/2012.11621}{{\tt 2012.11621}}.

\bibitem{Boto:2024tzp}
R.~Boto, P.~N. Figueiredo, J.~C. Rom\~ao and J.~P. Silva, \emph{{Novel two
  component dark matter features in the $Z_2 \times Z_2$ 3HDM}},
  \href{http://arxiv.org/abs/2407.15933}{{\tt 2407.15933}}.

\bibitem{Glashow:1976nt}
S.~L. Glashow and S.~Weinberg, \emph{{Natural Conservation Laws for Neutral
  Currents}}, \href{http://dx.doi.org/10.1103/PhysRevD.15.1958}{\emph{Phys.
  Rev. D} {\bf 15} (1977) 1958}.

\bibitem{Paschos:1976ay}
E.~A. Paschos, \emph{{Diagonal Neutral Currents}},
  \href{http://dx.doi.org/10.1103/PhysRevD.15.1966}{\emph{Phys. Rev. D} {\bf
  15} (1977) 1966}.

\bibitem{Borzumati:1998tg}
F.~Borzumati and C.~Greub, \emph{{2HDMs predictions for anti-B $\to$ X(s) gamma
  in NLO QCD}}, \href{http://dx.doi.org/10.1103/PhysRevD.58.074004}{\emph{Phys.
  Rev. D} {\bf 58} (1998) 074004},
  [\href{http://arxiv.org/abs/hep-ph/9802391}{{\tt hep-ph/9802391}}].

\bibitem{Ciuchini:1997xe}
M.~Ciuchini, G.~Degrassi, P.~Gambino and G.~Giudice, \emph{{Next-to-leading QCD
  corrections to $B \to X_s \gamma$: Standard model and two Higgs doublet
  model}}, \href{http://dx.doi.org/10.1016/S0550-3213(98)00244-2}{\emph{Nucl.
  Phys. B} {\bf 527} (1998) 21--43},
  [\href{http://arxiv.org/abs/hep-ph/9710335}{{\tt hep-ph/9710335}}].

\bibitem{Hermann:2012fc}
T.~Hermann, M.~Misiak and M.~Steinhauser, \emph{{$\bar{B}\to X_s \gamma$ in the
  Two Higgs Doublet Model up to Next-to-Next-to-Leading Order in QCD}},
  \href{http://dx.doi.org/10.1007/JHEP11(2012)036}{\emph{JHEP} {\bf 11} (2012)
  036}, [\href{http://arxiv.org/abs/1208.2788}{{\tt 1208.2788}}].

\bibitem{ParticleDataGroup:2022pth}
{\scshape Particle Data Group} collaboration, R.~L. Workman et~al.,
  \emph{{Review of Particle Physics}},
  \href{http://dx.doi.org/10.1093/ptep/ptac097}{\emph{PTEP} {\bf 2022} (2022)
  083C01}.

\bibitem{CMS:2021sdq}
{\scshape CMS} collaboration, A.~Tumasyan et~al., \emph{{Analysis of the $CP$
  structure of the Yukawa coupling between the Higgs boson and $\tau$ leptons
  in proton-proton collisions at $ \sqrt{s} $ = 13 TeV}},
  \href{http://dx.doi.org/10.1007/JHEP06(2022)012}{\emph{JHEP} {\bf 06} (2022)
  012}, [\href{http://arxiv.org/abs/2110.04836}{{\tt 2110.04836}}].

\bibitem{Engel:2013lsa}
J.~Engel, M.~J. Ramsey-Musolf and U.~van Kolck, \emph{{Electric Dipole Moments
  of Nucleons, Nuclei, and Atoms: The Standard Model and Beyond}},
  \href{http://dx.doi.org/10.1016/j.ppnp.2013.03.003}{\emph{Prog. Part. Nucl.
  Phys.} {\bf 71} (2013) 21--74}, [\href{http://arxiv.org/abs/1303.2371}{{\tt
  1303.2371}}].

\bibitem{Peskin:1990zt}
M.~E. Peskin and T.~Takeuchi, \emph{{A New constraint on a strongly interacting
  Higgs sector}},
  \href{http://dx.doi.org/10.1103/PhysRevLett.65.964}{\emph{Phys. Rev. Lett.}
  {\bf 65} (1990) 964--967}.

\bibitem{Peskin:1991sw}
M.~E. Peskin and T.~Takeuchi, \emph{{Estimation of oblique electroweak
  corrections}}, \href{http://dx.doi.org/10.1103/PhysRevD.46.381}{\emph{Phys.
  Rev. D} {\bf 46} (1992) 381--409}.

\bibitem{Maksymyk:1993zm}
I.~Maksymyk, C.~P. Burgess and D.~London, \emph{{Beyond S, T and U}},
  \href{http://dx.doi.org/10.1103/PhysRevD.50.529}{\emph{Phys. Rev. D} {\bf 50}
  (1994) 529--535}, [\href{http://arxiv.org/abs/hep-ph/9306267}{{\tt
  hep-ph/9306267}}].

\bibitem{Grimus:2007if}
W.~Grimus, L.~Lavoura, O.~M. Ogreid and P.~Osland, \emph{{A Precision
  constraint on multi-Higgs-doublet models}},
  \href{http://dx.doi.org/10.1088/0954-3899/35/7/075001}{\emph{J. Phys.} {\bf
  G35} (2008) 075001}, [\href{http://arxiv.org/abs/0711.4022}{{\tt
  0711.4022}}].

\bibitem{Grimus:2008nb}
W.~Grimus, L.~Lavoura, O.~M. Ogreid and P.~Osland, \emph{{The Oblique
  parameters in multi-Higgs-doublet models}},
  \href{http://dx.doi.org/10.1016/j.nuclphysb.2008.04.019}{\emph{Nucl. Phys.}
  {\bf B801} (2008) 81--96}, [\href{http://arxiv.org/abs/0802.4353}{{\tt
  0802.4353}}].

\bibitem{Asadi:2022xiy}
P.~Asadi, C.~Cesarotti, K.~Fraser, S.~Homiller and A.~Parikh, \emph{{Oblique
  lessons from the W-mass measurement at CDF II}},
  \href{http://dx.doi.org/10.1103/PhysRevD.108.055026}{\emph{Phys. Rev. D} {\bf
  108} (2023) 055026}, [\href{http://arxiv.org/abs/2204.05283}{{\tt
  2204.05283}}].

\bibitem{Ellis:1975ap}
J.~R. Ellis, M.~K. Gaillard and D.~V. Nanopoulos, \emph{{A Phenomenological
  Profile of the Higgs Boson}},
  \href{http://dx.doi.org/10.1016/0550-3213(76)90382-5}{\emph{Nucl. Phys. B}
  {\bf 106} (1976) 292}.

\bibitem{Shifman:1979eb}
M.~A. Shifman, A.~Vainshtein, M.~Voloshin and V.~I. Zakharov, \emph{{Low-Energy
  Theorems for Higgs Boson Couplings to Photons}}, {\emph{Sov. J. Nucl. Phys.}
  {\bf 30} (1979) 711--716}.

\bibitem{Gunion:1989we}
J.~F. Gunion, H.~E. Haber, G.~L. Kane and S.~Dawson, \emph{{The Higgs Hunter's
  Guide}}, vol.~80.
\newblock {Frontiers in Physics}, 2000.

\bibitem{Djouadi:2005gj}
A.~Djouadi, \emph{{The Anatomy of electro-weak symmetry breaking. II. The Higgs
  bosons in the minimal supersymmetric model}},
  \href{http://dx.doi.org/10.1016/j.physrep.2007.10.005}{\emph{Phys. Rept.}
  {\bf 459} (2008) 1--241}, [\href{http://arxiv.org/abs/hep-ph/0503173}{{\tt
  hep-ph/0503173}}].

\bibitem{Grinstein:1987pu}
B.~Grinstein and M.~B. Wise, \emph{{Weak Radiative B Meson Decay as a Probe of
  the Higgs Sector}},
  \href{http://dx.doi.org/10.1016/0370-2693(88)90227-4}{\emph{Phys. Lett. B}
  {\bf 201} (1988) 274--278}.

\bibitem{Hou:1988gv}
W.-S. Hou and R.~Willey, \emph{{Effects of Extended Higgs Sector on Loop
  Induced $B$ Decays}},
  \href{http://dx.doi.org/10.1016/0550-3213(89)90434-3}{\emph{Nucl. Phys. B}
  {\bf 326} (1989) 54--72}.

\bibitem{Grinstein:1990tj}
B.~Grinstein, R.~P. Springer and M.~B. Wise, \emph{{Strong Interaction Effects
  in Weak Radiative $\bar{B}$ Meson Decay}},
  \href{http://dx.doi.org/10.1016/0550-3213(90)90350-M}{\emph{Nucl. Phys. B}
  {\bf 339} (1990) 269--309}.

\bibitem{Buras:1993xp}
A.~Buras, M.~Misiak, M.~Munz and S.~Pokorski, \emph{{Theoretical uncertainties
  and phenomenological aspects of B $\to$ X(s) gamma decay}},
  \href{http://dx.doi.org/10.1016/0550-3213(94)90299-2}{\emph{Nucl. Phys. B}
  {\bf 424} (1994) 374--398}, [\href{http://arxiv.org/abs/hep-ph/9311345}{{\tt
  hep-ph/9311345}}].

\bibitem{Ciafaloni:1997un}
P.~Ciafaloni, A.~Romanino and A.~Strumia, \emph{{Two loop QCD corrections to
  charged Higgs mediated b $\to$ s gamma decay}},
  \href{http://dx.doi.org/10.1016/S0550-3213(98)00190-4}{\emph{Nucl. Phys. B}
  {\bf 524} (1998) 361--376}, [\href{http://arxiv.org/abs/hep-ph/9710312}{{\tt
  hep-ph/9710312}}].

\bibitem{Bobeth:1999ww}
C.~Bobeth, M.~Misiak and J.~Urban, \emph{{Matching conditions for $b \to s
  \gamma$ and $b \to s gluon$ in extensions of the standard model}},
  \href{http://dx.doi.org/10.1016/S0550-3213(99)00688-4}{\emph{Nucl. Phys. B}
  {\bf 567} (2000) 153--185}, [\href{http://arxiv.org/abs/hep-ph/9904413}{{\tt
  hep-ph/9904413}}].

\bibitem{Bobeth:1999mk}
C.~Bobeth, M.~Misiak and J.~Urban, \emph{{Photonic penguins at two loops and
  $m_t$ dependence of $BR[B \to X_s l^+ l^-]$}},
  \href{http://dx.doi.org/10.1016/S0550-3213(00)00007-9}{\emph{Nucl. Phys. B}
  {\bf 574} (2000) 291--330}, [\href{http://arxiv.org/abs/hep-ph/9910220}{{\tt
  hep-ph/9910220}}].

\bibitem{Gambino:2001ew}
P.~Gambino and M.~Misiak, \emph{{Quark mass effects in anti-B $\to$ X(s
  gamma)}}, \href{http://dx.doi.org/10.1016/S0550-3213(01)00347-9}{\emph{Nucl.
  Phys. B} {\bf 611} (2001) 338--366},
  [\href{http://arxiv.org/abs/hep-ph/0104034}{{\tt hep-ph/0104034}}].

\bibitem{Cheung:2003pw}
K.~Cheung and O.~C. Kong, \emph{{Can the two Higgs doublet model survive the
  constraint from the muon anomalous magnetic moment as suggested?}},
  \href{http://dx.doi.org/10.1103/PhysRevD.68.053003}{\emph{Phys. Rev. D} {\bf
  68} (2003) 053003}, [\href{http://arxiv.org/abs/hep-ph/0302111}{{\tt
  hep-ph/0302111}}].

\bibitem{Misiak:2004ew}
M.~Misiak and M.~Steinhauser, \emph{{Three loop matching of the dipole
  operators for $b \to s \gamma$ and $b \to s g$}},
  \href{http://dx.doi.org/10.1016/j.nuclphysb.2004.02.006}{\emph{Nucl. Phys. B}
  {\bf 683} (2004) 277--305}, [\href{http://arxiv.org/abs/hep-ph/0401041}{{\tt
  hep-ph/0401041}}].

\bibitem{Czakon:2006ss}
M.~Czakon, U.~Haisch and M.~Misiak, \emph{{Four-Loop Anomalous Dimensions for
  Radiative Flavour-Changing Decays}},
  \href{http://dx.doi.org/10.1088/1126-6708/2007/03/008}{\emph{JHEP} {\bf 03}
  (2007) 008}, [\href{http://arxiv.org/abs/hep-ph/0612329}{{\tt
  hep-ph/0612329}}].

\bibitem{Misiak:2015xwa}
M.~Misiak et~al., \emph{{Updated NNLO QCD predictions for the weak radiative
  B-meson decays}},
  \href{http://dx.doi.org/10.1103/PhysRevLett.114.221801}{\emph{Phys. Rev.
  Lett.} {\bf 114} (2015) 221801}, [\href{http://arxiv.org/abs/1503.01789}{{\tt
  1503.01789}}].

\bibitem{Misiak:2017bgg}
M.~Misiak and M.~Steinhauser, \emph{{Weak radiative decays of the B meson and
  bounds on $M_{H^\pm }$ in the Two-Higgs-Doublet Model}},
  \href{http://dx.doi.org/10.1140/epjc/s10052-017-4776-y}{\emph{Eur. Phys. J.
  C} {\bf 77} (2017) 201}, [\href{http://arxiv.org/abs/1702.04571}{{\tt
  1702.04571}}].

\bibitem{Misiak:2020vlo}
M.~Misiak, A.~Rehman and M.~Steinhauser, \emph{{Towards $ \overline{B}\to
  {X}_s\gamma $ at the NNLO in QCD without interpolation in m$_{c}$}},
  \href{http://dx.doi.org/10.1007/JHEP06(2020)175}{\emph{JHEP} {\bf 06} (2020)
  175}, [\href{http://arxiv.org/abs/2002.01548}{{\tt 2002.01548}}].

\bibitem{Inami:1980fz}
T.~Inami and C.~S. Lim, \emph{{Effects of Superheavy Quarks and Leptons in
  Low-Energy Weak Processes k(L) ---\ensuremath{>} mu anti-mu, K+
  ---\ensuremath{>} pi+ Neutrino anti-neutrino and K0
  \ensuremath{<}---\ensuremath{>} anti-K0}},
  \href{http://dx.doi.org/10.1143/PTP.65.297}{\emph{Prog. Theor. Phys.} {\bf
  65} (1981) 297}.

\bibitem{Misiak:2006ab}
M.~Misiak and M.~Steinhauser, \emph{{NNLO QCD corrections to the anti-B $\to$
  X(s) gamma matrix elements using interpolation in m(c)}},
  \href{http://dx.doi.org/10.1016/j.nuclphysb.2006.11.027}{\emph{Nucl. Phys. B}
  {\bf 764} (2007) 62--82}, [\href{http://arxiv.org/abs/hep-ph/0609241}{{\tt
  hep-ph/0609241}}].

\bibitem{Barr:1990vd}
S.~M. Barr and A.~Zee, \emph{{Electric Dipole Moment of the Electron and of the
  Neutron}}, \href{http://dx.doi.org/10.1103/PhysRevLett.65.21}{\emph{Phys.
  Rev. Lett.} {\bf 65} (1990) 21--24}.

\bibitem{Pilaftsis:2002fe}
A.~Pilaftsis, \emph{{Higgs mediated electric dipole moments in the MSSM: An
  application to baryogenesis and Higgs searches}},
  \href{http://dx.doi.org/10.1016/S0550-3213(02)00826-X}{\emph{Nucl. Phys. B}
  {\bf 644} (2002) 263--289}, [\href{http://arxiv.org/abs/hep-ph/0207277}{{\tt
  hep-ph/0207277}}].

\bibitem{ACME:2018yjb}
{\scshape ACME} collaboration, V.~Andreev et~al., \emph{{Improved limit on the
  electric dipole moment of the electron}},
  \href{http://dx.doi.org/10.1038/s41586-018-0599-8}{\emph{Nature} {\bf 562}
  (2018) 355--360}.

\bibitem{Roussy:2022cmp}
T.~S. Roussy et~al., \emph{{An improved bound on the electron\textquoteright{}s
  electric dipole moment}},
  \href{http://dx.doi.org/10.1126/science.adg4084}{\emph{Science} {\bf 381}
  (2023) adg4084}, [\href{http://arxiv.org/abs/2212.11841}{{\tt 2212.11841}}].

\bibitem{Haber:2022gsn}
H.~E. Haber, V.~Keus and R.~Santos, \emph{{P-even, CP-violating signals in
  scalar-mediated processes}},
  \href{http://dx.doi.org/10.1103/PhysRevD.106.095038}{\emph{Phys. Rev. D} {\bf
  106} (2022) 095038}, [\href{http://arxiv.org/abs/2206.09643}{{\tt
  2206.09643}}].

\bibitem{Hagiwara:1986vm}
K.~Hagiwara, R.~D. Peccei, D.~Zeppenfeld and K.~Hikasa, \emph{{Probing the Weak
  Boson Sector in e+ e- ---\ensuremath{>} W+ W-}},
  \href{http://dx.doi.org/10.1016/0550-3213(87)90685-7}{\emph{Nucl. Phys. B}
  {\bf 282} (1987) 253--307}.

\bibitem{ATLAS:2017pbb}
{\scshape ATLAS} collaboration, M.~Aaboud et~al., \emph{{Measurement of $WW/WZ
  \to \ell \nu q q^{\prime}$ production with the hadronically decaying boson
  reconstructed as one or two jets in $pp$ collisions at $\sqrt{s}=8$ TeV with
  ATLAS, and constraints on anomalous gauge couplings}},
  \href{http://dx.doi.org/10.1140/epjc/s10052-017-5084-2}{\emph{Eur. Phys. J.
  C} {\bf 77} (2017) 563}, [\href{http://arxiv.org/abs/1706.01702}{{\tt
  1706.01702}}].

\bibitem{CMS:2019ppl}
{\scshape CMS} collaboration, A.~M. Sirunyan et~al., \emph{{Search for
  anomalous triple gauge couplings in WW and WZ production in lepton + jet
  events in proton-proton collisions at $\sqrt{s} =$ 13 TeV}},
  \href{http://dx.doi.org/10.1007/JHEP12(2019)062}{\emph{JHEP} {\bf 12} (2019)
  062}, [\href{http://arxiv.org/abs/1907.08354}{{\tt 1907.08354}}].

\bibitem{CMS:2020gtj}
{\scshape CMS} collaboration, A.~M. Sirunyan et~al., \emph{{Measurements of
  ${\mathrm{p}} {\mathrm{p}} \rightarrow {\mathrm{Z}} {\mathrm{Z}} $ production
  cross sections and constraints on anomalous triple gauge couplings at
  $\sqrt{s} = 13\,\text {TeV} $}},
  \href{http://dx.doi.org/10.1140/epjc/s10052-020-08817-8}{\emph{Eur. Phys. J.
  C} {\bf 81} (2021) 200}, [\href{http://arxiv.org/abs/2009.01186}{{\tt
  2009.01186}}].

\bibitem{Grzadkowski:2016lpv}
B.~Grzadkowski, O.~M. Ogreid and P.~Osland, \emph{{CP-Violation in the $ZZZ$
  and $ZWW$ vertices at $e^+e^-$ colliders in Two-Higgs-Doublet Models}},
  \href{http://dx.doi.org/10.1007/JHEP05(2016)025}{\emph{JHEP} {\bf 05} (2016)
  025}, [\href{http://arxiv.org/abs/1603.01388}{{\tt 1603.01388}}].

\bibitem{Feng:2017uoz}
J.~L. Feng, I.~Galon, F.~Kling and S.~Trojanowski, \emph{{ForwArd Search
  ExpeRiment at the LHC}},
  \href{http://dx.doi.org/10.1103/PhysRevD.97.035001}{\emph{Phys. Rev. D} {\bf
  97} (2018) 035001}, [\href{http://arxiv.org/abs/1708.09389}{{\tt
  1708.09389}}].

\bibitem{Drechsel:2018mgd}
P.~Drechsel, G.~Moortgat-Pick and G.~Weiglein, \emph{{Prospects for direct
  searches for light Higgs bosons at the ILC with 250 GeV}},
  \href{http://dx.doi.org/10.1140/epjc/s10052-020-08438-1}{\emph{Eur. Phys. J.
  C} {\bf 80} (2020) 922}, [\href{http://arxiv.org/abs/1801.09662}{{\tt
  1801.09662}}].

\bibitem{Kling:2022uzy}
F.~Kling, S.~Li, H.~Song, S.~Su and W.~Su, \emph{{Light Scalars at FASER}},
  \href{http://dx.doi.org/10.1007/JHEP08(2023)001}{\emph{JHEP} {\bf 08} (2023)
  001}, [\href{http://arxiv.org/abs/2212.06186}{{\tt 2212.06186}}].

\bibitem{Robens:2023bzp}
T.~Robens, \emph{{A short overview on low mass scalars at future lepton
  colliders -- LCWS23 proceedings}},  in \emph{{International Workshop on
  Future Linear Colliders}}, 7, 2023.
\newblock \href{http://arxiv.org/abs/2307.15962}{{\tt 2307.15962}}.

\bibitem{CMS:2018cyk}
{\scshape CMS} collaboration, A.~M. Sirunyan et~al., \emph{{Search for a
  standard model-like Higgs boson in the mass range between 70 and 110 GeV in
  the diphoton final state in proton-proton collisions at $\sqrt{s}=$ 8 and 13
  TeV}}, \href{http://dx.doi.org/10.1016/j.physletb.2019.03.064}{\emph{Phys.
  Lett. B} {\bf 793} (2019) 320--347},
  [\href{http://arxiv.org/abs/1811.08459}{{\tt 1811.08459}}].

\bibitem{Biekotter:2019kde}
T.~Biek\"otter, M.~Chakraborti and S.~Heinemeyer, \emph{{A 96 GeV Higgs boson
  in the N2HDM}},
  \href{http://dx.doi.org/10.1140/epjc/s10052-019-7561-2}{\emph{Eur. Phys. J.
  C} {\bf 80} (2020) 2}, [\href{http://arxiv.org/abs/1903.11661}{{\tt
  1903.11661}}].

\bibitem{Heinemeyer:2021mnz}
S.~Heinemeyer, C.~Li, F.~Lika, G.~Moortgat-Pick and S.~Paasch, \emph{{A 96 GeV
  Higgs Boson in the 2HDMS: $e^+e^-$ collider prospects}},  in
  \emph{{International Workshop on Future Linear Colliders}}, 5, 2021.
\newblock \href{http://arxiv.org/abs/2105.11189}{{\tt 2105.11189}}.

\bibitem{Heinemeyer:2021msz}
S.~Heinemeyer, C.~Li, F.~Lika, G.~Moortgat-Pick and S.~Paasch,
  \emph{{Phenomenology of a 96~GeV Higgs boson in the 2HDM with an additional
  singlet}}, \href{http://dx.doi.org/10.1103/PhysRevD.106.075003}{\emph{Phys.
  Rev. D} {\bf 106} (2022) 075003},
  [\href{http://arxiv.org/abs/2112.11958}{{\tt 2112.11958}}].

\bibitem{Biekotter:2022jyr}
T.~Biek\"otter, S.~Heinemeyer and G.~Weiglein, \emph{{Mounting evidence for a
  95 GeV Higgs boson}},
  \href{http://dx.doi.org/10.1007/JHEP08(2022)201}{\emph{JHEP} {\bf 08} (2022)
  201}, [\href{http://arxiv.org/abs/2203.13180}{{\tt 2203.13180}}].

\bibitem{Biekotter:2022abc}
T.~Biek\"otter, S.~Heinemeyer and G.~Weiglein, \emph{{Excesses in the low-mass
  Higgs-boson search and the ${W}$-boson mass measurement}},
  \href{http://dx.doi.org/10.1140/epjc/s10052-023-11635-3}{\emph{Eur. Phys. J.
  C} {\bf 83} (2023) 450}, [\href{http://arxiv.org/abs/2204.05975}{{\tt
  2204.05975}}].

\bibitem{Biekotter:2023jld}
T.~Biek\"otter, S.~Heinemeyer and G.~Weiglein, \emph{{The CMS di-photon excess
  at 95 GeV in view of the LHC Run 2 results}},
  \href{http://dx.doi.org/10.1016/j.physletb.2023.138217}{\emph{Phys. Lett. B}
  {\bf 846} (2023) 138217}, [\href{http://arxiv.org/abs/2303.12018}{{\tt
  2303.12018}}].

\bibitem{Biekotter:2023oen}
T.~Biek{\"o}tter, S.~Heinemeyer and G.~Weiglein, \emph{{95.4~GeV diphoton
  excess at ATLAS and CMS}},
  \href{http://dx.doi.org/10.1103/PhysRevD.109.035005}{\emph{Phys. Rev. D} {\bf
  109} (2024) 035005}, [\href{http://arxiv.org/abs/2306.03889}{{\tt
  2306.03889}}].

\bibitem{Cao:2019ofo}
J.~Cao, X.~Jia, Y.~Yue, H.~Zhou and P.~Zhu, \emph{{96 GeV diphoton excess in
  seesaw extensions of the natural NMSSM}},
  \href{http://dx.doi.org/10.1103/PhysRevD.101.055008}{\emph{Phys. Rev. D} {\bf
  101} (2020) 055008}, [\href{http://arxiv.org/abs/1908.07206}{{\tt
  1908.07206}}].

\bibitem{Benbrik:2022azi}
R.~Benbrik, M.~Boukidi, S.~Moretti and S.~Semlali, \emph{{Explaining the 96 GeV
  Di-photon anomaly in a generic 2HDM Type-III}},
  \href{http://dx.doi.org/10.1016/j.physletb.2022.137245}{\emph{Phys. Lett. B}
  {\bf 832} (2022) 137245}, [\href{http://arxiv.org/abs/2204.07470}{{\tt
  2204.07470}}].

\bibitem{Azevedo:2023zkg}
D.~Azevedo, T.~Biek\"otter and P.~M. Ferreira, \emph{{2HDM interpretations of
  the CMS diphoton excess at 95 GeV}},
  \href{http://dx.doi.org/10.1007/JHEP11(2023)017}{\emph{JHEP} {\bf 11} (2023)
  017}, [\href{http://arxiv.org/abs/2305.19716}{{\tt 2305.19716}}].

\bibitem{LEPHiggsWorkingGroupforHiggsbosonsearches:2001dnp}
{\scshape LEP Higgs Working Group for Higgs boson searches, OPAL, ALEPH,
  DELPHI, L3} collaboration, \emph{{Search for the standard model Higgs boson
  at LEP}},  in \emph{{2001 Europhysics Conference on High Energy Physics}}, 7,
  2001.
\newblock \href{http://arxiv.org/abs/hep-ex/0107029}{{\tt hep-ex/0107029}}.

\bibitem{McNamara:2002nk}
P.~A. McNamara and S.~L. Wu, \emph{{The Higgs particle in the standard model:
  Experimental results from LEP}},
  \href{http://dx.doi.org/10.1088/0034-4885/65/4/201}{\emph{Rept. Prog. Phys.}
  {\bf 65} (2002) 465--528}.

\bibitem{CMS:2022goy}
{\scshape CMS} collaboration, A.~Tumasyan et~al., \emph{{Searches for
  additional Higgs bosons and for vector leptoquarks in $\tau\tau$ final states
  in proton-proton collisions at $\sqrt{s}$ = 13 TeV}},
  \href{http://dx.doi.org/10.1007/JHEP07(2023)073}{\emph{JHEP} {\bf 07} (2023)
  073}, [\href{http://arxiv.org/abs/2208.02717}{{\tt 2208.02717}}].

\bibitem{Iguro:2022dok}
S.~Iguro, T.~Kitahara and Y.~Omura, \emph{{Scrutinizing the 95\textendash{}100
  GeV di-tau excess in the top associated process}},
  \href{http://dx.doi.org/10.1140/epjc/s10052-022-11028-y}{\emph{Eur. Phys. J.
  C} {\bf 82} (2022) 1053}, [\href{http://arxiv.org/abs/2205.03187}{{\tt
  2205.03187}}].

\bibitem{Crivellin:2023zui}
A.~Crivellin and B.~Mellado, \emph{{Anomalies in particle physics and their
  implications for physics beyond the standard model}},
  \href{http://dx.doi.org/10.1038/s42254-024-00703-6}{\emph{Nature Rev. Phys.}
  {\bf 6} (2024) 294--309}, [\href{http://arxiv.org/abs/2309.03870}{{\tt
  2309.03870}}].

\bibitem{ATLAS-CONF-2023-035}
{\scshape ATLAS} collaboration, \emph{{Search for diphoton resonances in the 66
  to 110 GeV mass range using 140 fb$^{-1}$ of 13 TeV $pp$ collisions collected
  with the ATLAS detector}},  tech. rep., CERN, Geneva, 2023.

\bibitem{ATLAS:2024itc}
{\scshape ATLAS} collaboration, G.~Aad et~al., \emph{{ATLAS searches for
  additional scalars and exotic Higgs boson decays with the LHC Run~2
  dataset}}, \href{http://dx.doi.org/10.1016/j.physrep.2024.09.002}{\emph{Phys.
  Rept.} {\bf 1116} (2025) 184--260},
  [\href{http://arxiv.org/abs/2405.04914}{{\tt 2405.04914}}].

\bibitem{Hahn:2006hr}
T.~Hahn, \emph{{Routines for the diagonalization of complex matrices}},
  \href{http://arxiv.org/abs/physics/0607103}{{\tt physics/0607103}}.

\end{thebibliography}\endgroup

\end{document}